\documentclass[preprint,aps,tightenlines,eqsecnum,floatfix,showpacs]{revtex4}
\usepackage{dcolumn}
\usepackage{bm}
\usepackage{color}
\usepackage{graphicx}
\usepackage{pstricks}
\begin{document}
\title{Electromagnetic two-body currents of one- and two-pion range}
\author{S.\ Pastore$^{\,{\rm a}}$, R.\ Schiavilla$^{\,{\rm a,b}}$, and J.L.\ Goity$^{\,{\rm b,c}}$}
\affiliation{
$^{\rm a}$\mbox{Department of Physics, Old Dominion University, Norfolk, VA 23529, USA}\\
$^{\rm b}$\mbox{Jefferson Lab, Newport News, VA 23606, USA}\\
$^{\rm c}$\mbox{Department of Physics, Hampton University, Hampton, VA 23668, USA}
}

\date{\today}

\begin{abstract}
Nuclear electromagnetic currents are derived in time-ordered perturbation
theory within an effective-field-theory framework including explicit nucleons,
$\Delta$ isobars, and pions up to one loop, or N$^3$LO.  The currents obtained
at N$^2$LO, {\it i.e.}~ignoring loop corrections, are used in a study
of neutron radiative captures on protons and deuterons at thermal energies,
and of $A$=2 and 3 nuclei magnetic moments.  The wave functions for $A$=2
are derived from solutions of the Schr\"odinger
equation with the Argonne $v_{18}$ (AV18) or CD-Bonn (CDB) potentials, while
those for $A$=3 are obtained with the hyperspherical-harmonics-expansion method
from a realistic Hamiltonian including, in addition to the AV18 or CDB two-nucleon,
also a three-nucleon potential.  With the strengths of the $\Delta$-excitation
currents occurring at N$^2$LO determined to reproduce the $n$-$p$ cross section
and isovector combination of the trinucleon magnetic moments, we find that the
cross section and photon circular polarization parameter, measured in $n$-$d$
and $\vec{n}$-$d$ processes, are underpredicted by theory, for example
the cross section by (11--38)\% as the cutoff is increased from 500 to 800 MeV.
A complete analysis of the results, in particular their large cutoff
dependence, is presented.
\end{abstract}

\pacs{12.39.Fe, 13.40.-f, 25.10.+s, 25.40.Lw}

\maketitle
\section{Introduction}
\label{sec:intro}

The present work is the first stage of a research program aimed at studying
electromagnetic observables of light nuclei ($A \leq 8$), and particularly
radiative capture processes in the three- and four-nucleon systems, within
a theoretical approach in which many-body electromagnetic current operators
derived in chiral effective field theory ($\chi$EFT)~\cite{Weinberg90,vanKolck94,Epelbaum98}
are used in transition matrix elements between nuclear wave functions obtained
from realistic Hamiltonians with two- and three-body potentials.  This ``hybrid''
approach was adopted in a recent calculation of the astrophysical factor for
the $p$-$p$ and $p$-$^3$He fusion reactions by weak capture at the keV energies
relevant in the interior of the Sun ~\cite{Park03}.

Neutron and proton radiative captures on $^2$H, $^3$H and $^3$He are
particularly challenging from the standpoint of nuclear few-body
theory.  This can be appreciated by comparing the measured values
for the cross sections of thermal neutron radiative capture on
$^1$H, $^2$H, $^3$He.  Their respective values in mb are: ($332.6\pm 0.7$)~\cite{Mughabghab81},
($0.508 \pm 0.015$)~\cite{Jurney82}, and ($0.055\pm 0.003$)~\cite{Wolfs89}.
Thus, in going from $A$=2 to 4 the cross section has dropped by
almost four orders of magnitude.  These processes are induced by
magnetic dipole transitions between the initial two-cluster state
in relative S-wave and the final bound state.  The $^3$H and $^4$He
wave functions, respectively $\Psi_3$ and $\Psi_4$, are approximately
eigenfunctions of the magnetic dipole operator ${\bm \mu}$, namely
$\mu_z  \Psi_3 \simeq \mu_p \Psi_3$ and $\mu_z  \Psi_4 \simeq 0$,
where $\mu_p$=2.793 n.m.~is the proton magnetic moment---the experimental
value of the $^3$H magnetic moment is 2.979 n.m, while $^4$He has no
magnetic moment.  These relations would be exact, if the $^3$H
and $^4$He wave functions were to consist of the symmetric S-wave
term only.  In fact, tensor components in the nuclear potentials
generate significant D-state admixtures, that partially spoil
this eigenstate property.  To the extent that it is approximately
satisfied, though, the matrix elements $\langle\Psi_3\!\mid\!\mu_z\!\mid\!\Psi_{1+2}\rangle$
and $\langle\Psi_4\!\mid\!\mu_z\!\mid\!\Psi_{1+3}\rangle$ vanish
due to orthogonality between the initial and final states.
This orthogonality argument fails in the case of the deuteron,
since then $\mu_z\Psi_2\simeq (\mu_p-\mu_n) \,\phi_2(S)\, \chi^0_0\, \eta^1_0\,$,
where $\chi^S_{M_S}$ and $\eta^T_{M_T}$ are two-nucleon spin and isospin
states, respectively.  The magnetic dipole operator can therefore
connect the large S-wave component $\phi_2({\rm S})$ of the deuteron
to a $T$=1 $^1$S$_0$ $n$-$p$ state---the orthogonality between the latter
and the deuteron follows from the orthogonality between their respective
spin-isospin states.

As a result of this suppression, the $n$-$d$, $p$-$d$, $n$-$^3$He, and
$p$-$^3$H radiative (as well as $p$-$^3$He weak) captures are very
sensitive to small components in the wave functions, particularly
the D-state admixtures generated by tensor forces, and to many-body
terms in the electromagnetic (and weak) current operators.

There have been in the past several calculations of these processes
in the conventional framework---referred to as the standard nuclear
physics approach (SNPA) in Ref.~\cite{Park03}---see~\cite{Carlson98}
and references therein.  Some of these studies, in particular the recent
ones of Ref.~\cite{Marcucci05}, have used accurate (essentially exact) bound
and continuum wave functions corresponding to realistic Hamiltonians, which
provide an excellent description of $A$=3 and 4 binding energies and radii,
as well as of a variety of low-energy scattering observables
(see~\cite{Kievsky08} and references therein).  The electromagnetic
current operator includes, in addition to the standard convection
and spin-magnetization terms of individual protons and neutrons, also
two- and three-body terms, constructed from, respectively, the two-
and three-nucleon potentials so as to satisfy exactly current conservation
(CC) with them.  The method by which this is achieved has been improved 
over the years~\cite{Schiavilla89}, and its latest implementation is discussed
at length in Ref.~\cite{Marcucci05}---for an alternative formulation, though,
see Ref.~\cite{Buchmann85}.  It is not unique, since obviously the CC
relation puts no constraints on the transverse component of the
current.  Nevertheless, it does generate two- and three-body terms,
whose behavior, particularly at short range, is consistent with that
of the corresponding potentials.  This behavior in the latter is ultimately
``determined'' by reproducing a set of experimental two- and three-nucleon
scattering data and binding energies.

These currents have been shown to provide a very satisfactory description
of a variety of electronuclear properties, including, for example, $n$-$p$
capture~\cite{Marcucci05} and deuteron photodisintegration at low
energy~\cite{Schiavilla05}, and magnetic moments of $A$=3-7 nuclei~\cite{Marcucci08}.
Yet, they lead to $\sim 10$\% overestimate of the experimental cross
section in $n$-$d$ capture~\cite{Marcucci05}.  The discrepancy between
theory and experiment increases to $\sim 60$\% in the case of the $n$-$^3$He
capture cross section~\cite{Schiavilla92}, although this earlier study,
in contrast to that of Ref.~\cite{Marcucci05}, is not based on
wave functions derived from the latest generation of potentials.  The one-body
(impulse-approximation or IA) term of the magnetic dipole operator
gives, respectively, only 45\% and 10\% of the $n$-$d$ and $n$-$^3$He
cross-section experimental values because of the suppression mentioned above.

Electromagnetic currents up to one loop corrections have been derived
in $\chi$EFT within the heavy-baryon formalism by Park {\it et al.}
in Ref.~\cite{Park96}.  These currents were used in hybrid calculations of
the $n$-$p$ capture cross section~\cite{Park96,Song07}, spin observables
in $\vec{n}$-$\vec{p}$ capture~\cite{Park00}, and, more recently, magnetic
moments of the deuteron and trinucleons~\cite{Song07}.

In the present work, we derive the electromagnetic currents by including,
in addition to nucleon and pion, also $\Delta$-isobar degrees of freedom.
Thus we assume that the $\Delta$-nucleon mass difference, just as the pion
mass, is of the same order as the low momentum scale generically indicated
by $Q$.  Formal expressions up to one loop are obtained in time-ordered
perturbation theory by employing non-relativistic Hamiltonians derived
from the chiral Lagrangian formulation of Refs.~\cite{Weinberg90,vanKolck94,Epelbaum98}.
The present study is similar to that of Ref.~\cite{Park96}, albeit it uses a
different formalism.  Various aspects of the calculations are discussed in
considerable detail.  However, a discussion of renormalization is not given here:
it is deferred to a later publication~\cite{Goity08}.  It is nonetheless opportune
to comment on it.  There are two stages of regularization necessary in the one
loop calculations: the first is the usual regularization of the one-loop corrections
to the potential and to the currents, and the second is the regularization necessary
for solving the Schr\"odinger equation and for the calculation of the current matrix
elements.  All this must be followed by corresponding renormalization procedures.
In the calculations to follow at next-to-next-to-leading order (N$^2$LO), the only
loop corrections needed are those of the one-body current, which can be absorbed
into the proton and neutron magnetic moments and electromagnetic radii.  Therefore
the required regularization is in the calculation of the matrix elements of the
two-body components of the current.  The latter is implemented as usual through
a short-range cutoff parameter and, although a full fledged  renormalization is not
carried out at this point, we expect that a choice of the cutoff equal to that in
the input potentials will give realistic values for the matrix elements.

These N$^2$LO currents are used to calculate the magnetic
moments of $A$=2 and 3 nuclei, and thermal neutron radiative captures on
protons and deuterons.  Realistic two- and three-nucleon (for $A$=3) potentials
are used to generate the bound and continuum wave functions.  To have an
estimate of the model dependence arising from short-range
phenomena, the variation of the predictions is studied as function
of the short-range cutoff parameter mentioned above, which is used to regularize
the two-body operators, as well as of the input potentials---either
the Argonne $v_{18}$ (AV18)~\cite{Wiringa95} or CD-Bonn (CDB)~\cite{Machleidt01}
in combination with respectively the Urbana IX~\cite{Pudliner97} or
Urbana IX$^*$~\cite{Viviani07}---used to generate the wave functions
(the AV18 and CDB have rather different short-range behaviors).  

We find that the N$^2$LO calculations do not provide a satisfactory
description of the experimental data, particularly for the suppressed
process $^2$H($n,\gamma$)$^3$H.  This clearly points to the need of
including loop corrections.  However, it remains an interesting question
whether these corrections will resolve the present discrepancies between
theory and experiment.

This paper is organized into eight sections and four appendices.  In
Sec.~\ref{sec:prelims}, we list, after defining our notation and
conventions, the relevant strong- and electromagnetic-interaction
Hamiltonians, obtained from chiral Lagrangians with nucleons, $\Delta$
isobars, and pions, while in Sec.~\ref{sec:tree} we derive the nuclear
electromagnetic current up to N$^2$LO, {\it i.e.}~ignoring loop corrections,
in momentum space, and give the configuration-space representation
of its operators in Sec.~\ref{sec:r-sp}.  Section~\ref{sec:loop} consists of
two subsections: the first contains a derivation of one-loop
two-body currents, while the second lists the two-body currents at N$^3$LO,
implied by four-nucleon contact Lagrangians involving two gradients.
In Sec.~\ref{sec:jcons} we show that the currents up to N$^3$LO are conserved
when used in combination with $\chi$EFT potential including corrections
up to one loop.  In Sec.~\ref{sec:calce}, we present
and discuss results for the magnetic moments of the deuteron and trinucleons,
and for the radiative captures of thermal neutrons on protons and deuterons.
Finally, in Sec.~\ref{sec:concls} we summarize our conclusions
and outline the next stage in the research program initiated here.  A number
of details are relegated in the Appendices, including: expressions for the
vertices associated with the interaction Hamiltonians of interest (Appendix~\ref{app:vert});
a collection of formulae relevant for the configuration-space representation
of the N$^2$LO operators (Appendix~\ref{app:fns}); a listing of the
analytical expressions for the one-loop currents involving $\Delta$-isobars
in the intermediate states (Appendix~\ref{app:l2pid}); and, lastly, a listing of
the four-nucleon contact Hamiltonians (Appendix~\ref{app:ct2d}).

\section{Preliminaries}
\label{sec:prelims}

Before listing the interaction Hamiltonians,
it is useful to define our notation and conventions.
In the Schr\"odinger picture adopted in the present study, the
isospin triplet of pion fields $\pi_a({\bf x})$ and their canonical
conjugates $\Pi_a({\bf x})$, with $a=x,y,z$, are represented as
\begin{eqnarray}
\pi_a({\bf x}) &=& \sum_{\bf p} \frac{1}{\sqrt{2\,\omega_p}}
\left[ c_{{\bf p},a}\, {\rm e}^{i {\bf p}\cdot {\bf x} } +{\rm h.c.}\right] \ , \\
\Pi_a({\bf x}) &=& \sum_{\bf p} \, i\, \sqrt{ \frac{\omega_p}{ 2 } }\left[
c_{{\bf p},a}\, {\rm e}^{i {\bf p}\cdot {\bf x} } - {\rm h.c.}\right] \ ,
\end{eqnarray}
where the annihilation and creation operators $c_{{\bf p},a}$
and $c^\dagger_{{\bf p},a}$ satisfy standard commutation
relations, and $\omega_p \equiv (p^2+m_\pi^2)^{1/2}$.  The nucleon and $\Delta$-isobar
fields, respectively $N({\bf x})$ and $\Delta({\bf x})$, 
with their corresponding canonical conjugates $i\, N^\dagger({\bf x})$
and $i\, \Delta^\dagger({\bf x})$, are given, in the non-relativistic limit
of interest here, as 
\begin{eqnarray}
 N({\bf x})&=&\sum_{{\bf p},\sigma\tau} b_{{\bf p},\sigma\tau} 
 \, {\rm e}^{i {\bf p}\cdot {\bf x} } \chi_{\sigma\tau} \ , \\
 \Delta({\bf x})&=&\sum_{{\bf p},\sigma_\Delta \tau_\Delta} d_{{\bf p},\sigma_\Delta \tau_\Delta} 
 \, {\rm e}^{i {\bf p}\cdot {\bf x} } \chi_{\sigma_\Delta \tau_\Delta} \ ,
\end{eqnarray}
where the $b$'s and $d$'s are annihilation operators for nucleons
and $\Delta$ isobars, and $\chi_{\sigma\tau}$ and
$\chi_{\sigma_\Delta \tau_\Delta}$ denote their respective
spin-isospin states.  Again, the $b$'s and $b^\dagger$'s,
and similarly the $d$'s and $d^\dagger$'s, satisfy the standard
anticommutation relations, appropriate for fermionic fields.

Standard time-ordered perturbation theory is used to calculate the
transition amplitude:
\begin{equation}
 \langle N^\prime N^\prime \mid T\mid NN;\gamma\rangle= 
 \langle N^\prime N^\prime \mid H_1 \sum_{n=1}^\infty \left( 
 \frac{1}{E_i -H_0 +i\, \eta } H_1 \right)^{n-1} \mid NN;\gamma \rangle \ ,
\label{eq:pt}
\end{equation}
where $\mid NN;\gamma\rangle$ ($\mid N^\prime N^\prime\rangle$) represents, in
a schematic notation, an initial (final) state containing two nucleons and a
photon (two nucleons only) of energy $E_i$ ($E_f=E_i$), $H_0$ is the Hamiltonian
describing free pions, nucleons and $\Delta$-isobars, and $H_1$ is the Hamiltonian
describing interactions among these particles as well as their couplings to the
electromagnetic field.  The evaluation of this amplitude is carried out in practice
by inserting complete sets of eigenstates of $H_0$ between successive terms of
$H_1$ in the expansion above.  However, since we are only interested in
retaining {\it irreducible} contributions, we omit from these intermediate
states those that contain {\it only} two nucleons (this aspect of the
present calculations is elaborated on in Secs.~\ref{sec:canc} and~\ref{sec:lp2}).
Thus, the nuclear electromagnetic current ${\bf j}$ of interest here is
obtained from
\begin{equation}
 \langle N^\prime N^\prime \mid T\mid NN;
\gamma\rangle|_{\rm irreducible} = -
 \frac{\hat{\bf e}_{{\bf q} \lambda}}{\sqrt{2\, \omega}_q} 
\cdot {\bf j} \ ,
\label{eq:irre}
\end{equation}
where ${\bf q}$, $\omega_q$, and $\hat{\bf e}_{{\bf q} \lambda}$ denote respectively
the photon momentum, energy, and polarization, and only first-order contributions
in the electromagnetic interaction are retained in the evaluation of the transition
amplitude.

\subsection{Pion, nucleon, and $\Delta$-isobar interaction Hamiltonians}
\label{sec:h-s}

The interaction Hamiltonians involving nucleon, $\Delta$-isobar, and
pion fields are derived from the effective chiral Lagrangian approach formulated
in Refs.~\cite{Weinberg90,vanKolck94}.  Their explicit expressions are given by 
\begin{eqnarray}
H_{\pi NN} &=& \frac{g_A}{F_\pi} \int {\rm d}{\bf x}\, 
N^\dagger({\bf x})\, \left[ {\bm \sigma} \cdot \nabla
 \pi_a({\bf x})\right]\, \tau_a\, N({\bf x}) \ , \\
H_{\pi N\Delta} &=& \frac{h_A}{F_\pi} \int {\rm d}{\bf x}\,
\Delta^\dagger({\bf x})\, \left[{\bf S} \cdot 
\nabla \pi_a({\bf x})\right]\, T_a \,N({\bf x}) 
+{\rm h.c.}\ , \\
H_{\pi\pi NN}&=&\frac{1}{F_\pi^2}\int {\rm d}{\bf x}\,
N^\dagger({\bf x})\,\left[{\bm \pi}({\bf x}) \times {\bm \Pi}({\bf x}) \right] 
 \cdot {\bm \tau}  N({\bf x}) \ ,
\end{eqnarray}
where $g_A \simeq 1.25 $ and $F_\pi \simeq 186$ MeV are the nucleon axial
coupling constant and pion decay amplitude, respectively, $h_A$ is the $\pi N\Delta$
coupling constant, and $S_\alpha$ and $T_a$ are transition spin and
isospin operators, converting a nucleon into a $\Delta$ isobar and
satisfying
\begin{equation}
S^\dagger_\alpha \, S_\beta =\frac{2}{3} \delta_{\alpha\beta} -
\frac{i}{3} \epsilon_{\alpha\beta\gamma} \sigma_\gamma \ ,
\label{eq:stran}
\end{equation}
and similarly for $T^\dagger_a \, T_b$ (note that in Ref.~\cite{vanKolck94}
the isospin transition operator is half that defined here). 

In addition to these, there is a set of four-fermion contact interactions
described by
\begin{eqnarray}
H_{{\rm CT},1} &=& \sum_{\alpha=S,T} \frac{C_\alpha}{2}
\int {\rm d}{\bf x}\, \left[N^\dagger({\bf x})\, {\bm \Gamma}_\alpha\, N({\bf x})\right]
\cdot \left[N^\dagger({\bf x}) {\bm \Gamma}_\alpha N({\bf x})\right] \ , \\
H_{{\rm CT},2} &=& D_T \int {\rm d}{\bf x}\,
\left[N^\dagger({\bf x}) {\bm \sigma}\,\tau_a N({\bf x})\right]
\cdot \left[\Delta^\dagger({\bf x}) {\bf S}\,T_a N({\bf x})\right] 
+ {\rm h.c.} \ , \\
H_{{\rm CT},3} &=& \sum_{\alpha=S,T} C^\prime_\alpha \int {\rm d}{\bf x}\,
\left[N^\dagger({\bf x})\, {\bm \Gamma}_\alpha\, N({\bf x})\right]
\cdot \left[\Delta^\dagger({\bf x})\, {\bm \Gamma}^\prime_\alpha \, 
\Delta({\bf x})\right]  \ , \\
H_{{\rm CT},4} &=& D^\prime_T \int {\rm d}{\bf x}\,
\left[\Delta^\dagger({\bf x}) {\bf S}\,T_a N({\bf x})\right]
\cdot \left[\Delta^\dagger({\bf x}) {\bf S}\,T_a N({\bf x})\right]
 + {\rm h.c.} \ , \\
H_{{\rm CT},5} &=& D^{\prime\prime}_T \int {\rm d}{\bf x}\,
\left[\Delta^\dagger({\bf x}) {\bf S}\,T_a N({\bf x})\right]
\cdot \left[N^\dagger({\bf x}) {\bf S}^\dagger \,T^\dagger_a 
\Delta({\bf x})\right] \ ,
\end{eqnarray}
where we have defined 
\begin{equation}
{\bm \Gamma}_S = {\bm \Gamma}^\prime_S = 1 \ , \quad {\bm \Gamma}_T = {\bm \sigma}\ , \qquad
{\bm \Gamma}_T^\prime = {\bf \Sigma} \ ,
\label{eq:cntc}
\end{equation}
and ${\bm \Sigma}/2$ is the spin-3/2 operator.  As it will become clear below,
terms involving more than two $\Delta$-isobars are not needed in
the present study.
Finally, when discussing the renormalization of the two-body currents at tree
level in a later work~\cite{Goity08}, we shall also need to consider the
following Hamiltonians involving three- and four-pion interactions,
\begin{equation}
 H_{3\pi} =-\frac{g_A}{F^3_\pi}\! \int \!{\rm d}{\bf x}\, {\bm \pi}^2({\bf x}) \, 
N^\dagger({\bf x})\, \left[ {\bm \sigma} \cdot \nabla
 \pi_a({\bf x})\right]\, \tau_a\, N({\bf x}) \ ,
 \end{equation}
 \begin{equation}
 H_{4\pi} = \frac{1}{2\, F_\pi^2}\! \int\! {\rm d}{\bf x}\,
 \Bigg[ \! \left[ {\bm \pi}^2({\bf x}) \, {\bm \Pi}^2({\bf x}) \! 
 -\! {\bm \pi}^2({\bf x}) \, \nabla \pi_a({\bf x}) \cdot \nabla \pi_a({\bf x})
 \! + \!{\rm h.c.} \right] 
  \!-\! m_\pi^2 \left[ {\bm \pi}^2({\bf x})\right]^2 \! \Bigg]\ , 
\end{equation}
obtained by including corrections up to ${\bm \pi}^2({\bf x})/F_\pi^2$
in the expansion of $D^{-1}$ factors, where $D \equiv 1 +{\bm \pi}^2({\bf x})/F_\pi^2$,
entering the chiral Lagrangians~\cite{Weinberg90}.
 
\subsection{Electromagnetic interactions}
\label{sec:h-e}

The charged pion field is defined as
\begin{equation}
\pi_{\mp}({\bf x}) = \frac{1}{\sqrt{2}} 
\left[ \pi_x({\bf x}) \mp i\, \pi_y({\bf x}) \right] \ ,
\end{equation}
and minimal substitution,
\begin{equation}
\nabla \pi_{\mp}({\bf x}) \rightarrow \left[ \nabla
\mp i\, e {\bf A}({\bf x}) \right] \pi_{\mp}({\bf x}) \ ,
\end{equation}
in the pion-derivative couplings leads to the interaction Hamiltonians:
\begin{eqnarray}
H_{\gamma \pi NN} &=& -e \frac{g_A}{F_\pi}\epsilon_{abz} \int {\rm d}{\bf x}\, 
{\bf A}({\bf x})\cdot N^\dagger({\bf x})\, {\bm \sigma} \,
\tau_a  N({\bf x}) \, \pi_b({\bf x})\ , \\
H_{\gamma \pi N\Delta}&=& -e \frac{h_A}{F_\pi}\epsilon_{abz} \int {\rm d}{\bf x}\, 
{\bf A}({\bf x})\cdot \Delta^\dagger({\bf x})\, {\bf S} \,
T_a  N({\bf x}) \, \pi_b({\bf x}) + {\rm h.c.} \ , \\
H_{\gamma\pi\pi} &=& -e \, \epsilon_{abz}\int {\rm d}{\bf x}\, {\bf A}({\bf x})\cdot
\left[\nabla \pi_a({\bf x})\right] \pi_b({\bf x})  \ ,\\
H_{\gamma \pi\pi NN} &=&-\frac{e}{2\, m_N} \frac{1}{F_\pi^2}\int {\rm d}{\bf x}\,
{\bf A}({\bf x})\cdot \Bigg[N^\dagger({\bf x}) \Big[  i\, 
( \overrightarrow{\nabla}-\overleftarrow{\nabla})
+{\bm \sigma}\times ( \overrightarrow{\nabla}+\overleftarrow{\nabla})
 \Big] \tau_a N({\bf x})\Bigg] \nonumber \\
&&\qquad\qquad\qquad\qquad\qquad\qquad \times
 \left[ \pi_a({\bf x}) \pi_z({\bf x})-\delta_{a,z} {\bm \pi}^2({\bf x}) \right] \ , \\
H_{\gamma\,3\pi} &=& e \frac{g_A}{F^3_\pi}\epsilon_{abz} \int {\rm d}{\bf x}\, 
{\bf A}({\bf x})\cdot N^\dagger({\bf x})\, {\bm \sigma} \,
\tau_a  N({\bf x}) \, \pi_b({\bf x})\, {\bm \pi}^2({\bf x}) \ , \\
H_{\gamma\, 4\pi} &=& e\frac{2}{F_\pi^2} \, \epsilon_{abz}\int {\rm d}{\bf x}\, {\bf A}({\bf x})\cdot
\left[\nabla \pi_a({\bf x})\right] \pi_b({\bf x}) {\bm \pi}^2({\bf x})  \ ,
\end{eqnarray}
where $e (> 0)$ is the electric charge, and the transverse vector field 
${\bf A}({\bf x})$ (in Coulomb gauge) is expanded as
\begin{equation}
 {\bf A}({\bf x}) =\sum_{\bf p}\sum_{\lambda=1,2} \frac{1}{\sqrt{2\omega_p}}
\left[ a_{{\bf p},\lambda }\, {\rm e}^{i {\bf p}\cdot {\bf x} }\,
\hat{\bf e}_{{\bf p},\lambda} +{\rm h.c.}\right] \ .
\end{equation}
The linear polarization (unit) vectors $\hat{\bf e}_{{\bf p},1}$,
$\hat{\bf e}_{{\bf p},2}$ form along with $\hat{\bf p}$ a right-handed
orthonormal system of axes, $\hat{\bf e}_{{\bf p},1} \times
\hat{\bf e}_{{\bf p},2}=\hat{\bf p}$.

The interactions of individual nucleons and $\Delta$-isobars with the electromagnetic
field are described by the following Hamiltonians:
\begin{eqnarray}
H_{\gamma NN} &=& \frac{e}{2\, m_N} \int {\rm d}{\bf x}\, 
N^\dagger({\bf x})\Bigg[i\, e_N 
\left[ -\overleftarrow{\nabla}\cdot {\bf A}({\bf x})
+ {\bf A}({\bf x}) \cdot \overrightarrow{\nabla} \right] \nonumber \\
&-&\mu_N \, 
{\bm \sigma}\cdot  \nabla \times {\bf A}({\bf x})  \Bigg] N({\bf x}) 
\label{eq:hgn} \ ,\\
H_{\gamma N\Delta}   &=&-\frac{e\, \mu^*}{2\, m_N}
\int {\rm d}{\bf x}\, \Delta^\dagger({\bf x})\, {\bf S} \cdot 
\left[ \nabla \times {\bf A}({\bf x}) \right] \, T_z \, N({\bf x}) +{\rm h.c.} \ ,
\label{eq:hgd}
\end{eqnarray}
with
\begin{equation}
e_N = (1+\tau_z)/2 \ , \qquad 
\kappa_N =  (\kappa_S+ \kappa_V \tau_z)/2 \ , \qquad \mu_N = e_N+\kappa_N \ ,
\label{eq:ekm}
\end{equation}
where $\kappa_S$ and $\kappa_V$ are the isoscalar and isovector combinations
of the anomalous magnetic moments of the proton and neutron ($\kappa_S =-0.12$ n.m.
and $\kappa_V =3.706$ n.m.), and $\mu^*$ is the $N$$\Delta$-transition magnetic
moment ($\mu^* \simeq 3$ n.m.).  In the $\gamma$$N$$\Delta$ term we only take into account
the dominant magnetic dipole (M1) coupling, ignoring the much smaller Coulomb (C2) and
electric quadrupole (E2) couplings.  The
expressions in Eqs.~(\ref{eq:hgn}) and (\ref{eq:hgd}) result from considering
the non-relativistic limit of the effective Hamiltonians with non-minimal couplings
\begin{eqnarray}
H^{\rm R}_{\gamma NN} &=& e \int {\rm d}{\bf x}\, \overline{\psi}_N({\bf x}) \left[
e_N \, A^\mu({\bf x}) \gamma_\mu + \frac{\kappa_N}{4\, m_N} 
\sigma_{\mu \nu} F^{\mu \nu}({\bf x}) \right] \psi_N({\bf x}) \ , 
\label{eq:hg1r}\\
H^{\rm R}_{\gamma N\Delta} &=&-i \frac{e\, \mu^*}{2\, m_N} \int {\rm d}{\bf x}\,
 \overline{\psi}_\Delta^{\,\mu}({\bf x}) \, g_{\mu \lambda} \gamma_\nu \gamma_5 \, T_z \, \psi_N({\bf x})\,
F^{\nu \lambda}({\bf x}) + {\rm h.c.} 
\label{eq:hg1dr}\ , 
\end{eqnarray}
where $\psi_N({\bf x})$ and $\psi_\Delta^\mu({\bf x})$ are the
spinor and spinor-vector fields describing the nucleon and $\Delta$ isobar, and
$F^{\nu \lambda}({\bf x})$ is the electromagnetic field tensor.  The Bjorken and Drell conventions~\cite{Bjorken64} are used for relativistic four-vectors, $\gamma$-matrices, 
and Dirac spin-1/2 spinors, except that the latter are taken to be normalized
as $u^\dagger({\bf p},s) u({\bf p},s)=1$.  The Rarita-Schwinger spin-3/2 spinors
are defined as
\begin{equation}
 u^\mu({\bf p},s_\Delta) =\sum_{\lambda s} \langle 1 \lambda, 1/2 s \mid 3/2 s_\Delta\rangle
 \, \epsilon^\mu({\bf p},\lambda) \, u({\bf p},s) \ , 
\end{equation}
where in the particle rest-frame the four-vector $\epsilon^\mu$ is space-like,
$\epsilon^\mu = (0,\hat{\bm \epsilon}_\lambda)$, and $\lambda=\pm1,0$ denote spherical
components.  

\subsection{Power counting}
\label{sec:count}

We denote generically by $Q$ a ``small momentum'', {\it i.e.} $Q \ll M$, where
$M \simeq 1$ GeV is the typical hadronic mass scale, and consider as effective degrees
of freedom the nucleon, $\Delta$-isobar, and pion.  Thus, we assume that the pion
mass and the mass difference between the $\Delta$-isobar and nucleon are both of order
$Q$, $m_\pi \sim Q$ and $m_\Delta-m_N \sim Q$.  However, the photon energy $\omega_q$
is assumed to be suppressed by an additional factor $Q/M$ relative to this
small-momentum scale, {\it i.e.} $\omega_q \sim Q^2/M$.  A generic coupling
constant of dimensions (energy)$^\alpha$ is assumed to scale with $M$ as
$g=\widetilde{g}\, M^\alpha$ with the expectation that $\widetilde{g} \simeq 1$. 

Contributions to the transition amplitude in Eq.~(\ref{eq:irre}) can be organized
as an expansion in powers of $Q/M$~\cite{Weinberg90}.  The power counting implied
by the interaction Hamiltonians in the previous two sections can be easily inferred
by examining the structure of their associated vertices, listed in Appendix~\ref{app:vert}.
This power counting is summarized in Table~\ref{tb:tab1}.
\begin{table}[bth]
\begin{tabular}{c|c||c|c}
\hline
\hline
      &  $Q$-scaling  &       & $Q$-scaling \\
\hline
$H_{\pi NN}$ & $Q$ & $H_{\gamma \pi NN}$ & $e\,Q^0$ \\
$H_{\pi N\Delta}$ & $Q$ & $H_{\gamma \pi N\Delta}$ & $e\,Q^0$ \\
$H_{\pi\pi NN}$   & $Q$ & $H_{\gamma \pi \pi}$ & $e\,Q$\\
$H_{{\rm CT},1-5}$ & $Q^0$ & $H_{\gamma \pi\pi NN}$ & $e\,Q$\\
$H_{3\pi}$  &   $Q$ &  $H_{\gamma\, 3\pi}$ & $e\,Q^0$ \\
$H_{4\pi}$ & $Q^2$ & $H_{\gamma\, 4\pi}$ & $e\,Q$\\
& & $H_{\gamma NN}$ & $e\, Q$ \\
& & $H_{\gamma N\Delta}$ & $e\, Q$ \\
\hline
\hline
\end{tabular}
\caption{Powers of $Q$, the small momentum scale, associated with the vertices
from the strong- and electromagnetic-interaction Hamiltonians of
Secs.~\protect\ref{sec:h-s}--\protect\ref{sec:h-e}.}
\label{tb:tab1}
\end{table}

In the perturbative series, Eq.~(\ref{eq:pt}), a generic irreducible contribution
will be characterized by a certain number, say $N$, of vertices, each scaling as
$Q^{\alpha_i}\times Q^{-\beta_i/2}$ ($i$=$1,\dots,N$), where $\alpha_i$ is
the power counting in Table~\ref{tb:tab1} and $\beta_i$ is the number of
pions in and/or out of the vertex (this last factor is associated with the
$1/\sqrt{2\, \omega_p}$ included in the pion field), a corresponding $N$--1 number of energy
denominators, and possibly $L$ loops.  Each of the energy denominators
will involve pion energies and/or $\Delta$$N$ mass differences, both of
order $Q$, as well as kinetic energies of nucleons and/or $\Delta$ isobars,
which, however, are suppressed by a further $Q/M$ factor relative to $Q$. 
Loops on the other hand will contribute a factor $Q^3$ each, since they involve
integrations over intermediate three momenta.  Hence the power counting
associated with such a contribution is 
\begin{equation}
{\rm irreducible\,\, contribution}=\left(\prod_{i=1}^N  Q^{\alpha_i-\beta_i/2}
\right)\times Q^{-(N-1)}
\times Q^{3L} \ .
\label{eq:count}
\end{equation}
When one expands the nucleon propagators in powers of the kinetic energy around
the static limit, terms with higher powers of $Q$ appear as the kinetic energy is order $Q^2$.

The power counting of Eq.~(\ref{eq:count}) can also be obtained by considering the
Feynman diagram where the loop integrals are carried out in four dimensions.
The power of $Q$ of an irreducible diagram is then given by $4 \, L-2 \, n_\pi-
n_N+\sum_{i=1}^{N} \alpha_i$, where $n_\pi$ is the number of pion propagators,
and $n_N$ the number of nucleon propagators.  This equation together with the 
topological relation $n_\pi+n_N=L+N-1$ leads to the power counting of Eq.~(\ref{eq:count}).

\begin{figure}[bthp]
\centerline{
\includegraphics[width=4.7in]{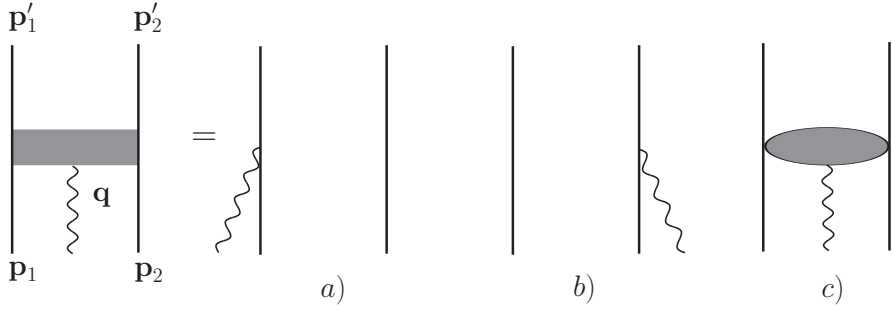}}
\caption{Schematic representation of the disconnected, a) and b), and connected, c),
contributions to the $NN\gamma \rightarrow NN$ amplitude.
Solid and wavy lines denote nucleons and photons, respectively.}
\label{fig:fig1}
\end{figure}

Finally, the transition amplitude in Eq.~(\ref{eq:irre}) can be represented
diagrammatically as in Fig.~\ref{fig:fig1}.  The disconnected contributions
in panels a) and b) will each contain a $\delta$-function in the initial
and final three momenta of one of the two particles, for example panel a)
$\propto \delta({\bf p}_2^\prime -{\bf p}_2)$, and will therefore be
enhanced by a factor $Q^{-3}$ relative to the connected (and irreducible)
contributions in panel c).  The power counting of diagrams a) and b)
is then $e\, Q \times Q^{-3}=e\, Q^{-2}$.  In fact, these diagrams
are the leading contributions to the nuclear electromagnetic current.
This fact certainly fits in well with what is known empirically about, for
example, magnetic moments of nuclei or radiative captures, such as
the $^1$H$(n,\gamma)$$^2$H process considered later in this work. 
\section{Currents up to N$^2$LO}
\label{sec:tree}

\begin{figure}[bthp]
\centerline{
\includegraphics[width=4.8in]{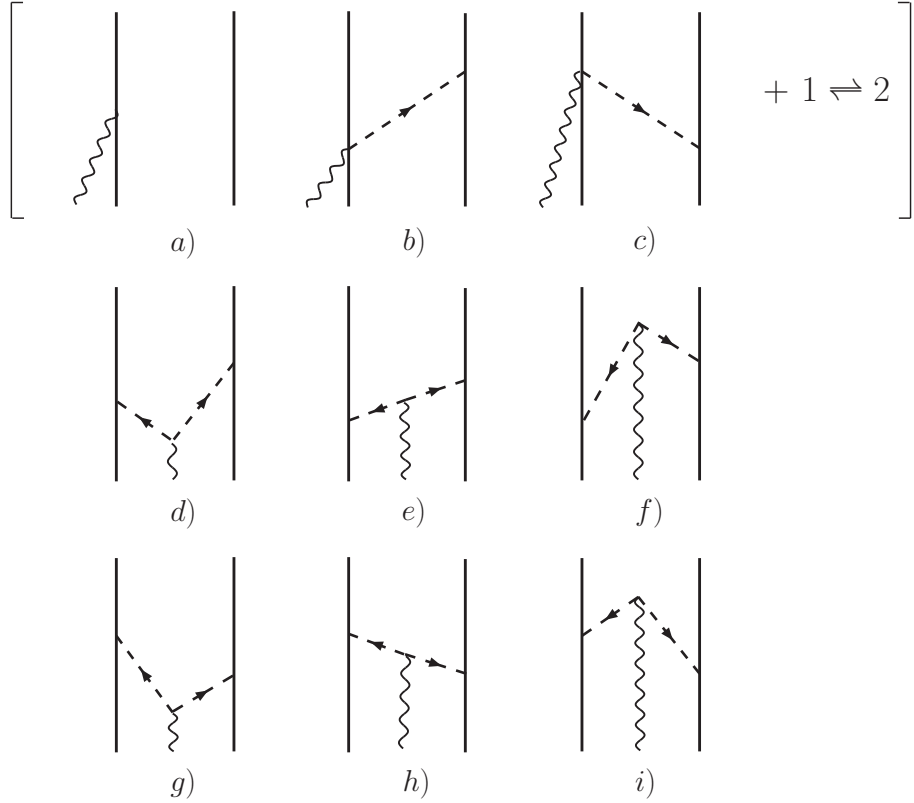}}
\caption{The complete set of time-ordered diagrams contributing at LO and NLO.  Nucleons, pions, and
photons are denoted by solid, dashed, and wavy lines, respectively.}
\label{fig:fig2}
\end{figure}

\begin{figure}[thp]
\centerline{
\includegraphics[width=4.8in]{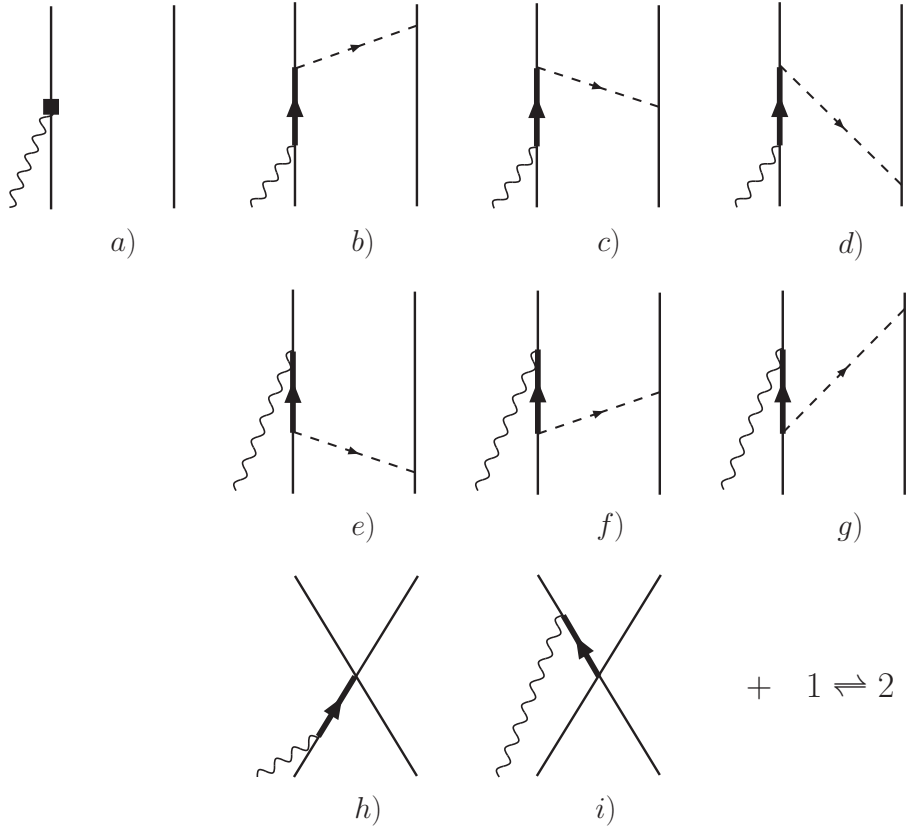}}
\caption{The complete set of time-ordered diagrams contributing at N$^2$LO.
The square represents the $(Q/M)^2$ correction to the LO one-body current,
while $\Delta$ isobars are denoted by thick solid lines, otherwise the notation
is as in Fig.~\protect\ref{fig:fig2}.}
\label{fig:fig3}
\end{figure}

In this section we derive the nuclear electromagnetic currents up to
next-to-next-to-leading order (N$^2$LO), that is $e\, Q^0$.  The relevant
contributions are illustrated in Figs.~\ref{fig:fig2} and~\ref{fig:fig3}.
The expressions below are given in momentum space, configuration-space
representations are discussed in Sec.~\ref{sec:r-sp}.  The momenta are
defined as
\begin{equation}
{\bf k}_i = {\bf p}_i^\prime -{\bf p}_i \ , \qquad {\bf K}_i =({\bf p}_i^\prime +{\bf p}_i)/2 \ ,
\label{eq:defs}
\end{equation}
where ${\bf p}_i$ and ${\bf p}_i^\prime$ are the initial and final momenta of nucleon $i$.
The leading order (LO), $e\, Q^{-2}$, is given by the one-body current,
panel a) in Fig.~\ref{fig:fig2},
\begin{equation}
{\bf j}^{\rm LO}_{\rm a}=\frac{e}{2\, m_N}
\Big[ 2\, e_{N,1} \, {\bf K}_1 +i\,\mu_{N,1}\, {\bm \sigma}_1\times {\bf q }\Big] +
 1 \rightleftharpoons 2\ ,
 \label{eq:jlo}
\end{equation}
where ${\bf q}$ is the photon momentum, ${\bf q}={\bf k}_i$.  

The contributions to next-to-leading order (NLO), $e\, Q^{-1}$, are
represented by diagrams b)-c) and d)-i) in Fig.~\ref{fig:fig2}.
A straightforward evaluation of these diagrams in the static limit
leads to the expressions:
\begin{eqnarray}
 {\bf j}^{\rm NLO}_{\rm b-c}&=& -i\, e\frac{g^2_A}{F^2_\pi}\,
 ({\bm \tau}_1 \times {\bm \tau}_2)_z \frac{1}{k_2^2+m_\pi^2}  
 {\bm \sigma}_1 \, ({\bm \sigma}_2\cdot {\bf k}_2)+ 1 \rightleftharpoons 2 \ , 
 \label{eq:nlo1} \\
  {\bf j}^{\rm NLO}_{\rm d-i}&=& i\, e\frac{g^2_A}{F^2_\pi}\,
 ({\bm \tau}_1 \times {\bm \tau}_2)_z 
\frac{{\bf k}_1-{\bf k}_2}{(k^2_1+m_\pi^2) (k_2^2+m_\pi^2)}
  ({\bm \sigma}_1\cdot {\bf k}_1)\, ({\bm \sigma}_2\cdot {\bf k}_2) \ ,
\label{eq:nlo2}
\end{eqnarray}
where the momenta transferred to nucleons 1 and 2 add up to ${\bf q}$, 
${\bf k}_1+{\bf k}_2={\bf q}$.

\begin{figure}[thp]
\centerline{
\includegraphics[width=2.2in]{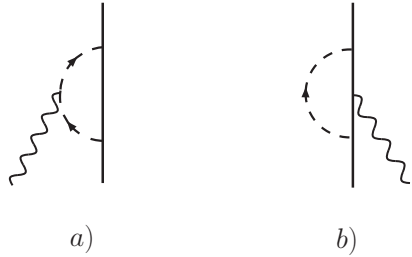}
}
\caption{NLO and N$^2$LO corrections to the one-body current.
Notation is as in Fig.~\protect\ref{fig:fig2}.}
\label{fig:fjose}
\end{figure}

At next-to-next-to-leading order (N$^2$LO), $e\, Q^0$, there
are two distinct contributions, illustrated in Fig.~\ref{fig:fig3}.
The first is due to $(Q/M)^2$ corrections to the one-body current
in Eq.~(\ref{eq:jlo}).  These are easily derived from a non-relativistic
expansion of Eq.~(\ref{eq:hg1r}):
\begin{eqnarray}
 {\bf j}^{\rm N^2LO}_{\rm a}=&-&\frac{e}{8 \, m_N^3}
 e_{N,1}\, \Bigg[ 
2\, \left( K_1^2 +q^2/4 \right) 
 \left( 2\, {\bf K}_1+i\, {\bm \sigma}_1\times {\bf q } \right)
+ {\bf K}_1\cdot {\bf q}\,
 \left({\bf q} +2i\, {\bm \sigma}_1\times {\bf K }_1 \right)\Bigg]
 \nonumber \\ 
 &-& \frac{i\,e}{8 \, m_N^3}
 \kappa_{N,1}\, \Bigg[ {\bf K}_1\cdot {\bf q}\, 
 \left( 4\, {\bm \sigma}_1\times {\bf K}_1 -i\, {\bf q}\right) 
 - \left(  2\, i\, {\bf K}_1 -{\bm \sigma}_1\times {\bf q} \right)\, q^2/2 \nonumber \\
 && \qquad\qquad \qquad +2\, \left({\bf K}_1\times {\bf q}\right) \, {\bm \sigma}_1\cdot {\bf K}_1
 \Bigg] + 1 \rightleftharpoons 2  \ .
\label{eq:j1rc}
\end{eqnarray}
At this point we should comment on the one-loop corrections to the one-body
current, see Fig.~\ref{fig:fjose}.  They occur at NLO and N$^2$LO, and are
absorbed into the anomalous magnetic moments and electromagnetic radii
of the proton and neutron.

The second class of N$^2$LO contributions, represented by diagrams b)-i),
involve $\Delta$-isobar excitation, and therefore vertices from $H_{\gamma N\Delta}$,
$H_{\pi N\Delta}$, and $H_{\rm CT, 2}$.  In the static limit, we find:
\begin{eqnarray}
 {\bf j}^{\rm N^2LO}_{\rm b-g}&=& i\,\frac{ e\, \mu^*}{9 \, m_N}\,
 \frac{g_A\, h_A}{\Delta\, F_\pi^2} \frac{{\bm \sigma}_2 \cdot {\bf k}_2}{k_2^2+m_\pi^2} 
 \Big[ 4\, \tau_{2,z}\, {\bf k}_2-({\bm \tau}_1\times {\bm \tau}_2)_z 
 \,{\bm \sigma}_1 \times{\bf k}_2\Big]
 \times {\bf q} + 1 \rightleftharpoons 2 \ , 
\label{eq:nlo2d}\\
 {\bf j}^{\rm N^2LO}_{\rm h-i}&=& -i\,\frac{ e\, \mu^*}{9 \, m_N}\, \frac{D_T}{\Delta} 
  \Big[ 4\, \tau_{2,z}\, {\bm \sigma}_2-
 ({\bm \tau}_1\times {\bm \tau}_2)_z\, 
 {\bm \sigma}_1 \times{\bm \sigma}_2 \Big]\times {\bf q} + 1 \rightleftharpoons 2\ ,
\end{eqnarray}
where $\Delta$ is the $\Delta$-$N$ mass difference,
$\Delta$=$m_\Delta -m_N$, and use has been made of the
identities in Eq.~(\ref{eq:stran}) to eliminate spin-
and isospin-transition operators in favor of Pauli
spin and isospin matrices.

We conclude this section by noting that the expressions in
Eqs.~(\ref{eq:nlo1})--(\ref{eq:nlo2}) and~(\ref{eq:nlo2d})
are the well known pion seagull and in-flight, and $\Delta$-excitation
currents commonly used in the literature (see, for example, the classic
work of Ref.~\cite{Riska72}).

\subsection{Recoil corrections: cancellations at N$^2$LO}
\label{sec:canc}

\begin{figure}[bthp]
\centerline{
\includegraphics[width=4.8in]{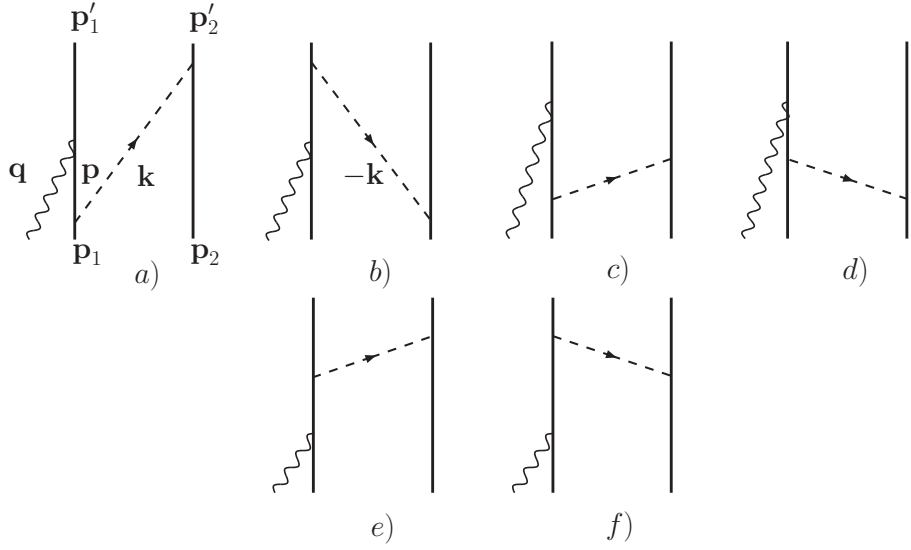}}
\caption{Time-ordered diagrams illustrating the cancellation of the irreducible
contributions a) and b) by the recoil corrections to the LO diagrams c)-f).  Notation
is as in Fig.~\protect\ref{fig:fig2}. }
\label{fig:fig4}
\end{figure}

In the present formulation based on time-independent
perturbation theory, there are in principle the additional N$^2$LO 
contributions represented by diagrams a) and b) in Fig.~\ref{fig:fig4}.
However, these are exactly canceled by recoil corrections, also entering
at N$^2$LO, to the reducible diagrams c)-f).  For example, the contribution
of diagrams c)+d) is given by
\begin{eqnarray}
\label{eq:recoil1}
 {\rm c)+d)}&=&\frac{V_{\gamma NN}(1)\, V_{\pi NN}(1)\,  V_{\pi NN}(2)}
{E_i\!-\!E_p-\!E_2^{\,\prime}\! -\!\omega_q+i\eta}\Bigg[\frac{1}
{E_i\!-\!E_p\!-E_2\! -\omega_q\!-\omega_k+i\, \eta}\nonumber \\
&&\qquad \qquad +\frac{1}{E_i\!-\!E_1\!-\! E^{\, \prime}_2\! -\omega_q\!-\omega_k+i\, \eta} \Bigg] \ ,
\end{eqnarray}
where the $V(i)$'s denote the vertices from the
interaction Hamiltonians relative to nucleon $i$, and the labeling
of momenta is as illustrated in the figure.  The initial and
final energies $E_i$ and $E_f$ ($E_i=E_f$) are $E_i$=$E_1+E_2+\omega_q$ 
and $E_f$=$E_1^{\, \prime}+E_2^{\, \prime}$, while $E_p$ is the energy of
the intermediate nucleon.  These energies are all suppressed by $Q/M$
relative to $\omega_k \sim Q$, and therefore the denominators in square brackets
can be expanded as
\begin{equation}
 \Bigg[ \dots \Bigg] \simeq  -\frac{2}{\omega_k} - \frac{E_i-E_p-E_2^{\,\prime} 
 -\omega_q}{\omega_k^2} \ ,
\end{equation}
so that the contribution of diagrams c+d) now reads
\begin{equation}
 {\rm c)+d)} = V_{\gamma NN}(1) \frac{1}{E_i\!-\!E_p-\!E_2^{\,\prime}\! -\!\omega_q+i\, \eta}
v^\pi({\bf k})\left[ 1+ \frac{E_i\!-\!E_p-\!E_2^{\,\prime}\! -\!\omega_q}{2\,\omega_k}\right] \ ,
\label{eq:iter}
\end{equation}
where $v^\pi({\bf k})$ is the static one-pion-exchange potential
(OPEP), defined as
\begin{equation}
 v^\pi({\bf k}) \equiv -\frac{2}{\omega_k}\, V_{\pi NN}(1)\, 
 V_{\pi NN}(2)  = -\frac{g^2_A}{F_\pi^2} 
 \frac{{\bm \sigma}_1 \cdot {\bf k}\,\, {\bm \sigma}_2 \cdot {\bf k}}{k^2+m_\pi^2}
 {\bm \tau}_1 \cdot {\bm \tau}_2 \ .
\label{eq:vpi}
\end{equation}
The second term on the r.h.s.~exactly cancels the contribution due to
(the irreducible) diagram a) in Fig.~\ref{fig:fig4}, as can be easily
surmised by noting that, in the static limit, the two energy denominators
occurring in this diagram are each given by $(-1/\omega_k)$, and therefore
\begin{equation}
 {\rm a)} = -\frac{V_{\gamma NN}(1)\, v^\pi({\bf k})}{2\, \omega_k}  \ .
\end{equation}

The discussion above becomes more delicate when the intermediate states describe fully
interacting rather than free particles.  Let $\delta v^\pi$ be the recoil correction to the
static OPEP $v^\pi$, and let $\mid\! \!\varphi\rangle$ denote a bound or continuum state
corresponding to $H_0+v^\pi$ with energy $E$.  To first order in $\delta v^\pi$, the
perturbed state $\mid\! \psi \rangle$ is
\begin{equation}
 \mid \!\psi \rangle= \mid\! \varphi \rangle +
\frac{1}{E-H_0-v^\pi}\delta v^\pi\mid\! \varphi\rangle 
\equiv \mid\! \varphi \rangle + \mid \! \delta \varphi \rangle \ ,
\end{equation}
and the matrix element of the current operator ${\bf j}$ between initial and final states
$\mid\! \psi_i\rangle$ and $\mid\! \psi_f\rangle$ can be expressed as
\begin{eqnarray}
\langle \psi_f\mid {\bf j}\mid \psi_i\rangle&=& \langle \varphi_f \mid {\bf j}\mid \varphi_i\rangle
+\langle \varphi_f\mid {\bf j}^{\rm LO}\mid \delta\varphi_i\rangle+\langle 
\delta\varphi_f\mid {\bf j}^{\rm LO}\mid \varphi_i\rangle \nonumber \\
&=&\langle \varphi_f\mid {\bf j}\mid \varphi_i\rangle+
\langle\varphi_f\mid {\bf j}^{\rm LO}\frac{1}{E_i-H_0-v^{\pi}}\delta v^{\pi}+{\rm h.c.}\mid\varphi_i\rangle \ , 
\end{eqnarray}
where we have dropped terms of order ($\delta\varphi$)$^2$ as well as corrections beyond LO to the current
in the matrix elements between $\mid \!\varphi_{i,f}\rangle$ and $\mid \!\delta\varphi_{f,i}\rangle$.  In the
analysis following Eq.~(\ref{eq:recoil1}), in which the nucleonic intermediate states are free particles,
the recoil correction can be written as
\begin{equation}
 \delta v^{\pi}\! \mid_{\rm free}=(E_i-H_0)\frac{v^{\pi}}{2\,\omega} + {\rm h.c.} \ ,
\end{equation}
where $\omega$ is the pion energy.  In momentum space the expression for $\delta v^{\pi}\!\!\mid_{\rm free}$
coincides with that implied by Eq.~(\ref{eq:iter}), namely
\begin{equation}
 \delta v^{\pi}({\bf k},{\bf K}_1,{\bf K}_2)\mid_{\rm free} =
\frac{v^{\pi}({\bf k})}{2\,\omega_k}\,\frac{{\bf k}\cdot({\bf K}_1-{\bf K}_2)}{m_N}\ ,
\end{equation}
with ${\bf K}_{\it i}$ defined as in Eq.~(\ref{eq:defs}).

If we assume that the nucleonic intermediate states describe fully interacting particles,
{\it i.e.}~they are eigenstates of $H_0+v^{\pi}$, then it is plausible that the
correction $\delta v^{\pi}$ should be expressed as
\begin{equation}
 \delta v^{\pi}=(E_i -H_0-v^{\pi})\frac{v^{\pi}}{2\,\omega} + {\rm h.c.} \ ,
\end{equation}
from which it follows that
\begin{equation}
\langle \psi_f \mid {\bf j}\mid \psi_i \rangle = \langle \varphi_f \mid {\bf j}^{\rm LO}+{\bf j}^{\rm NLO}-\left(
\frac{v^{\pi}}{2\, \omega}\,{\bf j}^{\rm LO}+{\rm h.c.} \right)\mid \varphi_i\rangle
+\langle\varphi_f \mid \frac{v^{\pi}}{2\, \omega}\,{\bf j}^{\rm LO}+{\rm h.c.}\mid\varphi_i\rangle \ . 
\end{equation}
The last two terms exactly cancel the two-body current contribution represented in 
Fig.~\ref{fig:fig4} panels a)-b), namely the terms in brackets.  Thus, if OPEP is taken in the static limit, Eq.~(\ref{eq:vpi}),
as is the case for the calculations reported below, then the contributions of diagrams a) and b)
should not be retained, since they are canceled by recoil corrections to OPEP.

\section{Currents in configuration space}
\label{sec:r-sp}

The calculations of electromagnetic observables reported in Sec.~\ref{sec:calce}
are carried out in configuration space, and hence configuration-space representations
of the current operators are needed.  Those of the one-body operators in Eqs.~(\ref{eq:jlo})
and~(\ref{eq:j1rc}), generically denoted as ${\bf j}^{(1)}$, follow from
\begin{equation}
{\bf j}^{(1)}({\bf q}) = \int_{{\bf k}_1} \int_{{\bf K}_1} 
{\rm e}^{i {\bf k}_1\cdot ({\bf r}^\prime_1+{\bf r}_1)/2}\,
{\rm e}^{i {\bf K}_1\cdot ({\bf r}^\prime_1-{\bf r}_1)}\,\,
\overline{\delta}({\bf k}_1-{\bf q})\,\,
{\bf j}^{(1)}({\bf k}_1,{\bf K}_1) \ ,
\end{equation}
while those for the two-body current operators ${\bf j}^{(2)}$
are derived from
\begin{equation}
 {\bf j}^{(2)}({\bf q}) = \int_{{\bf k}_1}\int_{{\bf k}_2}
{\rm e}^{i {\bf k}_1\cdot {\bf r}_1}\,
{\rm e}^{i {\bf k}_2\cdot {\bf r}_2}\,\,
\overline{\delta}({\bf k}_1+{\bf k}_2-{\bf q})\,\,
{\bf j}^{(2)}({\bf k}_1,{\bf k}_2) \ ,
\end{equation}
where the momenta ${\bf k}_i$ and ${\bf K}_i$ are defined 
as in Eq.~(\ref{eq:defs}), and we have introduced the notation
\begin{equation}
 \int_{{\bf p}} \equiv \int\frac{{\rm d}{\bf p}}{(2\pi)^3} \ ,
\qquad \overline{\delta}(\dots) \equiv (2\pi)^3\delta(\dots) \ .
\end{equation}
Note that ${\bf K}_i \rightarrow -i \nabla_i^\prime 
\delta({\bf r}_i^\prime-{\bf r}_i)$, {\it i.e.}
the configuration-space representation of the momentum
operator.  The LO current is then found to have the standard
expression associated with the nucleon's convection and
spin-magnetization currents,
\begin{equation}
{\bf j}_{\rm a}^{\rm LO}({\bf q})=\frac{e}{2\, m_N}
\Big( e_{N,1} \, \left[ -i \nabla_1\, ,\,  {\rm e}^{i {\bf q}\cdot {\bf r}_1}\right]_+
+i\,\mu_{N,1}\, {\bm \sigma}_1\times {\bf q }\,\,
{\rm e}^{i {\bf q}\cdot {\bf r}_1} \Big) +
 1 \rightleftharpoons 2\ ,
 \label{eq:jlor}
\end{equation}
where $\left[\dots \, , \, \dots\right]_+$ denotes the anticommutator---this notation
is also used in Eq.~(\ref{eq:j1rcc}) below.

At NLO and N$^2$LO, however, the operators have $1/r^2$ and $1/r^3$ singularities
($r$ is the interparticle separation), which need to be regularized in order
to avoid divergencies in the matrix elements of these operators between
nuclear wave functions.  We adopt a simple regularization procedure~\cite{vanKolck94},
{\it i.e.}~a momentum-space cutoff.  While its precise functional form is arbitrary,
the choice made here of a Gaussian cutoff function,
\begin{equation}
C_\Lambda(p) = {\rm e}^{-(p/\Lambda)^2} \ ,
\end{equation}
with the parameter $\Lambda \leq M \simeq 1$ GeV, is merely dictated
by convenience, since it leads to analytical expressions for the Fourier
transforms below.  It is expected that this arbitrariness be of little
relevance, since the dependence of theoretical predictions on variations
in the cutoff is (or should be) largely removed by a renormalization of the
theory free parameters, which are fixed by reproducing a given set of
observables.  We shall return to this issue later, in
Secs.~\ref{sec:calce} and~\ref{sec:concls}.

The two-body currents at NLO are obtained as
\begin{eqnarray}
{\bf j}^{\rm NLO}_{\rm b-c}({\bf q})\!&=&\! \frac{e\, g_A^2}{F^2_\pi}
\, \left({\bm \tau}_1 \times {\bm \tau}_2\right)_z\, {\rm e}^{i\,{\bf q}\cdot {\bf r}_1}
\, {\bm \sigma}_1 \, \left({\bm \sigma}_2 \cdot \nabla\right) f_\Lambda(r)   
+ 1 \rightleftharpoons 2  
\label{eq:nlor1} \ , \\
{\bf j}^{\rm NLO}_{\rm d-i}({\bf q})\!&=&\! \frac{e\, g_A^2}{F^2_\pi}\,
\left({\bm \tau}_1 \times {\bm \tau}_2\right)_z\, {\rm e}^{i\,{\bf q}\cdot {\bf R}}
\left[{\bm \sigma}_1 \cdot (\nabla + i{\bf q}/2)\right] \,
\left[{\bm \sigma}_2 \cdot (\nabla -i{\bf q}/2)\right] \nabla g_\Lambda({\bf r},{\bf q}) \ ,
\label{eq:nlor2}
\end{eqnarray}
where ${\bf r}$ and ${\bf R}$ denote the relative and center-of-mass position
vectors, respectively ${\bf r}={\bf r}_1 -{\bf r}_2$ and
${\bf R}=({\bf r}_1 +{\bf r}_2)/2$, and the
functions $f_\Lambda(r)$ and $g_\Lambda({\bf r},{\bf q})$ are defined as
\begin{eqnarray}
 f_\Lambda(r) &=& \int_{\bf p} {\rm e}^{-i\,{\bf p}\cdot {\bf r}} \frac{C_\Lambda(p)}
 {p^2+m_\pi^2} \ , \label{eq:flb} \\
 g_\Lambda({\bf r},{\bf q})&=& \int_{-1}^{+1} {\rm d}x\, 
 {\rm e}^{-i\,x\, {\bf q}\cdot {\bf r}/2} \int_{\bf p} 
{\rm e}^{-i\,{\bf p}\cdot {\bf r}}
  \frac{C_\Lambda(p)}
 {\left[ p^2+L^2(q;x)\right]^2} 
 \label{eq:glb}\ ,
\end{eqnarray}
with 
\begin{equation}
L(q;x)=\sqrt{m_\pi^2+(1- x^2)\,q^2/4} \ .
\label{eq:lfnt}
\end{equation}
Note that standard Feynman parametrization techniques have been employed
to express $g_\Lambda$ in the form given above.  We defer
to Appendix~\ref{app:fns} for a listing of the formulae resulting
from application of the gradients to $f_\Lambda(r)$ and $g_\Lambda({\bf r},{\bf q})$. 

Finally, at N$^2$LO the one-body term, Eq.~(\ref{eq:j1rc}), reads 
\begin{eqnarray}
{\bf j}^{\rm N^2LO}_{\rm a}({\bf q})\!\!&=&\!\!-\frac{e}{16 \, m_N^3}
 e_{N,1}\, \Bigg[ 
-2\,i \left( -\nabla_1^2 +q^2/4 \right) 
 \left(2\, \nabla_1-{\bm \sigma}_1\times {\bf q } \right) \nonumber \\
&-&i\, \nabla_1\cdot {\bf q}\,
 \left({\bf q} +2 \,{\bm \sigma}_1\times \nabla_1 \right)\, ,\, 
 {\rm e}^{i {\bf q}\cdot {\bf r}_1}\Bigg]_+
 - \frac{i\,e}{16 \, m_N^3}
 \kappa_{N,1}\, \Bigg[ -\nabla_1\cdot {\bf q}\, 
 \left( 4\, {\bm \sigma}_1\times \nabla_1 +{\bf q}\right) \nonumber \\ 
 &-& \left(  2\, \nabla_1 -{\bm \sigma}_1\times {\bf q} \right)\, q^2/2
   -2 \left(\nabla_1\times {\bf q}\right) \, {\bm \sigma}_1\cdot \nabla_1
 \, , \, {\rm e}^{i {\bf q}\cdot {\bf r}_1}\Bigg]_+ + 1 \rightleftharpoons 2  \ ,
 \label{eq:j1rcc}
\end{eqnarray} 
while the two-body terms are given by
\begin{equation}
  {\bf j}^{\rm N^2LO}_{\rm b-g}({\bf q})= i\,\frac{ e\, \mu^*}{9 \, m_N}\,
 \frac{g_A\, h_A}{\Delta\, F_\pi^2}\, {\rm e}^{i\,{\bf q}\cdot {\bf r}_1}\, {\bf q} \times
 \Big[ 4\, \tau_{2,z}\, \nabla -({\bm \tau}_1\times {\bm \tau}_2)_z 
 \,{\bm \sigma}_1 \times \nabla \Big] {\bm \sigma}_2 
 \cdot \nabla f_\Lambda(r) +1 \rightleftharpoons 2 \ ,
 \label{eq:j1dlt}
 \end{equation}
 \begin{equation}
   {\bf j}^{\rm N^2LO}_{\rm h-i}({\bf q})= i\,\frac{ e\, \mu^*}{9 \, m_N}\, \frac{D_T}{\Delta}
   \, {\rm e}^{i\,{\bf q}\cdot {\bf r}_1}\,
{\bf q} \times   \Big[ 4\, \tau_{2,z}\, {\bm \sigma}_2-
 ({\bm \tau}_1\times {\bm \tau}_2)_z\, 
 {\bm \sigma}_1 \times{\bm \sigma}_2 \Big] h_\Lambda(r) + 1 \rightleftharpoons 2 \ ,
 \label{eq:j2dlt}
\end{equation}
where $h_\Lambda(r)$ is simply the Fourier transform of the
Gaussian cutoff function,
\begin{equation}
 h_\Lambda(r)= \int_{\bf p} {\rm e}^{-i\,{\bf p}\cdot {\bf r}}
 C_\Lambda(p)= \frac{\Lambda^3}{(4\, \pi)^{3/2}} \, {\rm e}^{-(\Lambda \, r/2)^2} \ .
\end{equation}
In the limit $\Lambda \rightarrow \infty$, $h_\Lambda(r)$ reduces
to $\delta({\bf r})$.  In this limit, as discussed later in Sec.~\ref{sec:calce},
the magnetic dipole operator derived from ${\bf j}^{\rm N^2LO}_{\rm h-i}({\bf q})$
gives no contribution to nuclear electromagnetic observables---this follows from
the antisymmetry of two-nucleon states.  Smearing the $\delta$-function as
in Eq.~(\ref{eq:j2dlt}) is effectively including corrections of higher order
than N$^2$LO.  We shall return to this issue in Sec.~\ref{sec:calce}.
\section{Beyond N$^2$LO: loop corrections}
\label{sec:loop}

At N$^3$LO ($e\, Q$), there are four classes of contributions:
i) one-loop two-body currents, \hbox{ii) currents} from four-nucleon
contact interactions involving two gradients of the nucleon fields,
iii) one-loop renormalization corrections to tree-level two-body
currents, and iv) $(Q/M)^2$ relativistic corrections to the NLO
currents resulting from the non-relativistic reduction
of the vertices.  We now turn to a derivation of the contributions
in the first two classes.  Those in the last two will be
derived in a later work~\cite{Goity08}.

\subsection{One-loop two-body currents}
\label{sec:lp2}

\begin{figure}[bthp]
\includegraphics[width=3.8in]{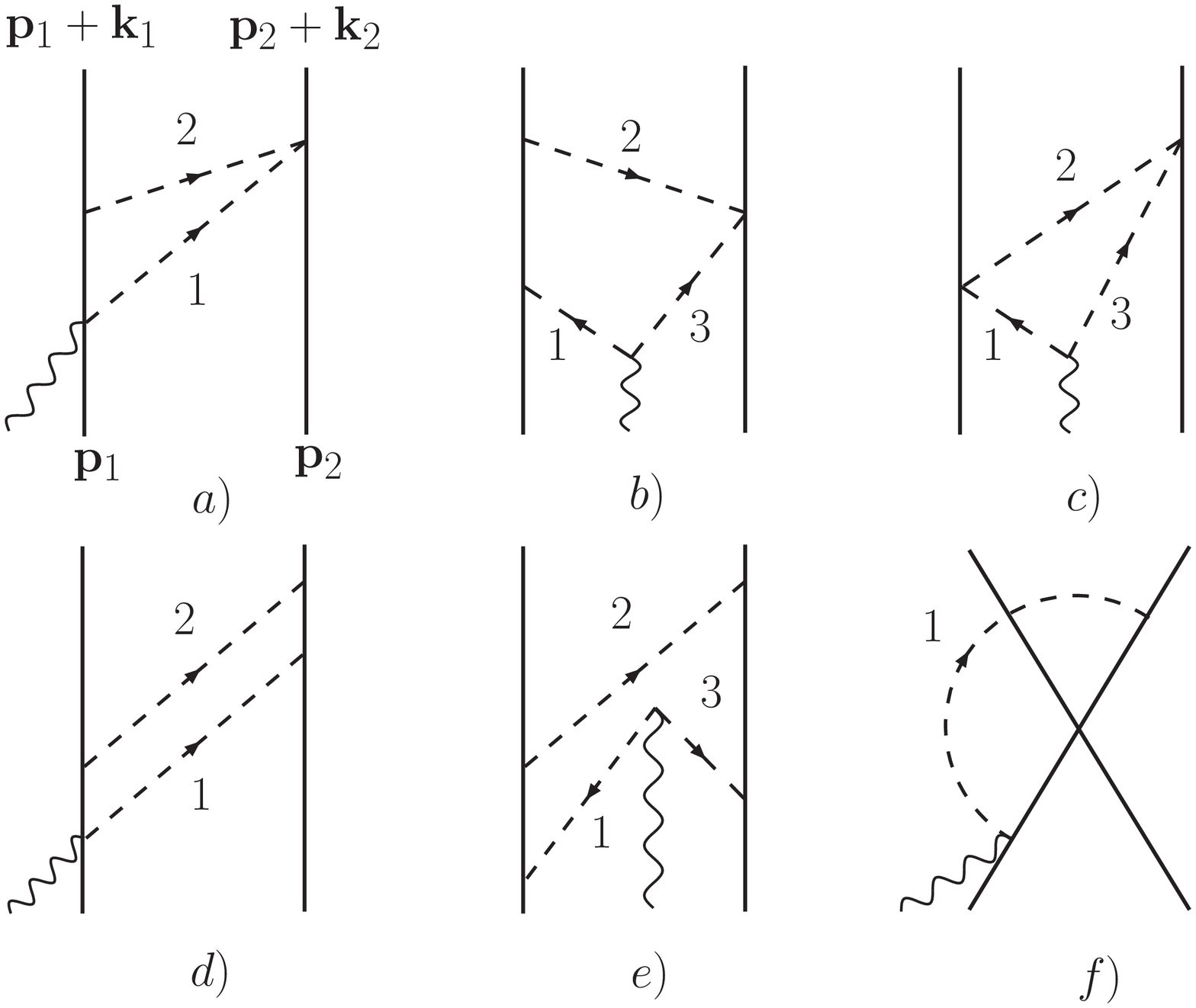}
\includegraphics[width=3.8in]{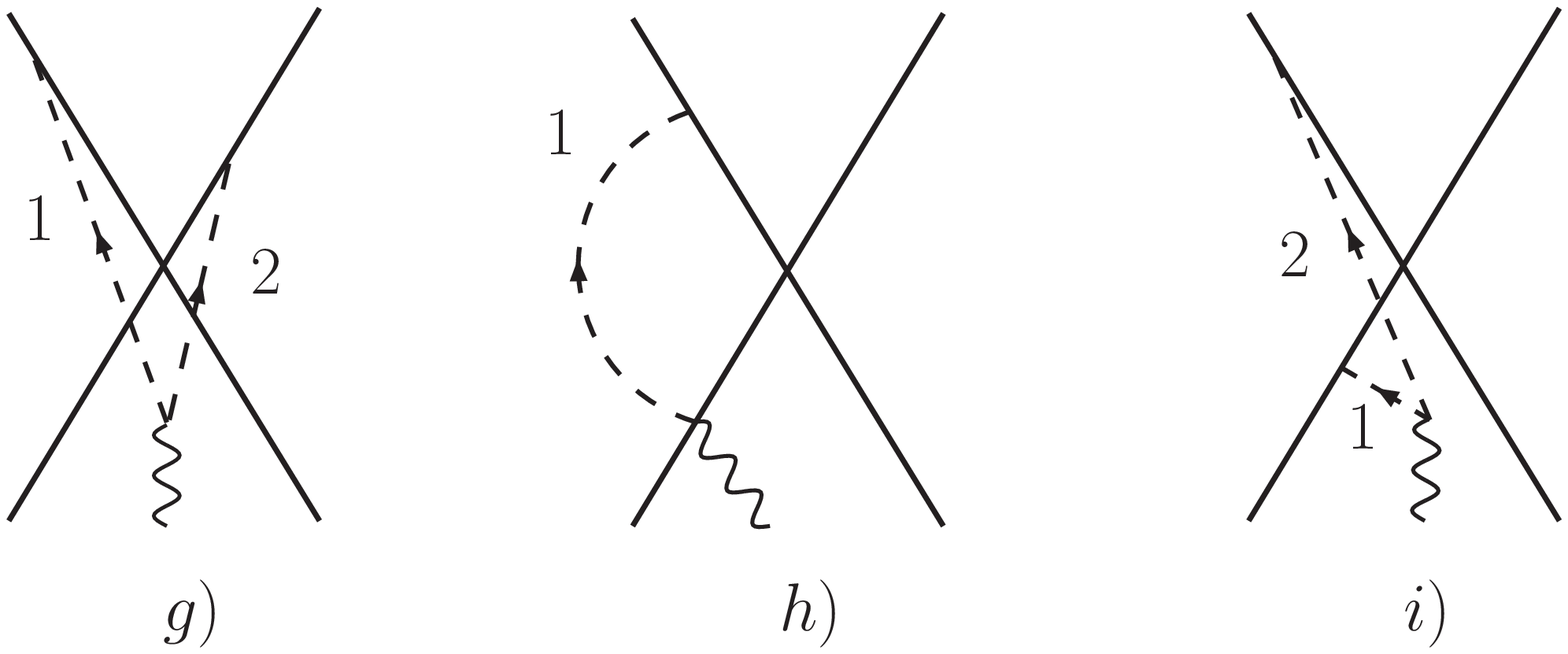}
\caption{Diagrams illustrating one-loop two-body currents.  Only one among the
possible time orderings is shown.  Notation is as in Fig.~\protect\ref{fig:fig2}.}
\label{fig:fig9}
\end{figure}

In this section we consider one-loop two-body currents.  Those
involving pions and nucleons only in the intermediate states are
illustrated by the diagrams in Fig.~\ref{fig:fig9}, where we show
only one among all possible time orderings.
Referring to this figure, we find
\begin{equation}
{\rm type\,\, a)}=-2\,i\frac{e\, g^2_A}{F^4_\pi} \,
 \int \frac{\tau_{2,z}\,({\bm \sigma}_1 \times {\bf q}_2) \, 
 +({\bm \tau}_1\times{\bm \tau}_2)_z\, {\bf q}_2}
{\omega_1 \, \omega_2 (\omega_1 +\omega_2) }+ 1 \rightleftharpoons 2   \ ,
 \label{eq:dia_a}
 \end{equation}
where the ${\bf q}_i$'s and $\omega_i=(q_i^2+m_\pi^2)^{1/2}$ denote the momenta
(with the flow as indicated in the figure) and energies of the exchanged pions, and
the integration is on any one of the ${\bf q}_i$'s, the remaining ${\bf q}_j$'s with $j\not= i$
being fixed by momentum-conserving $\delta$-functions.  Type b) diagrams give
\begin{eqnarray}
{\rm type\,\, b)}&=& 2\, i\frac{e\,g^2_A}{F^4_\pi} \,
 \int \frac{{\bf q}_1-{\bf q}_3}{\omega_1\, \omega_2\, \omega_3}
\frac{\omega_1+\omega_2+\omega_3}
  {(\omega_1+\omega_2)(\omega_1+\omega_3)(\omega_2+\omega_3)} \Big[({\bm \tau}_1\times{\bm \tau}_2)_z
\,{\bf q}_1\cdot{\bf q}_2 \nonumber\\
&-&\tau_{2,z}\,{\bm \sigma}_1\cdot({\bf q}_1\times{\bf q}_2)\Big]
+ 1 \rightleftharpoons 2 \ .
\end{eqnarray}
Next, the contributions of type c)-e) diagrams are:
\begin{equation}
{\rm type\,\, c)}=- i \frac{e}{2\, F^4_\pi} \, 
\left({\bm \tau}_1 \times {\bm \tau}_2\right)_z
 \int \frac{{\bf q}_1-{\bf q}_3}{\omega_1\,\omega_3} 
\frac{\omega_2(\omega_1+\omega_2+\omega_3)-3\, \omega_1\,\omega_3}
{(\omega_1+\omega_2)(\omega_1+\omega_3)(\omega_2+\omega_3)} \ ,
\end{equation}
\begin{eqnarray}
{\rm type\,\, d)}=\!\!&-&\!\!2\, i\frac{e\,g^4_A}{F^4_\pi}
 \int \frac{\omega_1^2+\omega_2^2 +\omega_1 \omega_2}{\omega_1^3\,\omega_2^3\,(\omega_1+\omega_2)}\,
 \Big[ ({\bm \tau}_1\times{\bm \tau}_2)_z \,{\bf q}_2\, ({\bf q}_1\cdot{\bf q}_2)  
 + 2 \, \tau_{2,z}\, {\bf q}_1\cdot{\bf q}_2 \, ({\bm \sigma}_1\times{\bf q}_2) \nonumber \\
 \!\!& + &\!\! 2 \, \tau_{1,z}\,{\bf q}_2\, {\bm \sigma}_2\cdot ({\bf q}_1\times{\bf q}_2) \Big] + 1 \rightleftharpoons 2  \ ,
 \label{eq:dia_e}
\end{eqnarray}
\begin{eqnarray}
\!\!\!\!\!\!\!\!\!\!\!\!\!{\rm type \,\, e)}\!\!&=&\!\!2\, i\frac{e\,g^4_A}{F^4_\pi}
 \int \,({\bf q}_1-{\bf q}_3)
  f(\omega_1,\omega_2,\omega_3)\,
 \Big[ ({\bm \tau}_1\times{\bm \tau}_2)_z \, ({\bf q}_1\cdot{\bf q}_2) ({\bf q}_2\cdot{\bf q}_3) \nonumber \\
 &+& 2 \, \tau_{2,z}\, ({\bf q}_2\cdot{\bf q}_3) \, {\bm \sigma}_1\cdot ({\bf q}_2\times{\bf q}_1) 
  +  2 \, \tau_{1,z}\, ({\bf q}_1\cdot{\bf q}_2) \, {\bm \sigma}_2\cdot ({\bf q}_3\times{\bf q}_2) \Big] \ ,
\label{eq:dia_f}
\end{eqnarray}
where the function $f(\omega_1,\omega_2,\omega_3)$ containing the pion energy
factors from field normalizations and energy denominators for diagrams of
type e) is defined as
\begin{eqnarray}
\label{eq:effe}
f(\omega_1,\omega_2,\omega_3)&=&  \frac{1}{\omega_1\, \omega_2\, \omega_3
(\omega_1+\omega_2)(\omega_1+\omega_3)(\omega_2+\omega_3)} 
\Bigg[\frac{\omega_1\,\omega_2+\omega_2\,\omega_3+\omega_1\,\omega_3}{\omega_1\,\omega_2\,\omega_3} \nonumber \\
&+&\frac{ (\omega_1+\omega_2)\,(\omega_2+\omega_3)\,(\omega_1^2+\omega_3^2)}
 {\omega_1^2\, \omega_2\, \omega_3^2}
+\frac{\omega_2}{\omega_1\,\omega_3} 
+\frac{\omega_1+\omega_2+\omega_3}{\omega_2^2}\Bigg] \ .
\end{eqnarray}
Lastly, diagrams of type f) and h) vanish, since the integrand (in the
static limit) is an odd function of the loop momentum ${\bf q}_1$,
\begin{equation}
{\rm type\,\, f)\,\, {\rm and}\,\, h)}\propto \int \frac{{\bf q}_1}{\omega_1^3} \times
({\rm spin\!\!-\!\!isospin\,\, structure}) \ .
\end{equation}
However, the contributions of type g) and i) diagrams read:
\begin{eqnarray}
{\rm type \,\, g)}&=&2\, i\frac{e\, g^2_A\, C_T}{F^2_\pi}\, 
({\bm \tau}_1\times{\bm \tau}_2)_z\,
 \int \frac{{\bf q}_1-{\bf q}_2}{\omega_1^3\, \omega_2^3} \frac{\omega_1^2+\omega_1\, \omega_2+\omega_2^2}
 {\omega_1+\omega_2}({\bm \sigma}_1
\cdot{\bf q}_2)({\bm \sigma}_2\cdot{\bf q}_1)  \ ,
\end{eqnarray}
\begin{eqnarray}
\!\!\!\!\!\!{\rm type \,\, i)}\!\!&=&\!\!-\frac{e\,g^2_A}{2\, F^2_\pi}\, 
({\bm \tau}_1\times{\bm \tau}_2)_z\,
 \int \frac{{\bf q}_1-{\bf q}_2}{\omega_1^3\, \omega_2^3} \frac{\omega_1^2+\omega_1\, \omega_2+\omega_2^2}
 {\omega_1+\omega_2} \Bigg[ C_{S}\, {\bm \sigma}_1\cdot({\bf q}_1\times{\bf q}_2) \nonumber \\
&+&i\, C_{T}\Big[ ({\bm \sigma}_1\cdot{\bf q}_2)\,({\bm \sigma}_2\cdot{\bf q}_1) 
+({\bm \sigma}_1\cdot{\bf q}_1)({\bm \sigma}_2\cdot{\bf q}_2)
+ i\, {\bm \sigma}_2\cdot({\bf q}_1\times{\bf q}_2)\Big] \Bigg] 
+ 1 \rightleftharpoons 2 \ .
\end{eqnarray} 
\begin{figure}[bthp]
\centerline{
\includegraphics[width=3.8in]{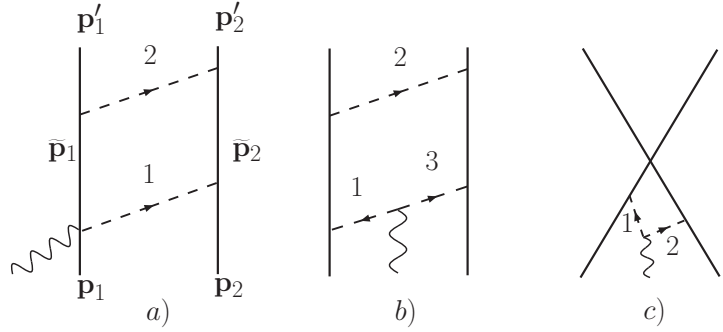}
}
\caption{Diagrams illustrating the reducible one-loop two-body currents.  Only one among the
possible time orderings is shown.  Notation is as in Fig.~\protect\ref{fig:fig2}.}
\label{fig:figred}
\end{figure}
A few comments are now in order.  Firstly, the evaluation of the current
operators resulting from the diagrams of Fig.~\ref{fig:fig9} panels d)-e) and g)
is carried out by including the recoil corrections of order $e\,Q$ to the reducible
diagrams shown in Fig.~\ref{fig:figred} panels a)-c).  As an example, we consider
the irreducible and reducible contributions represented in Fig.~\ref{fig:fig9} d)
and \hbox{Fig.~\ref{fig:figred} a)}, respectively.  We follow the procedure adopted
in Sec.~\ref{sec:canc}, and expand, in the reducible diagrams, the energies
of the intermediate nucleonic states, which are suppressed by a factor $Q/M$ with
respect to the pionic energies $\omega_i \sim Q$. Up to order $e\, Q$, the current
operator ${\bf j}_{\rm red}$ associated with the reducible box diagrams then reads
 \begin{eqnarray}
\label{eq:red}
{\bf j}_{\rm red }&=& \int
v^{\pi}({\bf q}_2)\, \frac{1}{E_i-{\widetilde E}_1-{\widetilde E}_2+i\eta }\,
{\bf j}^{\rm NLO}({\bf q}_1) \nonumber \\
&-& \int 2 \,  \frac{\omega_1+\omega_2}{\omega_1\,\omega_2}\,
V_{\pi NN}(2,{\bf q}_2)\,V_{\pi NN}(2,{\bf q}_1)\,
V_{\pi NN}(1,{\bf q}_2)\, V_{\gamma \pi NN}(1,{\bf q}_1) \ ,
\end{eqnarray}
where $v^{\pi}({\bf q}_2)$ and ${\bf j}^{\rm NLO}({\bf q}_1)$
are the OPEP and pion-seagull current operators defined in Eqs.~(\ref{eq:vpi})
and~(\ref{eq:nlo1}), respectively.  The $V(i,{\bf q}_{\,j})$ denotes the vertex
from the interaction Hamiltonian relative to nucleon $i$ and a pion
with momentum ${\bf q}_{\, j}$, and $E_i$ is the initial energy of the system,
while $\widetilde E_1$ and  $\widetilde E_2$ are the energies of the
intermediate nucleons.  The first term of Eq.~(\ref{eq:red}) is then embedded
in the iterated solution of the Lippmann-Schwinger equation, whereas 
the second term due to recoil corrections is retained
and added to the irreducible contribution, ${\bf j}_{\rm irr}$, which is given by
\begin{eqnarray}
\label{eq:irr}
{\bf j}_{\rm irr }&=& \int
   \frac{2}{\omega_1\, \omega_2(\omega_1+\omega_2)}\,
V_{\pi NN}(2,{\bf q}_2)\,V_{\pi NN}(2,{\bf q}_1)\,
V_{\pi NN}(1,{\bf q}_1)\, V_{\gamma \pi NN}(1,{\bf q}_1) \nonumber \\
&+&\int
  2\, \frac{\omega_1^2+\omega_2^2+\omega_1\,\omega_2}{\omega_1\, \omega_2(\omega_1+\omega_2)}\,
V_{\pi NN}(2,{\bf q}_1)\,V_{\pi NN}(2,{\bf q}_2)\,
V_{\pi NN}(1,{\bf q}_2)\, V_{\gamma \pi NN}(1,{\bf q}_1) \ .
\end{eqnarray}
The first term above comes from the irreducible direct diagrams (in which,
with reference to Fig.~\ref{fig:fig9} d), pion $1$ is absorbed before pion $2$), while the second term
is from the crossed diagrams (in which pion $1$ is absorbed after pion $2$).
Equation~(\ref{eq:irr}) can be further simplified expressing the product
$V_{\pi NN}(2,{\bf q}_1)\,V_{\pi NN}(2,{\bf q}_2)$ as
\begin{equation} 
V_{\pi NN}(2,{\bf q}_1)\,V_{\pi NN}(2,{\bf q}_2)=
[V_{\pi NN}(2,{\bf q}_1),V_{\pi NN}(2,{\bf q}_2)]_-+V_{\pi NN}(2,{\bf q}_2)\,V_{\pi NN}(2,{\bf q}_1) \ , 
\end{equation}
to obtain
\begin{eqnarray}
\label{eq:irr1}
{\bf j}_{\rm irr }&=& \int
  2 \frac{\omega_1+\omega_2}{\omega_1\, \omega_2}\,
V_{\pi NN}(2,{\bf q}_2)\,V_{\pi NN}(2,{\bf q}_1)\,
V_{\pi NN}(1,{\bf q}_2)\, V_{\gamma \pi NN}(1,{\bf q}_1) \nonumber \\
&+&\int
  2\, \frac{\omega_1^2+\omega_2^2+\omega_1\,\omega_2}{\omega_1\, \omega_2(\omega_1+\omega_2)}\!
\left[V_{\pi NN}(2,{\bf q}_1),V_{\pi NN}(2,{\bf q}_2)\right]_-\!\!
V_{\pi NN}(1,{\bf q}_2) V_{\gamma \pi NN}(1,{\bf q}_1) \ .
\end{eqnarray}
The complete current of type d) is then
\begin{equation}
\label{eq:typed}
{\rm type \,\, d)} = \int  2\, \frac{\omega_1^2+\omega_2^2+\omega_1\,\omega_2}{\omega_1\, \omega_2(\omega_1+\omega_2)}
\left[V_{\pi NN}(2,{\bf q}_1)\,,\,V_{\pi NN}(2,{\bf q}_2)\right]_- 
V_{\pi NN}(1,{\bf q}_2) V_{\gamma \pi NN}(1,{\bf q}_1) \, - {\rm h.c.} \ ,
\end{equation}
where the h.c.~term corresponds to including the diagrams in which the
photon hooks up to the pion with momentum ${\bf q}_2$.
Note that the recoil corrections exactly cancel the first term of
Eq.~(\ref{eq:irr1}), leaving the term proportional to the energy
factor associated with the crossed diagrams only.
We find it interesting that these cancellations are also obtained for
the current of type e), which reads
\begin{eqnarray}
\label{eq:typee}
{\rm type \,\, e)}&=& \int
  4\,f(\omega_1,\omega_2,\omega_3)
\left[V_{\pi NN}(2,{\bf q}_3),V_{\pi NN}(2,{\bf q}_2)\right]_- \nonumber \\
&\times& V_{\pi NN}(1,{\bf q}_2) V_{\pi NN}(1,{\bf q}_1)V_{\gamma \pi \pi}({\bf q}_1,{\bf q}_3)
\, - {\rm h.c.} \  ,
\end{eqnarray}
and it is therefore tempting to conjecture that they persist at
higher orders.  However,  this statement has not been proven.

\begin{figure}[bthp]
\centerline{
\includegraphics[width=6.3in]{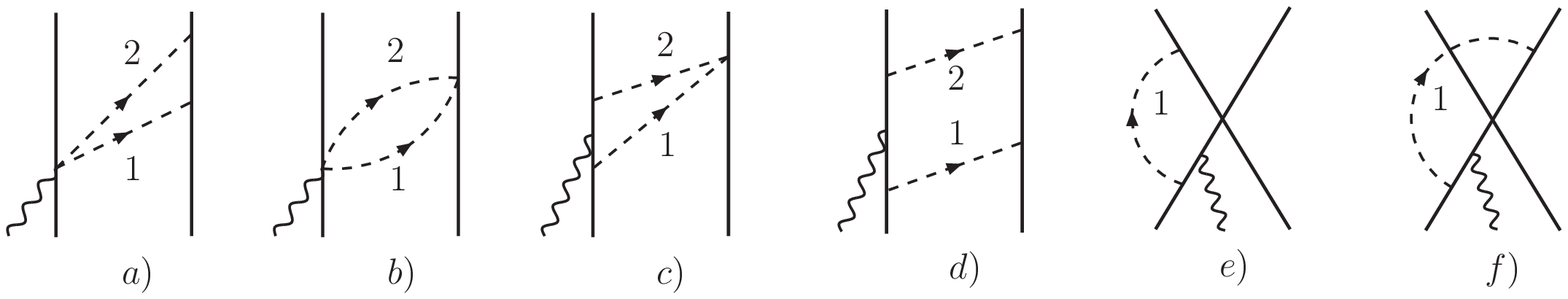}}
\caption{Diagrams illustrating N$^4$LO contributions not included
in the present work. Only one among the
possible time orderings is shown. Notation is as in Fig.~\protect\ref{fig:fig2}.}
\label{fig:fig10}
\end{figure}

Secondly, we observe that diagrams of the type shown in Fig.~\ref{fig:fig10}
are suppressed by an extra power of $Q$ relative to those considered in this
section, {\it i.e.}~they are of order $e\, Q^2$.  For example, the diagrams of
type a) give rise to the following current operator
\begin{equation}
{\rm type\,\, a) \,\, in\,\, Fig.\,8}=
\frac{e}{m}\frac{g_A^2}{F_{\pi}^4}\tau_{z,1}(2\,{\bf K}_1+i{\bm \sigma}_1\times{\bf k}_1)
\int \frac{{\bf q}_1\cdot{\bf q}_2}{\omega_1^2\,\omega_2^2} +1 \rightleftharpoons 2 \ ,
\end{equation}
where the momentum ${\bf K}_i$ is as given in Eq.~(\ref{eq:defs}), while those
of type b) vanish, since they are proportional $(\delta_{az}\, \tau_{1,b}+\delta_{bz}\,
 \tau_{1,a}-2\,\delta_{ab}\, \tau_{1,z}) \, \epsilon_{abc}\, \tau_{2,c}
=(\epsilon_{zbc}+\epsilon_{bzc})\,\tau_{1,b}\, \tau_{2,c}$.

Lastly, as a check, we have re-derived the nucleon-nucleon potential at
the one-loop level (both with and without the inclusion of explicit
$\Delta$-isobar degrees of freedom), and have explicitly verified
that it is in agreement with that obtained in Refs.~\cite{vanKolck94}
and~\cite{Epelbaum98}.  In particular, we note that if recoil corrections
to the reducible diagrams, for example box diagrams, are retained along with
the contributions of irreducible diagrams, the resulting potential is in agreement with
that derived in Ref.~\cite{Epelbaum98} with the method of unitary
transformations (in this respect, see Sec.~\ref{sec:jcons}).

\begin{figure}[bthp]
\centerline{
\includegraphics[width=4.8in]{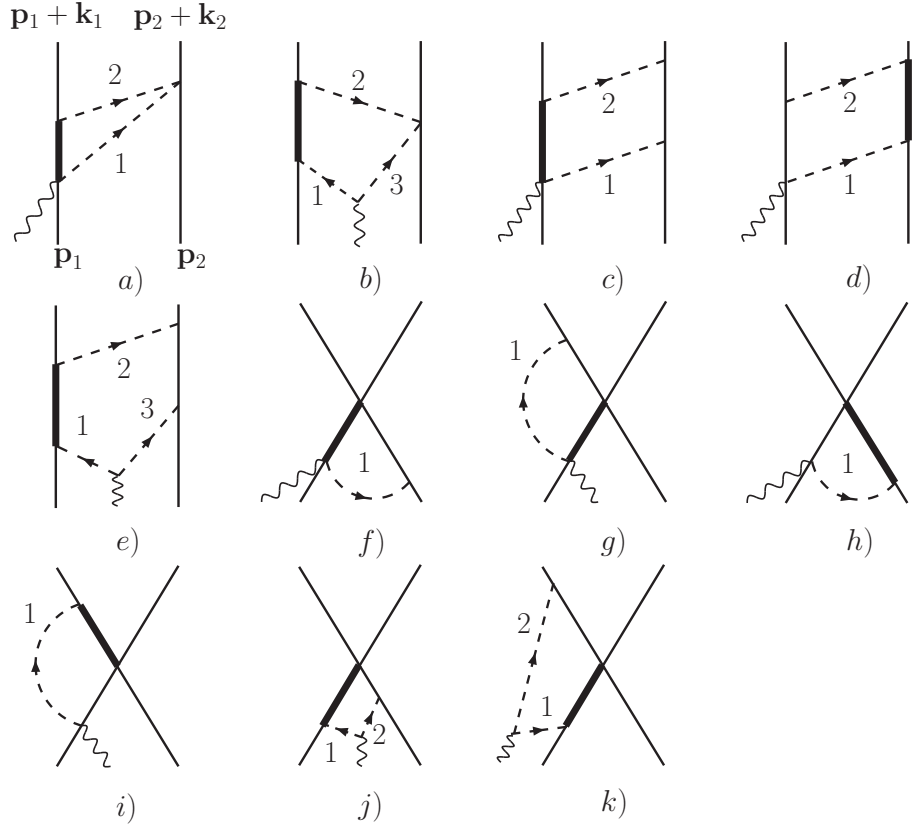}}
\caption{Diagrams illustrating one-loop two-body currents with a single $\Delta$
isobar in the intermediate states.  Only one among the possible time orderings
is shown.  Thin, thick, dashed, and wavy lines denote nucleons,
$\Delta$ isobars, pions, and photons, respectively.}
\label{fig:fig11}
\end{figure}

\begin{figure}[bthp]
\includegraphics[width=3.8in]{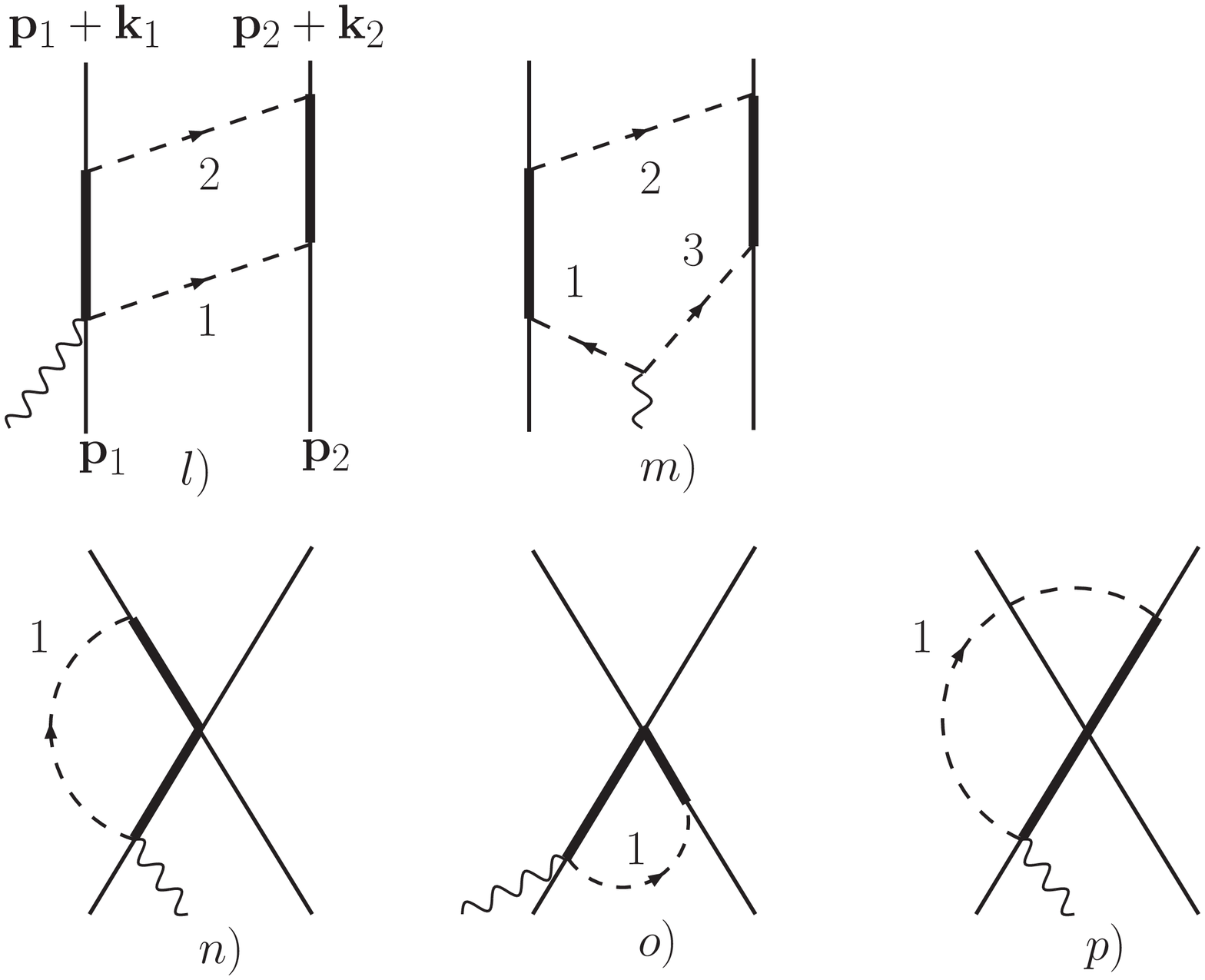}
\includegraphics[width=3.8in]{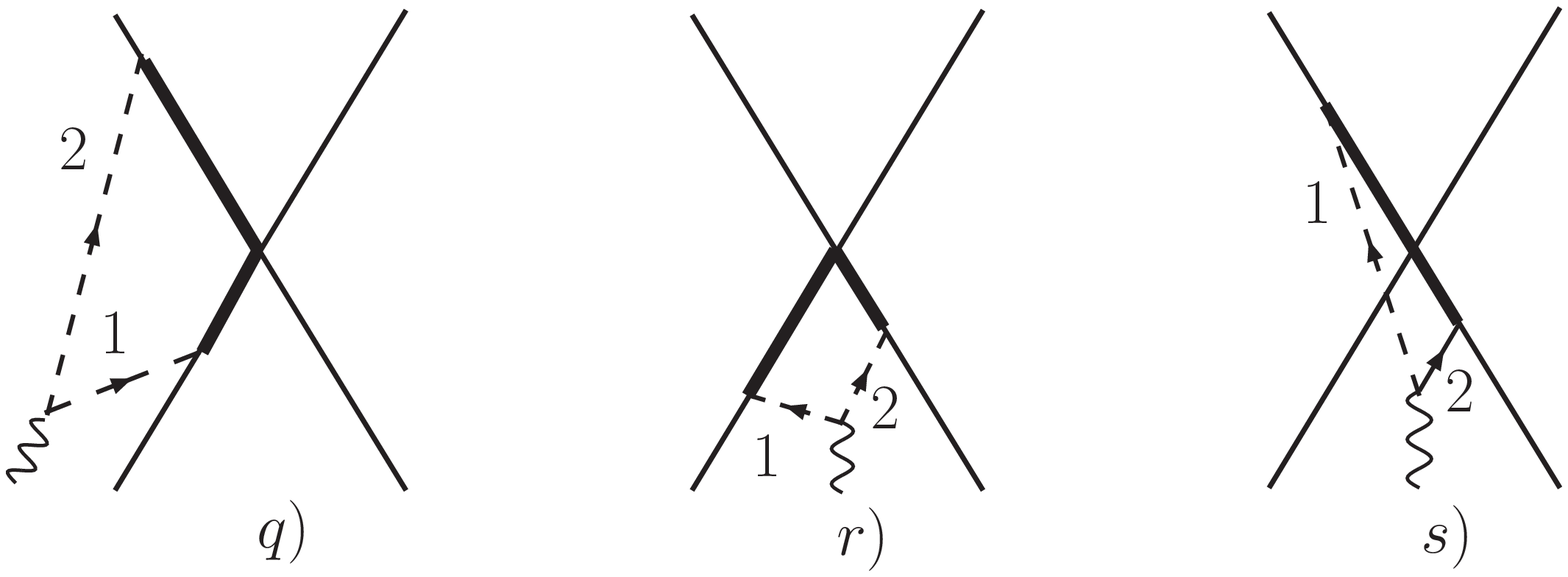}
\caption{Diagrams illustrating one-loop two-body currents with two $\Delta$
isobars in the intermediate states.  Only one among the possible time orderings
is shown.  Notation is as in Fig.~\protect\ref{fig:fig11}.}
\label{fig:fig13}
\end{figure}

We conclude this section by showing in Figs.~\ref{fig:fig11} and~\ref{fig:fig13} the
one-loop two-body currents involving one- and two-$\Delta$ intermediate
states.  A listing of the explicit expressions is given in Appendix~\ref{app:l2pid}.

\subsection{Currents from four-nucleon contact interactions}
\label{sec:c4cnt}

In this section we report the N$^3$LO contributions to the
current operator from the contact electromagnetic interactions of
Eqs.~(\ref{eq:ct2d1g})--(\ref{eq:ct2d2g}).  We find:
\begin{eqnarray}
{\bf j}_{\rm CT\gamma}^{\rm N^3LO}&=& -e \, e_1\bigg[
2\, \left(2\, C^{\prime}_{1}-C_{2}^{\prime}\right)\, {\bf K}_{2}+4\, C^{\prime}_{3}\,{\bf K}_{1}
+i\,C_{4}^{\prime}\, \left({\bm \sigma}_{1}+
{\bm \sigma}_{2}\right)\times{\bf k}_{2}
 + i \,C^{\prime}_{5}\, {\bm \sigma}_1\times{\bf k}_1 \nonumber \\
&-&i \,C^{\prime}_{6}\,
{\bm \sigma}_2\times{\bf k}_1 + 2\, \left(2\, C^{\prime}_7-
C^{\prime}_{10}\right)({\bf K}_{2}\cdot{\bm \sigma}_2)\, {\bm \sigma}_1 + 
  2\, \left(2\, C^{\prime}_8-C^{\prime}_{11}\right)({\bf K}_{2}
 \cdot{\bm \sigma}_1)\, {\bm \sigma}_2  \nonumber \\
&-&2\, C^{\prime}_{13}\, \left[({\bf K}_1\cdot{\bm \sigma}_1)\,
{\bm \sigma}_2+ ({\bf K}_1\cdot {\bm \sigma}_2)\, {\bm \sigma}_1 \right] 
+ 2\, \left(2\, C^{\prime}_{9}-C^{\prime}_{12}\right){\bf K}_2\,
({\bm \sigma}_1\cdot{\bm \sigma}_2)\nonumber \\
&-&4\, C^{\prime}_{14}\,{\bf K}_1\,
({\bm \sigma}_1\cdot{\bm \sigma}_2) \bigg] + 1 \rightleftharpoons 2 \ .
\end{eqnarray}
Again, we note that the nucleon-nucleon potential generated
by the contact interactions of Eqs.~(\ref{eq:ct2d1})--(\ref{eq:ct2d2})
is in agreement with that obtained in Ref.~\cite{Epelbaum98}.
\section{Current conservation up to N$^3$LO}
\label{sec:jcons}
The nuclear electromagnetic current operator is related to the Hamiltonian
through the continuity equation, which in the momentum space reads
\begin{equation}
\label{eq:conty}
{\bf q}\cdot {\bf j}=\left[\, \frac{p^{\,2}_1}{2\, m_N}+\frac{p^{\,2}_2}{2\, m_N}+v_{12} \, ,\, \rho \,\right ]_{-} \ ,
\label{eq:cc}
\end{equation}
where $\left[\dots\, ,\, \dots\right]_-$ denotes the commutator, {\bf q} is momentum transfer,
and $\rho$ is the charge operator given to LO, in the notation of Eq.~(\ref{eq:ekm}), by
\begin{equation}
\rho=e\, (e_1+e_2) \ .
\end{equation}
\begin{figure}[bthp]
\centerline{
\includegraphics[width=2.2in]{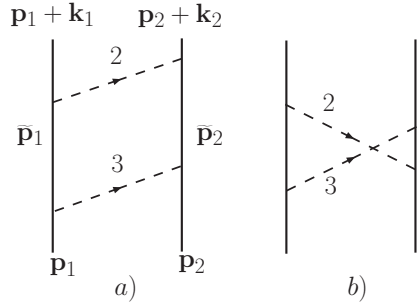}
}
\caption{Diagrams illustrating the reducible, panel a), and irreducible,
panel b), two-body box potential.  Only one among the possible time
orderings is shown.  Notation is as in Fig.~\protect\ref{fig:fig2}.}
\label{fig:figredv}
\end{figure}
It is well known (and easily verified) that the ${\rm LO}$ and ${\rm NLO}$ currents
satisfy the continuity equation with, respectively, the kinetic energy terms and the
LO ($Q^0$) contribution to the potential, {\it i.e.}~OPEP.  The ${\rm N^2LO}$ currents
involving $\Delta$ excitation are purely transverse, and therefore do not enter the
continuity equation, while those arising from relativistic corrections to the LO one-body
term require the inclusion of these corrections also in the charge operator, in order for
the continuity equation to be satisfied.  We will not discuss them further here.

As an internal check, we have explicitly verified that the ${\rm N^3LO}$ current operators
obtained in the previous section satisfy the continuity equation with the N$^2$LO ($Q^2$)
contributions to the potential, induced by the interaction Hamiltonians given in Sec.~\ref{sec:h-s}
and Appendix~\ref{sub:ct2d}.  For the purpose of illustration, we give more details on the
calculation carried out for the currents of type d) and e) of Fig.~\ref{fig:fig9}.
The potential generated by the box diagrams shown in Fig.~\ref{fig:figredv} is given by
\begin{equation}
\label{eq:box}
v_{\rm box}=-\frac{g_A^4}{F_{\pi}^4}\int
\frac{\omega_2^2+\omega_3^2+\omega_2\,\omega_3}{\omega_2^3\, \omega_3^3\,(\omega_2+\omega_3)}\bigg[2\,{\bm \tau}_1\cdot{\bm \tau}_2\,
({\bf q}_2\cdot{\bf q}_3)^2 +3\,
{\bm \sigma}_1\cdot({\bf q}_2\times{\bf q}_3)\,{\bm \sigma}_2\cdot({\bf q}_2\times{\bf q}_3)\bigg]  \ ,
\end{equation}
where ${\bf q}_2+{\bf q}_3={\bf k}_2$ or $-{\bf k}_1$.  We note that
the recoil corrections to the reducible box diagrams have been included, consistently with
our treatment in the previous section.

Evaluation of the commutator of $v_{\rm box}$ with the charge operator gives
\begin{eqnarray}
\label{eq:comm}
\left[\, v_{\rm box}\, ,\, \rho \, \right]= -i\,e \frac{2\, g_A^4}{F_{\pi}^4}({\bm \tau}_1\times{\bm \tau}_2)_z
\int\frac{\omega_2^2+\omega_3^2+\omega_2\,\omega_3}{\omega_2^3\omega_3^3(\omega_2+\omega_3)}
({\bf q}_2\cdot{\bf q}_3)^2 + 1 \rightleftharpoons 2 \ ,
\end{eqnarray}
where ${\bf k}_1+{\bf k}_2 = {\bf q}$ and ${\bf q}_2+{\bf q}_3={\bf k}_2$.  On the other hand,
the l.h.s.~of Eq.~(\ref{eq:cc}) for the currents of type d) and e) in Fig.~\ref{fig:fig9} reads
\begin{eqnarray}
\label{eq:jeq}
{\bf q}\cdot{\bf j}_{\rm d}&=&-i\,e\frac{2 \,g^4_A}{F^4_\pi}
 \int \frac{\omega_2^2+\omega_3^2 +\omega_2 \omega_3}{\omega_2^3\,\omega_3^3\,(\omega_2+\omega_3)}\,
 \Bigg[ ({\bm \tau}_1\times{\bm \tau}_2)_z \, {\bf q}_2 \, ({\bf q}_2\cdot{\bf q}_3) 
 + 2 \, \tau_{2,z}\, ({\bf q}_2\cdot{\bf q}_3) \,({\bm \sigma}_1\times{\bf q}_2) \nonumber \\
 &+&  2 \, \tau_{1,z}\,{\bf q}_2\, {\bm \sigma}_2\cdot ({\bf q}_3\times{\bf q}_2) \Bigg] \cdot{\bf q} + 1 \rightleftharpoons 2  \ ,
\end{eqnarray} 
\begin{eqnarray}
\label{eq:jfq}
{\bf q}\cdot{\bf j}_{\rm e}&=& i\,e\frac{2\,g_A^4}{F_{\pi}^4}\int
\Bigg[ \frac{\omega_2^2+\omega_3^2+\omega_2\,\omega_3}{\omega_2^3\omega_3^3(\omega_2+\omega_3)} -
 \frac{\omega_1^2+\omega_2^2+\omega_1\,\omega_2}{\omega_1^3\omega_2^3(\omega_1+\omega_2)}   \Bigg]
\Bigg[ ({\bm \tau}_1\times{\bm \tau}_2)_z \, ({\bf q}_1\cdot{\bf q}_2)({\bf q}_2\cdot{\bf q}_3) \nonumber \\
 &&\qquad \qquad + 2 \, \tau_{2,z}\, ({\bf q}_2\cdot{\bf q}_3)\,{\bm \sigma}_1\cdot ({\bf q}_2\times{\bf q}_1) 
  +  2 \, \tau_{1,z}\, ({\bf q}_1\cdot{\bf q}_2) \, {\bm \sigma}_2\cdot ({\bf q}_3\times{\bf q}_2) \Bigg] \ ,
\end{eqnarray}  
where ${\bf q}_1={\bf q}_2+{\bf k}_1$, and  the factor
${\bf q}\cdot({\bf q}_1-{\bf q}_3)\, f(\omega_1, \omega_2,\omega_3)=(\omega_1^2-\omega_3^2)f(\omega_1,\omega_2,\omega_3)$
has been written as in the square brackets of the last equation. 
Combining Eqs.~(\ref{eq:jeq})--(\ref{eq:jfq}), we obtain
\begin{eqnarray}
&&\!\!\!\! {\bf q} \cdot \left( {\bf j}_{\rm d}+{\bf j}_{\rm e} \right)  =-i\,e\frac{2 \,g^4_A}{F^4_\pi}
 \int\frac{\omega_2^2+\omega_3^2 +\omega_2 \omega_3}{\omega_2^3\,\omega_3^3\,(\omega_2+\omega_3)}\,
 \Bigg[ ({\bm \tau}_1\times{\bm \tau}_2)_z \, ({\bf q}_2\cdot{\bf q}_3)^2\nonumber \\
&& + 2 \, \tau_{2,z}\, ({\bf q}_2\cdot{\bf q}_3) \,{\bm \sigma}_1\cdot ({\bf q}_2\times{\bf q}_3) 
 +  2 \, \tau_{1,z}\, ({\bf q}_2\cdot{\bf q}_3) \,{\bm \sigma}_2\cdot ({\bf q}_3\times{\bf q}_2)\Bigg]  + 1 \rightleftharpoons 2  \ ,
\end{eqnarray}
and the last two terms of the previous equation vanish.  This is easily seen
by changing ${\bf q}_2 \rightarrow {\bf k}_2/2 +{\bf q}_2$
(implying ${\bf q}_3 = {\bf k}_2/2 -{\bf q}_2$), and observing that the integrands
 are odd under ${\bf q}_2\rightarrow-{\bf q}_2$.
Hence we are left with the first term which is equal to Eq.~(\ref{eq:comm}), showing
that the continuity equation is indeed satisfied.
\begin{figure}[bthp]
\centerline{
\includegraphics[width=2.4in]{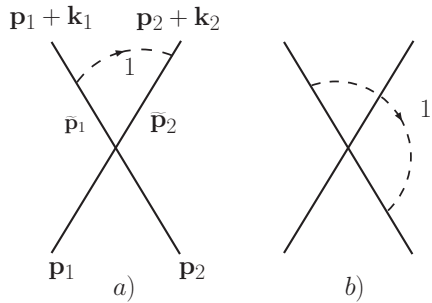}
}
\caption{Diagrams illustrating the reducible, panel a), and irreducible, panel b),
two-body one-loop contact potential.  Only one among the possible time orderings
is shown.  Notation is as in Fig.~\protect\ref{fig:fig2}.}
\label{fig:figredvCT}
\end{figure}

Two closing remarks are in order.  First, ignoring the recoil corrections in both the potential
and currents leads to a violation of the continuity equation.  This remains valid
also for the current of type g) and  the potential $v_{\rm ct}$, generated by the diagrams 
shown in Fig.~\ref{fig:figredvCT} and given explicitly by
\begin{equation}
v_{\rm ct}=\frac{4}{3}\frac{g_A^2}{F_{\pi}^2}\,{\bm \tau}_1\cdot{\bm \tau}_2\, \,{\bm \sigma}_1\cdot{\bm \sigma}_2
\int \frac{{\bf q}_1^2}{\omega_1^3}  \ .
\end{equation}

Second, in hybrid calculations, such as those reported below, current conservation
is not strictly satisfied.  Assuming, however, that differences between the $\chi$EFT
and realistic potentials occur at orders higher than N$^2$LO,  the N$^3$LO
currents derived here are then approximately conserved.

\section{Electromagnetic observables at N$^2$LO in $A$=2--3 systems}
\label{sec:calce}

We present results obtained for a number of low-energy
electromagnetic observables in the $A$=2 and 3 nuclei using
the current operators derived at N$^2$LO.  In the tables to
follow, we denote respectively with LO, NLO, N$^2$LO-RC,
N$^2$LO-$\Delta$, and N$^2$LO-$\Delta_c$ the contributions
calculated with the one-body current, or impulse-approximation
(IA), of Eq.~(\ref{eq:jlor}), the one-pion exchange two-body current
at tree level of Eqs.~(\ref{eq:nlor1})--(\ref{eq:nlor2}), the relativistic
correction to the LO current of Eq.~(\ref{eq:j1rcc}), the single $\Delta$-isobar
excitation current of Eq.~(\ref{eq:j1dlt}), and lastly the two-body
current of Eq.~(\ref{eq:j2dlt}), due to the contact $(NN)(\Delta N)$
interaction.  In the long-wavelength limit of interest in the present
work, the LO and N$^2$LO-RC currents are completely determined by the
experimental values of the proton and neutron magnetic moments, respectively
+2.793 and --1.913 in units of nuclear magnetons (n.m.).  The NLO current
involves the combination $g_A/F_\pi$, for which we adopt the value
$(m_\pi\, g_A/F_\pi)^2/(4\pi)$=0.075 as inferred from the Nijmegen analysis of
nucleon-nucleon elastic scattering data~\cite{Stoks94}.  The $N\Delta$
transition magnetic moment is taken to be $\mu^*$=3 n.m.~from an analysis
of $\gamma N$ data at resonance~\cite{Burkert08}.  The coupling constant $h_A$
in the N$^2$LO-$\Delta$ term is fixed by reproducing the width of the
$\Delta$ resonance, $h_A = 2.191\, g_A$ and $g_A=1.267$, while the (unknown)
coupling constant $D_T$ in the N$^2$LO-$\Delta_c$ term is expressed as
\begin{equation}
D_T= f^\prime \, \frac{g_A h_A}{F_\pi^2} \ ,
\label{eq:dtc}
\end{equation}
and the parameter $f^\prime$ is determined as described below.

In order to have a realistic estimate of the model dependence of the
results, we use cutoff values $\Lambda$ in the range 500--800 MeV and
wave functions corresponding to two different nuclear Hamiltonians.
The wave functions for $A$=2 are derived from solutions of
the Schr\"odinger equation with the Argonne $v_{18}$ (AV18)~\cite{Wiringa95}
or CD-Bonn (CDB)~\cite{Machleidt01} two-nucleon potentials, while those
for $A$=3 are obtained with the hyperspherical harmonics (HH) expansion
method (see Ref.~\cite{Kievsky08} and references therein) from a Hamiltonian
including, in addition to the AV18 or CDB two-nucleon, also a three-nucleon
potential, the Urbana-IX (UIX) model~\cite{Pudliner97}.  The AV18/UIX and
CDB/UIX$^*$ combinations both reproduce the experimental $^3$H binding energy.
The former also reproduces most of the measured low-energy $N$-$d$
scattering observables~\cite{Kievsky08}, with the notable exception of the vector
analyzing power in $N$-$d$ elastic scattering.  Unfortunately, HH continuum wave
functions for the CDB/UIX$^*$ combination are not yet available.  The UIX$^*$ model~\cite{Viviani07}
is a slightly modified version of the original UIX~\cite{Pudliner97} (in the UIX$^*$,
the parameter $U_0$ of the central repulsive term has been reduced by the factor $0.812$).

We consider the following two- and three-nucleon observables:
the $^1$H($n,\gamma$)$^2$H cross section at thermal energies, the
deuteron magnetic moment, the isoscalar and isovector combinations
of the trinucleon magnetic moments, the cross section and
photon circular polarization parameter $R_c$ measured
in the radiative capture of (polarized, in the case of $R_c$)
neutrons on deuterons at thermal energies.  At N$^2$LO there
are no three-body currents.  We also observe that at this order
the only isoscalar terms are from the (one-body) LO and
N$^2$LO-RC operators, which are independent of the cutoff $\Lambda$.
In Tables~\ref{tb:tab2} and~\ref{tb:tab7}
\begin{table}[bth]
\begin{tabular}{c|c|c}
\hline
\hline
         & AV18 & CDB \\
\hline
LO          &0.8469 &0.8521   \\
N$^2$LO-RC  &--0.0082 &--0.0080  \\
\hline
\hline
\end{tabular}
\caption{Contributions in n.m.~to the deuteron magnetic moment, obtained with the
AV18 and CDB potential models.  The experimental value is 0.8574 n.m..}
\label{tb:tab2}
\end{table}
\begin{table}[bth]
\begin{tabular}{c|c|c}
\hline
\hline
         & AV18/UIX & CDB/UIX$^*$ \\
\hline
LO          &  +0.4104 &  +0.4183  \\
N$^2$LO-RC  &--0.0045  &  --0.0052  \\
\hline
\hline
\end{tabular}
\caption{Contributions in n.m.~to the isoscalar combination of the trinucleon magnetic 
moments, obtained with the AV18/UIX and CDB/UIX$^*$ Hamiltonian models.  The experimental
value is 0.4257 n.m..}
\label{tb:tab7}
\end{table}
we list their contributions
to the deuteron magnetic moment and isoscalar combination of the
$^3$He and $^3$H magnetic moments.  The N$^2$LO-RC correction is (in magnitude)
about 1\% of the LO contribution but of opposite sign, so that its
inclusion increases the difference between the measured and calculated
values.  As a result, the experimental deuteron and trinucleon isoscalar magnetic
moments are underpredicted by theory at the (1.6--2.1)\% and
(3.0--4.7)\% levels, respectively, depending on whether the CDB
and CDB/UIX$^*$ or AV18 and AV18/UIX combinations are adopted in the
$A$=2 and $A$=3 calculations.  However, a recent calculation
of these same observables~\cite{Song07}, based on variational Monte
Carlo (VMC) wave functions corresponding to the AV18/UIX Hamiltonian, finds the
magnitude of the N$^2$LO-RC correction somewhat smaller in $A$=2 (--0.0069 n.m.)
and significantly larger in $A$=3 (--0.012 n.m.) than obtained here.  The magnetic
moment operator is derived via
\begin{equation}
{\bm \mu}=-\frac{i}{2} \nabla_q \times {\bf j}({\bf q})\Big|_{q=0} \ ,
\end{equation}
and from Eq.~(\ref{eq:j1rcc}) the N$^2$LO-RC term follows as 
\begin{equation}
{\bm \mu}^{\rm N^2LO}_{\rm RC} = -\frac{e}{8\, m_N^3}\sum_{i=1}^A 
\Bigg[ \left[ p_i^2 \, ,\, e_{N,i}\, {\bf L}_i + \mu_{N,i} \, {\bm \sigma}_i\right]_+
+ e_{N,i}\, {\bf p}_i\times({\bm \sigma}_i \times {\bf p}_i) \Bigg] \ ,
\label{eq:murc}
\end{equation}
where ${\bf p}_i=-i\nabla_i$ and ${\bf L}_i$ are the linear momentum and
angular momentum operators of particle $i$, and $[\dots\, ,\, \dots]_+$ denotes
the anticommutator.  The expression for this correction is different from
that given in Ref.~\cite{Song07}, which is, in turn, different from that listed in
an earlier work~\cite{Park00} by some of the same authors of Ref.~\cite{Song07}.
When compared to Eq.~(\ref{eq:murc}) above, for example, the term with the anticommutator
is missing in Ref.~\cite{Park00}, since the authors of that
work normalize the spinors as $\overline{u} u=1$ rather than $u^\dagger u=1$ as in
the present work.  These differences might partly explain the different
contributions calculated here and in Ref.~\cite{Song07} for the N$^2$LO-RC
correction.

Next, we examine the $^1$H($n,\gamma$)$^2$H radiative capture at thermal neutron
energies.  Various aspects of the calculations, which will not be discussed here,
were reviewed most recently in Ref.~\cite{Schiavilla04}.  The calculated values
for the cross section are listed in Table~\ref{tb:tab3}.
\begin{table}[bth]
\begin{tabular}{c|c|c|c||c|c|c}
 &  \multicolumn{3}{c}{AV18} & \multicolumn{3}{c}{CDB} \\
\hline
\hline
 $\Lambda$ (MeV)               & 500 & 600 & 800 & 500  & 600 & 800 \\
\hline
LO               & 304.6    & 304.6    & 304.6    & 306.6    &  306.6    & 306.6    \\
NLO              & 319.1    & 319.6    & 319.9    & 321.3    &  321.7    & 321.9    \\
N$^2$LO-RC       & 317.4    & 317.9    & 318.2    & 319.9    &  320.3    & 320.5    \\
N$^2$LO-$\Delta$ & 321.9    & 323.8    & 326.3    & 323.8    &  325.3    & 327.1    \\
\hline
\hline
\end{tabular}
\caption{Cumulative contributions in mb to the $^1$H($n,\gamma$)$^2$H cross section at thermal neutron
energy, obtained with the AV18 and CDB potential models and cutoff values in the range 500--800 MeV.
The LO and N$^2$LO-RC contributions are cutoff independent, while the matrix element of the N$^2$LO-$\Delta_c$ operator vanishes. The experimental value is 332.6(0.7) mb from
Ref.~\protect\cite{Mughabghab81}.}
\label{tb:tab3}
\end{table}
As remarked earlier in Sec.~\ref{sec:r-sp}, we note that the N$^2$LO-$\Delta_c$ current does not contribute, since
the associated magnetic moment operator,
\begin{equation}
   {\bm \mu}_{\Delta_c}^{\rm N^2LO}= -\,\frac{2\, e\, \mu^*}{9 \, m_N}\, \frac{D_T}{\Delta}\sum_{i<j=1}^A
\Big[ 2\, (\tau_{i,z}\, {\bm \sigma}_i + \tau_{j,z}\, {\bm \sigma}_j)
-({\bm \tau}_i\times {\bm \tau}_j)_z\,
 {\bm \sigma}_i \times{\bm \sigma}_j \Big] h_\Lambda(r_{ij})  \ ,
 \label{eq:m2dlt}
\end{equation}
is readily seen to vanish when acting on the $^1$S$_0$ $n$-$p$ state~\cite{Park03}.  If
$P_{ij}^r$, $P_{ij}^\sigma$, and $P_{ij}^\tau$ denote respectively the space,
spin, and isospin exchange operators, where
\begin{equation}
P_{ij}^\sigma =\frac{1+{\bm \sigma}_i\cdot {\bm \sigma}_j}{2} \ ,
\end{equation} 
and similarly for $P^\tau_{ij}$, then $P^r_{ij}\, P^\sigma_{ij}\, P^\tau_{ij}=-1$ 
for a two-nucleon state.  The identity 
\begin{equation}
{\bm \sigma}_i \times {\bm \sigma}_j=i\, ({\bm \sigma}_i-{\bm \sigma}_j)\, P^\sigma_{ij} \ ,
\end{equation}
and the analogous one for ${\bm \tau}_i\times{\bm \tau}_j$ allow one to express
the magnetic dipole operator in Eq.~(\ref{eq:m2dlt}) as 
\begin{equation}
   {\bm \mu}_{\Delta_c}^{\rm N^2LO}= -\,\frac{2\, e\, \mu^*}{9 \, m_N}\, \frac{D_T}{\Delta}\sum_{i<j=1}^A
 h_\Lambda(r_{ij})\, \Big[ 2\,(\tau_{i,z}\, {\bm \sigma}_i + \tau_{j,z}\, {\bm \sigma}_j)
-(\tau_{i,z}-\tau_{j,z})\, ({\bm \sigma}_i-{\bm \sigma}_j)\, P^r_{ij} \Big]\ .
\end{equation}
When acting on a two-nucleon state of even relative orbital angular
momentum, the square bracket in the equation above reduces to
$(\tau_{i,z}+\tau_{j,z})\, ({\bm \sigma}_i+{\bm \sigma}_j)$, and therefore
vanishes, since this state will have either $S$=0 and $T$=1 or $S$=1 and $T$=0.  Indeed,
in the limit $h_\Lambda(r_{ij}) \rightarrow \delta({\bf r}_{ij})$ the operator
in Eq.~(\ref{eq:m2dlt}) gives no contribution.  It is in this sense that one
can interpret contributions at finite $\Lambda$ as representing
corrections beyond N$^2$LO.

The cutoff dependence and the different short-range behaviors of the AV18 and CDB wave functions
lead to a cross section of ($324.5 \pm 2.6$) mb.  Thus, at N$^2$LO the experimental
value, ($332.6\pm 0.7$) mb~\cite{Mughabghab81}, is underpredicted by roughly 2.5\%.
The LO and NLO contributions calculated here are in agreement with those obtained for
the AV18 in Refs.~\cite{Song07}---which uses the same form for the cutoff
function---up to tiny differences presumably due to numerics.  The N$^2$LO-RC contribution,
however, is found to be considerably larger (in magnitude) here than in Ref.~\cite{Song07},
although both studies predict the same sign for it, opposite to the LO contribution.  Reference~\cite{Song07}
suggests that corrections from current operators at N$^3$LO might resolve the present
discrepancy between experiment and theory (at N$^2$LO), and possibly reduce the model dependence
in the latter.  This issue will be investigated in future work.  Instead, rather than using
the $\Delta$ width and $\gamma$-$N$ data to fix the values of the coupling
constants $h_A$ and $\mu^*$ entering the N$^2$LO-$\Delta$ current, Eq.~(\ref{eq:j1dlt}),
we determine the combination $\mu^* \, h_A$ by reproducing the $n$-$p$ radiative capture
cross section.  In fact, we make the replacement $\mu^*\, h_A \rightarrow
f\, \mu^*\, h_A$, take $\mu^*=3$ n.m. and $h_A=2.191\, g_A$ (as before) and fix
$f$ accordingly.  The latter is listed in Table~\ref{tb:tab4}
\begin{table}[bth]
\begin{tabular}{c|c|c}
\hline
\hline
$\Lambda$ (MeV)  &  AV18 & CDB \\
\hline
500 & 3.352 &  3.234 \\
600 & 2.471 &  2.447 \\
800 & 1.772 &  1.814 \\
\hline
\hline
\end{tabular}
\caption{The parameter $f$ required to reproduce, for a given value of
the cutoff $\Lambda$, the experimental $^1$H($n,\gamma$)$^2$H
cross section at thermal neutron energy with the AV18 or CDB potential models.
See text for notation.}
\label{tb:tab4}
\end{table}
for the various combinations of potentials and cutoffs considered in this work.
The resulting N$^2$LO contribution becomes then as large as the NLO---a somewhat unsettling
feature of the present procedure.  At NLO there is a significant cancellation between the
contributions of the seagull, Eq.~(\ref{eq:nlor1}), and pion-in-flight, Eq.~(\ref{eq:nlor2}),
currents.  This destructive interference persists also for the three-nucleon observables
considered next.  Lastly, it is worth noting that conventional calculations of the
$^1$H($n,\gamma$)$^2$H cross section based on two-body currents constructed to satisfy
current conservation with the potential used to generate the wave
functions, accurately reproduce the measured value~\cite{Marcucci05}.

Results for the isovector combination of the trinucleon magnetic moments are
presented in Table~\ref{tb:tab5}.  Note that the row labeled N$^2$LO-$\Delta$ lists
the contributions obtained with the strength of the N$^2$LO-$\Delta$ current determined
as in Table~\ref{tb:tab4}.  Consequently, they are significantly larger (in magnitude) and
much less cutoff-dependent than those at NLO.  The NLO contribution calculated
\begin{table}[bth]
\begin{tabular}{c|c|c|c||c|c|c}
 &  \multicolumn{3}{c}{AV18/UIX} & \multicolumn{3}{c}{CDB/UIX$^*$} \\
\hline
\hline
 $\Lambda$ (MeV)               & 500 & 600 & 800 & 500  & 600 & 800 \\
\hline
LO               &   --2.159 & --2.159     &  --2.159  & --2.180   & --2.180 & --2.180  \\
NLO              &   --0.156 & --0.197     &  --0.238  & --0.113   & --0.156 & --0.200  \\
N$^2$LO-RC       &    +0.029 &  +0.029     &   +0.029  &  +0.024   &  +0.024 &  +0.024  \\
N$^2$LO-$\Delta$ &   --0.258 & --0.253     &  --0.250  & --0.205   & --0.202 & --0.200  \\
\hline
Sum              &   --2.544 & --2.580    &   --2.618  & --2.474   & --2.514 & --2.556  \\
\hline
\hline
\end{tabular}
\caption{Contributions in units of n.m.~to the isovector combination of the trinucleon
magnetic moments, obtained with the AV18/UIX and CDB/UIX$^*$ Hamiltonian models and
cutoff values in the range 500--800 MeV.  The LO and N$^2$LO-RC contributions
are cutoff independent. The experimental value is --2.553 n.m..}
\label{tb:tab5}
\end{table}
in Ref.~\cite{Song07} with VMC wave functions and a cutoff of 600 MeV
is --0.205 n.m., which is 4\% larger than obtained here.  This is most likely
due to differences in the wave functions (we note, incidentally, that VMC wave
functions are less accurate than HH ones).
We determine the strength of the N$^2$LO-$\Delta_c$ current to reproduce, for
a given cutoff $\Lambda$ and Hamiltonian model, the experimental isovector
magnetic moment.  The resulting values for the parameter $f^\prime$, defined
in Eq.~(\ref{eq:dtc}), are listed in Table~\ref{tb:tab6}.  The violent change
of $f^\prime$ as the cutoff $\Lambda$ is increased is due to the fact that
the short-range behavior of the N$^2$LO-$\Delta_c$ current is governed by
a Gaussian of half-width 2/$\Lambda$.
\begin{table}[bth]
\begin{tabular}{c|c|c}
\hline
\hline
$\Lambda$ (MeV)  &  AV18/UIX & CDB/UIX$^*$ \\
\hline
500 & --3.036 & --38.57  \\
600 & +11.86 &  --25.25 \\
800 & +51.13 &  +3.485 \\
\hline
\hline
\end{tabular}
\caption{The parameter $f^\prime$ required to reproduce, for a given value of
the cutoff $\Lambda$, the experimental isovector combination
of the trinucleon magnetic moments with the AV18/UIX or CDB/UIX$^*$ Hamiltonian
models.}
\label{tb:tab6}
\end{table}

However, with the values of the parameters $f$ and $f^\prime$ fixed as discussed
above, the current up to N$^2$LO is now completely determined.  We can therefore
use it to make predictions for the cross section $\sigma_T$ and photon circular
polarization parameter $R_c$ measured in the reaction $^2$H($n,\gamma$)$^3$H.
At thermal energies this process proceeds through S-wave capture predominantly via
magnetic dipole transitions from the initial doublet $J$=1/2 and quartet $J$=3/2 $n$-$d$
scattering states.  In addition, there is a small contribution due to an electric quadrupole
transition from the initial quartet state.  We adopt here the notation and conventions of
Ref.~\cite{Viviani96} and define
\begin{equation}
m_{22} = \widetilde{M}_1^{\, 0\,1/2\,1/2} \ , \qquad m_{44}=\widetilde{M}_1^{\, 0\,3/2\,3/2} \ ,
\qquad e_{44}=\widetilde{E}_2^{\, 0\,3/2\,3/2} \ ,
\end{equation}
where $\widetilde{M}_\ell^{LSJ}$ and $\widetilde{E}_\ell^{LSJ}$ are the reduced
matrix elements (RME's) of the magnetic and electric multipole operators of order
$\ell$, normalized as in Eq.~(6.3) of Ref.~\cite{Viviani96}.  In terms of these
RME's, the capture total cross section is given by
\begin{equation}
\sigma_T=\frac{2}{9} \frac{\alpha}{v_{\rm rel}} \frac{q^3}{4 \, m_N^2}
\left( |m_{22}|^2 +|m_{44}|^2+|e_{44}|^2 \right) \ ,
\end{equation}
where $\alpha=e^2/(4\pi)$ is the fine structure constant, $v_{\rm rel}$ is
the $d$-$n$ relative velocity, $q$ is the energy of the emitted $\gamma$ ray,
and $m_N$ is the nucleon mass.  Similarly, the circular polarization
$P_\Gamma$ resulting from S-wave capture of a neutron polarized along
the direction ${\bf P}_n$ is proportional to the parameter $R_c$~\cite{Viviani96},
{\it i.e.} $P_\Gamma =R_c \, {\bf P}_n \cdot \hat{\bf q}$,  where
\begin{equation}
R_c=-\frac{1}{3} \left[ 1-\frac{(7/2)|m_{44}|^2+\sqrt{8}\, {\rm Re}(m_{22}m_{44}^*)
+(5/2)|e_{44}|^2+\sqrt{24}\, {\rm Im}(m_{22} e_{44}^*)-\sqrt{3}\, {\rm Im}(m_{44}e_{44}^*)}
{ |m_{22}|^2 +|m_{44}|^2+|e_{44}|^2} \right] \ .
\label{eq:rc_o}
\end{equation}
The predicted RME's are listed in Table~\ref{tb:tab8}, the cross section and parameter $R_c$
in Table~\ref{tb:tab9}.  Note that only results corresponding to the AV18/UIX Hamiltonian model
are available at this time for the reason explained earlier.
\begin{table}[bth]
\begin{tabular}{c|c|c|c||c|c|c||c|c|c}
 &  \multicolumn{3}{c}{$m_{22}$} & \multicolumn{3}{c}{$m_{44}$}& \multicolumn{3}{c}{$e_{44}$} \\
\hline
\hline
 $\Lambda$ (MeV)                   & 500 & 600 & 800 & 500  & 600 & 800 & 500 & 600 & 800  \\
\hline
LO                               & --10.6 & --10.6 & --10.6 & 13.5 & 13.5 & 13.5 & --0.14 & --0.14 & --0.14 \\
LO+NLO                           & --13.2 & --12.5 & --11.5 & 13.3 & 13.4 & 13.5 &  +0.02 &  +0.02 &  +0.02 \\
LO$+ \cdots +$N$^2$LO-RC         & --12.0 & --11.3 & --10.3 & 13.5 & 13.6 & 13.7 &  +0.02 &  +0.02 &  +0.02 \\
LO$+ \cdots +$N$^2$LO-$\Delta$   & --20.2 & --19.4 & --18.4 & 12.6 & 12.7 & 12.8 &  +0.02 &  +0.02 &  +0.02 \\
LO$+ \cdots +$N$^2$LO-$\Delta_c$ & --20.6 & --18.3 & --15.6 & 12.6 & 12.6 & 12.7 &  +0.02 &  +0.02 &  +0.02 \\
\hline
\hline
\end{tabular}
\caption{Cumulative contributions (in fm$^{3/2}$) to the reduced matrix elements
(RME's) of the $^2$H($n,\gamma$)$^3$H reaction at thermal energies, obtained with
the AV18/UIX Hamiltonian model and cutoff values in the range 500-800 MeV.  See text
for notation.  The $m_{22}$ and $m_{44}$ RME's are purely imaginary, while the
$e_{44}$ RME is purely real.}
\label{tb:tab8}
\end{table}
At LO the quartet $m_{44}$ is, in absolute value, about 27\% larger than
the doublet $m_{22}$.  However, the contributions at NLO and N$^2$LO are large
and interfere constructively with those at LO for $m_{22}$, while they are much
smaller and interfere destructively for $m_{44}$.  Consequently, the doublet
$m_{22}$ at N$^2$LO is found to be larger than the quartet $m_{44}$ by
63\%--23\% as the cutoff $\Lambda$ is increased from 500 MeV to 800 MeV.
The calculation of the RME's is carried out with the Monte Carlo (MC)
integration techniques of Ref.~\cite{Viviani96}, and the results reported in
Table~\ref{tb:tab8} are obtained from a random walk consisting
of a large number (of the order of 2M) configurations.  The statistical errors
associated with these MC integrations are typically less than 2\% for $m_{22}$ and
much less than 1\% for $m_{44}$.  However, they are 25\% for $e_{44}$ at LO, and
indeed much larger than the central value at N$^2$LO, so that beyond LO the value
of this RME is consistent with zero.  We note that in S-wave capture the $e_{44}$
RME is predominantly due to transitions S($^2$H) $\rightarrow$ D($^3$H) and
D($^2$H) $\rightarrow$ S($^3$H), where S and D denote S- and D-wave components
in the $^2$H and $^3$H ground states.  In the case of the AV18/UIX Hamiltonian,
the contributions associated with these transitions interfere destructively,
thus producing a small $e_{44}$.  This cancellation was found to be significantly
model dependent in Ref.~\cite{Viviani96}, and use of CDB/UIX$^*$ wave functions would
presumably produce somewhat different results for this RME in view of the considerably
weaker tensor force of CDB relative to AV18 at intermediate and short range.
\begin{table}[bth]
\begin{tabular}{c|c|c|c||c|c|c}
 &  \multicolumn{3}{c}{$\sigma_T$} & \multicolumn{3}{c}{$R_c$}  \\
\hline
\hline
 $\Lambda$ (MeV)   & 500 & 600 & 800 & 500  & 600 & 800  \\
\hline
LO                             & 0.229 & 0.229 & 0.229 & --0.060 & --0.060 & --0.060 \\
LO+NLO                         & 0.272 & 0.260 & 0.243 & --0.218 & --0.182 & --0.123 \\
LO$+\cdots+$N$^2$LO-RC         & 0.252 & 0.241 & 0.226 & --0.152 & --0.109 & --0.041 \\
LO$+\cdots+$N$^2$LO-$\Delta$   & 0.438 & 0.416 & 0.389 & --0.432 & --0.418 & --0.397 \\
LO$+\cdots+$N$^2$LO-$\Delta_c$ & 0.450 & 0.382 & 0.315 & --0.437 & --0.398 & --0.331 \\
\hline
\hline
\end{tabular}
\caption{Cumulative contributions to the cross section $\sigma_T$ (in mb) and photon
polarization parameter $R_c$ of the reaction $^2$H($n,\gamma$)$^3$H at thermal
energies, obtained with the AV18/UIX Hamiltonian model and cutoff values in
the range 500-800 MeV.  The experimental values for $\sigma_T$ and $R_c$ are
respectively ($0.508 \pm 0.015$) mb from Ref.~\cite{Jurney82} and $-0.42 \pm0.03$
from Ref.~\cite{Konijnenberg88}.}
\label{tb:tab9}
\end{table}

At N$^2$LO the cross section is underpredicted by theory by (11--38)\% as
the cutoff is increased from 500 MeV to 800 MeV.  This rather drastic cutoff
dependence is mostly due to the contribution of the N$^2$LO-$\Delta_c$ current.
Indeed removing it leads to a much weaker variation of the cross section---roughly
$\pm 5$\% about the value obtained with $\Lambda=600$ MeV (next to last
row of Table~\ref{tb:tab9}).  It will be interesting to see to what extent, if any,
loop corrections at N$^3$LO will improve the present predictions, and in
particular reduce the cutoff dependence.

The photon polarization parameter is very sensitive to contributions of
NLO and N$^2$LO currents, which produce more than a sixfold increase,
in absolute value, of the LO result, and bring it into much closer agreement
with the measured value.  All results listed in Table~\ref{tb:tab9} for $R_c$
(and $\sigma_T$) include the small $e_{44}$ RME, although it only has a significant
effect for the LO prediction ($R_c$=--0.060 versus --0.072 depending
on whether $e_{44}$ is retained or not).  The cause of the $R_c$ sensitivity
to corrections beyond LO becomes clear by examining the expression
given in Eq.~(\ref{eq:rc_o}).  Neglecting $e_{44}$, it reads
\begin{equation}
1+3\, R_c = \frac{7/2-\sqrt{8}\, |m_{22}/m_{44}|}{1+|m_{22}/m_{44}|^2} \ , 
\end{equation}
and therefore the value attained by $R_c$ is driven
by the ratio $|m_{22}/m_{44}|$, which is about 0.79
at LO and ranges from 1.63 to 1.23 for $\Lambda=500$--800 MeV.

We conclude this section by remarking that recent
calculations of $n$-$d$ capture observables~\cite{Sadeghi06},
based on an effective field theory formulated in terms of
nucleon, deuteron, and triton fields with gradient couplings,
seem to lead to predictions which are in agreement with data.
However, it should be stressed that such a theory cannot be
applied to other processes, for example the $n$-$^3$He capture,
without including additional degrees of freedom.  It is in this
sense of a more limited scope than the approach advocated in the
present work.
\section{Conclusions and Outlook}
\label{sec:concls}

The goals of the present work were twofold:  firstly, to
derive the nuclear electromagnetic current up to one loop,
{\it i.e.}~up to N$^3$LO, within an effective-field-theory
approach including explicit nucleons, $\Delta$-isobars, and
pions, and secondly to implement this formalism in the
calculation of a number of few-nucleon electromagnetic
observables at low energy by using accurate
wave functions---the so-called hybrid approach, advocated,
for example, in Refs.~\cite{Park03,Park96}.
This last objective has been partially carried out here, since results
have been reported only at N$^2$LO, {\it i.e.}~ignoring
loop corrections.

Up to this order, the only isoscalar terms are those
generated in a non-relativistic expansion of the
one-body current: they provide a (cutoff-independent) 1\% correction---relative
to LO---to the deuteron and isoscalar combination of the trinucleon
magnetic moments.  This correction is of opposite sign to the LO
contribution, and therefore increases the underprediction of the corresponding
experimental values from $(0.9\pm 0.3)$\% for the deuteron and $(2.7 \pm 0.9)$\%
for the trinucleons at LO to, respectively, $(1.9 \pm 0.3)$\% and $(3.8 \pm 0.8)$\%
at N$^2$LO.  The spread reflects differences in the short-range behavior of
the AV18 and CDB potentials, in particular the weaker tensor components of
the latter relative to the former in this range.

At NLO, isovector terms arise from the pion seagull and in-flight contributions,
while at N$^2$LO, in addition to the relativistic corrections mentioned above,
isovector terms due to $\Delta$-isobar excitation are also obtained.  The value 
for the combination of coupling constants $(g_A/F_\pi)^2$ entering the NLO two-body
currents is that inferred by an analysis of nucleon-nucleon elastic scattering
data~\cite{Stoks94}.  However, the strengths of the N$^2$LO two-body $\Delta$-excitation
currents, {\it i.e.} the combinations $\mu^*\, g_A\, h_A/F_\pi^2$ and
$\mu^*\, D_T$ in Eqs.~(\ref{eq:j1dlt}) and~(\ref{eq:j2dlt}), respectively, have
been determined, as functions of the cutoff $\Lambda$ and for the
Hamiltonian model of interest, by reproducing the cross section for
$n$-$p$ capture and the isovector combination of the
trinucleon magnetic moments.  This current has then been used to make
predictions---with the AV18/UIX model only---for the cross section $\sigma_T$ and photon
circular polarization parameter $R_c$.  The experimental $\sigma_T$
($|R_c|$) is found to be underestimated by 11\% (overestimated by 4\%) for $\Lambda$=500 MeV
and 38\% (underestimated by 21\%) for $\Lambda$=800 MeV.  We note that the
parameter $R_c$ is mostly sensitive to the ratio of doublet to quartet magnetic
dipole transition matrix elements $|m_{22}/m_{44}|$ (the cross section is proportional
to $|m_{22}|^2+|m_{44}|^2$).

The results display a significant cutoff dependence, particularly so for the
N$^2$LO contributions associated with $\Delta$ isobar degrees of freedom.
Indeed these contributions are much larger than those at NLO.  This is partly
due to the fact that the two NLO (pion seagull and in-flight) terms interfere
destructively.  For example, the seagull (in-flight) contributions to $m_{22}$
and $m_{44}$, in units of fm$^{3/2}$ and for $\Lambda$=500 MeV, are respectively
$-9.1\, i$ ($+6.5\, i$) and $-0.8\, i$ ($+0.6\, i$), which add up to the NLO values $-2.6\, i$ and $-0.2\, i$
from Table~\ref{tb:tab8}.  As a result $\sigma_T=0.425$ mb and $R_c=-0.425$
at LO+NLO (seagull only), which should be compared to $\sigma_T=0.272$ mb and $R_c=-0.218$
at LO+NLO (seagull+in-flight) from the second row of Table~\ref{tb:tab9}.

The relatively large $\Delta$-excitation contributions also point to the
need for including loop corrections at N$^3$LO, which these N$^2$LO
currents, because of the procedure adopted here to determine their strength,
are implicitly making up for.  This is also evident by examining the results
for the $n$-$p$ capture cross section in Table~\ref{tb:tab3}.  Had we chosen
to fix the $N$-$\Delta$ transition axial coupling constant $h_A$ and magnetic
moment $\mu^*$ via, respectively, the $\Delta$ width and $\gamma$-$N$ data at
resonance, the contribution of the N$^2$LO-$\Delta$ current would have been
considerably smaller than that at NLO, and would have fallen more in line
with naive expectations.

The N$^3$LO corrections will presumably reduce the cutoff dependence in the
$n$-$d$ predictions, and hopefully bring theory into more satisfying agreement
with experiment.  For the time being, we only observe that calculations~\cite{Marcucci05}
based on the AV18/UIX Hamiltonian with leading two- and three-body currents
constructed consistently---in the sense of being exactly conserved--- with,
respectively, the AV18 two-nucleon and UIX three-nucleon potentials overestimated
both $\sigma_T$ and $|R_c|$ by about 10\% in the $^2$H($n,\gamma$)$^3$H process
at thermal energies, while at the same time providing an excellent description
of cross section data for the $p$-$d$ radiative capture in the energy range from a few to 80 keV,
and, in particular, of the astrophysical factor at zero energy extrapolated from these
data.

Thus, as already emphasized in the Introduction, very low-energy radiative
(and weak) capture reactions involving three- and four-body nuclei
constitute a crucial testing ground for the models used to describe
the ground- and scattering-state wave functions---and indirectly,
the underlying interactions which generate these wave functions---and the
many-body electroweak current operators~\cite{Carlson98}.  

The next stage in the research program we have undertaken is to incorporate
the N$^3$LO operators derived here into the calculations of the captures and
magnetic moments involving light nuclei (with mass number $A \leq 8$), and
indeed to extend these calculations to also include $p$-$d$ capture at
energies up to a few MeV's, and possibly four-nucleon processes, in particular
$^3$He($n,\gamma$)$^4$He at thermal energies.  Of course, at N$^3$LO
three-body currents also occur, and will need to be derived.  Work along these
lines is being pursued vigorously.
\section*{Acknowledgments}
We would like to thank L.\ Girlanda,
A.\ Kievsky, L.E.\ Marcucci, and M.\ Viviani
for discussions and for letting us use their trinucleon wave functions,
E.\ Epelbaum and S.\ K\"olling for an interesting conversation and for
sharing with us some of their results on the same topic before publication,
and V.\ Cirigliano for discussions in the early phase of this work.
One of the authors (R.S.) would also like to thank the Physics Department of
the University of Pisa, the INFN Pisa branch, and especially the Pisa group
for the support and warm hospitality extended to him on several occasions.

The work R.S.\ is supported by the U.S.~Department of Energy,
Office of Nuclear Physics, under contract DE-AC05-06OR23177, and that of J.L.G.\ by
the National Science Foundation under grant PHY-0555559. The calculations were
made possible by grants of computing time from the National Energy Research
Supercomputer Center.
%
%

\appendix
\section{Vertices}
\label{app:vert}

The interaction Hamiltonians in Secs.~\ref{sec:h-s} and~\ref{sec:h-e} are assumed to be
normal-ordered.  Explicit expressions for the associated vertices are easily derived (these
expressions include the $1/\sqrt{2\omega_{k_i}}$ factors from pion fields) :
\vspace{0.5cm}

\noindent i) Strong-interaction vertices:
\begin{eqnarray}
\label{eq:a1}
 \langle {\bf p}^\prime ,\chi^\prime;{\bf k},a\mid 
 H_{\pi NN}\mid {\bf p},\chi\rangle &=&
 -i\frac{g_A}{F_\pi} \frac{ {\bm \sigma}\cdot {\bf k}}
 {\sqrt{2\, \omega}_k}\, \tau_a \ ,  \\
 \langle {\bf p}_\Delta^\prime ,\chi^\prime_\Delta;{\bf k},a\mid 
 H_{\pi N\Delta}\mid {\bf p},\chi\rangle &=& 
 -i\frac{h_A}{F_\pi} \frac{ {\bf S}\cdot {\bf k}} {\sqrt{2\, \omega}_k}\, T_a 
 \label{eq:pnd}\ , \\
 \langle {\bf p}^\prime ,\chi^\prime;{\bf k}_1,a; {\bf k}_2,b \mid 
 H_{\pi \pi NN}\mid {\bf p},\chi \rangle &=& 
-\frac{i}{F_\pi^2} \frac{\omega_{k_1}-\omega_{k_2}}
 {\sqrt{4 \, \omega_{k_1} \, \omega_{k_2}}} \, \epsilon_{abc} \tau_c
\label{eq:a3} \ , 
\end{eqnarray}
\begin{eqnarray}
 \langle {\bf p}_1^\prime ,\chi_1^\prime; 
         {\bf p}_2^\prime, \chi_2^\prime\mid 
 H_{{\rm CT},1}\mid {\bf p}_1 ,\chi_1; 
                    {\bf p}_2 ,\chi_2\rangle  &=& \sum_{\alpha=S,T} C_\alpha\,
{\bm \Gamma}_{1 \alpha} \cdot {\bm \Gamma}_{2 \alpha} \ , \\
\langle {\bf p}^\prime ,\chi^\prime; 
         {\bf p}^\prime_\Delta ,\chi^\prime_\Delta \mid 
 H_{{\rm CT},2}\mid {\bf p}_1 ,\chi_1; 
         {\bf p}_2 ,\chi_2 \rangle &=& D_T \,\, 
{\bm \tau}_1 \cdot {\bf T}_2 \, {\bm \sigma}_1 \cdot {\bf S}_2 \ , \\
\langle {\bf p}^\prime ,\chi^\prime; 
         {\bf p}^\prime_\Delta ,\chi^\prime_\Delta \mid 
 H_{{\rm CT},3}\mid {\bf p} ,\chi; 
         {\bf p}_\Delta ,\chi_\Delta \rangle &=&
	  \sum_{\alpha=S,T} C^\prime_\alpha\,
{\bm \Gamma}_{1 \alpha} \cdot {\bm \Gamma}^\prime_{2 \alpha} \ ,
\end{eqnarray}
\begin{eqnarray}
\langle {\bf p}_{1,\Delta}^\prime ,\chi_{1,\Delta}^\prime; 
{\bf p}^\prime_{2,\Delta} ,\chi^\prime_{2,\Delta} \mid 
 H_{{\rm CT},4}\mid {\bf p}_1 ,\chi_1;
 {\bf p}_2 ,\chi_2 \rangle &=& D_T^\prime\,\,
{\bf T}_1 \cdot {\bf T}_2 \, {\bf S}_1 \cdot {\bf S}_2 \ , \\
\langle {\bf p}_\Delta^\prime ,\chi_\Delta^\prime; 
{\bf p}^\prime ,\chi^\prime \mid 
 H_{{\rm CT},5}\mid {\bf p} ,\chi;
 {\bf p}_\Delta ,\chi_\Delta \rangle &=& D_T^{\prime\prime}\,\,
{\bf T}_1 \cdot {\bf T}^\dagger_2 \, {\bf S}_1 \cdot {\bf S}^\dagger_2 \ , 	  
\end{eqnarray}
\begin{eqnarray}
\langle{\bf p}^\prime,\chi^\prime; {\bf k}_1,a; {\bf k}_2,b;{\bf k}_3,c \mid 
 H_{3\pi}\mid {\bf p} ,\chi \rangle &=&  
 \frac{2\, i \, g_A}{F_\pi^3} \frac{1}{\sqrt{8\, \omega_{k_1} \, \omega_{k_2} \, 
 \omega_{k_3}} } 
 \Big( {\bm \sigma}\cdot {\bf k}_1 \, \tau_a\, \delta_{bc} \nonumber \\ 
        &+&{\bm \sigma}\cdot {\bf k}_2 \, \tau_b\, \delta_{ca} 
        +{\bm \sigma}\cdot {\bf k}_3 \, \tau_c\, \delta_{ab} \Big) \ ,
\label{eq:a9}
\end{eqnarray}
\begin{eqnarray}
 \langle {\bf k}_1,a; 
{\bf k}_2,b; {\bf k}_3,c; {\bf k}_4,d\mid  H_{4\pi}\mid 0 \rangle &=& - 
 \frac{4}{F_\pi^2} \frac{1}{\sqrt{16\, \omega_{k_1} \, \omega_{k_2} \, 
 \omega_{k_3}\, \omega_{k_4} } } \Big[ 
 \delta_{ab} \, \delta_{cd} 
\left(  k_{1\mu} k_2^\mu + k_{3\mu} k_4^\mu + m^2_\pi \right)
 \nonumber \\
 &+&\delta_{ac} \, \delta_{bd} 
\left(  k_{1\mu} k_3^\mu + k_{2\mu} k_4^\mu + m^2_\pi \right) \nonumber \\
 &+&\delta_{ad} \, \delta_{bc} 
\left(  k_{1\mu} k_4^\mu + k_{2\mu} k_3^\mu + m^2_\pi \right)\Big] \ ;
\label{eq:4pi2}
\end{eqnarray}

\noindent ii) Electromagnetic-interaction vertices:
\begin{eqnarray}
 \langle {\bf p}^\prime ,\chi^\prime;{\bf k},a\mid 
 H_{\gamma  \pi NN}\mid {\bf p},\chi; {\bf q},\lambda\rangle &=&
 e\frac{g_A}{F_\pi}\frac{{\bm \sigma}}{\sqrt{2\, \omega}_k} \cdot
 \frac{\hat{\bf e}_{{\bf q} \lambda}}
{\sqrt{2\, \omega}_q}\, \epsilon_{zab} \tau_b \ ,	\\
 \langle {\bf p}^\prime ,\chi^\prime_\Delta ;{\bf k},a\mid 
 H_{\gamma \pi N\Delta}\mid {\bf p},\chi; {\bf q},\lambda\rangle &=&
 e\frac{h_A}{F_\pi}\frac{\bf S\frac{}{}}{\sqrt{2\, \omega}_k} \cdot
 \frac{\hat{\bf e}_{{\bf q} \lambda}}{\sqrt{2\, \omega}_q}\, \epsilon_{zab} T_b 
 \label{eq:gpnd} \ , 
 \end{eqnarray}
 \begin{equation}
 \langle {\bf k}_1,a; {\bf k}_2,b \mid 
 H_{\gamma \pi\pi}\mid {\bf q},\lambda\rangle = i \, e \, \frac{{\bf k}_1-{\bf k}_2}
 {\sqrt{4 \, \omega_{k_1} \, \omega_{k_2}}} 
\cdot \frac{\hat{\bf e}_{{\bf q} \lambda}}{\sqrt{2\, \omega}_q}
 \, \epsilon_{zab} \ , 
\label{eq:a13}
\end{equation}
\begin{eqnarray}
 \langle {\bf p}^\prime ,\chi^\prime;{\bf k}_1,a;{\bf k}_2,b\mid 
 H_{\gamma  \pi \pi NN}\mid {\bf p},\chi; {\bf q},\lambda\rangle &=&
 -\frac{e}{F_\pi^2} \frac{1}{\sqrt{4\, \omega_{k_1}\, \omega_{k_2}}} \nonumber \\
\frac{\hat{\bf e}_{{\bf q} \lambda}}{\sqrt{2\, \omega}_q} \cdot \frac
 {({\bf p}^\prime+{\bf p}) +i\, {\bm \sigma}\times 
({\bf p}^\prime -{\bf p})}{2\, m_N}&&\!\!\!\!\!\!\!
 (\delta_{az}\, \tau_b +\delta_{bz}\, \tau_a -2\, \delta_{ab}\, \tau_z) \ ,
 \end{eqnarray} 
\begin{eqnarray}
\langle {\bf p}^\prime ,\chi^\prime;{\bf k}_1,a;{\bf k}_2,b; {\bf k}_3,c \mid  H_{\gamma\,3\pi}
\mid {\bf p},\chi; {\bf q},\lambda\rangle &=& -
2\, e \frac{g_A}{F^3_\pi}\frac{{\bm \sigma}}{\sqrt{8\, \omega_{k_1}\, \omega_{k_2}\,
\omega_{k_3} } }\cdot \frac{\hat{\bf e}_{{\bf q} \lambda}}{\sqrt{2\, \omega}_q} \nonumber \\
&&\tau_d\, \left( \epsilon_{zad} \, \delta_{bc} +\epsilon_{zbd} \, \delta_{ca} +
\epsilon_{zcd} \, \delta_{ab} \right)  \ ,
\end{eqnarray}
\begin{eqnarray}
 \langle  {\bf k}_1,a; {\bf k}_2,b;{\bf k}_3,c;{\bf k}_4,d \mid  H_{\gamma\, 4\pi}
\mid  {\bf q},\lambda\rangle &=& -i\, e\, \frac{4}{F_\pi^2} 
\frac{1}{\sqrt{ 16\,\omega_{k_1} \, \omega_{k_2} \, \omega_{k_3}\, \omega_{k_4} } }
\frac{\hat{\bf e}_{{\bf q} \lambda}}{\sqrt{2\, \omega}_q} \nonumber \\
\Big[\!\!\!\!\!&&\delta_{cd}\, \epsilon_{zab} ({\bf k}_1-{\bf k}_2)
     +\delta_{ab}\, \epsilon_{zcd} ({\bf k}_3-{\bf k}_4)  \nonumber \\
      &+& \delta_{bd}\, \epsilon_{zac} ({\bf k}_1-{\bf k}_3)
+\delta_{ad}\, \epsilon_{zbc} ({\bf k}_2-{\bf k}_3) \nonumber \\
      &+&\delta_{ac}\, \epsilon_{zbd} ({\bf k}_2-{\bf k}_4)
+\delta_{bc}\, \epsilon_{zad} ({\bf k}_1-{\bf k}_4)\Big] \ ,
\label{eq:a16}
\end{eqnarray}
\begin{eqnarray}
 \langle {\bf p}^\prime ,\chi^\prime \mid  H_{\gamma NN}
\mid {\bf p},\chi; {\bf q},\lambda\rangle &=&
-\frac{e}{2\, m_N} \frac{\hat{\bf e}_{{\bf q} \lambda}}{\sqrt{2\, \omega}_q} \cdot
\Big[ e_N \, ({\bf p}^\prime+{\bf p}) +i\,\mu_n\, {\bm \sigma}\times {\bf q }\Big] \ , \\
\langle {\bf p}_\Delta^\prime ,\chi_\Delta^\prime \mid  H_{\gamma N\Delta}
\mid {\bf p},\chi; {\bf q},\lambda\rangle &=&
-i \frac{e\, \mu^*}{2\, m_N} \frac{\hat{\bf e}_{{\bf q} \lambda}}{\sqrt{2\, \omega}_q} 
\cdot  {\bf S}\times {\bf q } \, T_z \ .
\end{eqnarray}
In these expressions ${\bf p}$ and ${\bf p}_\Delta$ denote nucleon and
$\Delta$-isobar momenta in spin-isospin states specified by $\chi$ and
$\chi_\Delta$ respectively, while the ${\bf k}$'s and $a,b,\dots$ denote 
pion momenta in isospin states $a,b,\dots$, and ${\bf q}$ and $\lambda$
the photon momentum and polarization state.  For  brevity, on the r.h.s.~of
the equations above the spin-isospin states of the nucleon and $\Delta$
isobar as well as the $\delta$-functions enforcing three-momentum
conservation, are not shown explicitly.  In Eq.~(\ref{eq:4pi2}), the
notation $k_i^\mu k_{j \mu}$ denotes the combination
$\omega_{k_i} \omega_{k_j}-{\bf k}_i \cdot {\bf k}_j$.  Finally,
vertices involving $\Delta$-isobar deexcitation into a nucleon are obtained
by replacing ${\bf S}$ and ${\bf T}$ by their adjoint operators ${\bf S}^\dagger$
and ${\bf T}^\dagger$, while vertices in which
one or more pions are in the initial state are obtained from those listed
in Eqs.~(\ref{eq:a1})--(\ref{eq:a3}), (\ref{eq:a9}), (\ref{eq:4pi2}), (\ref{eq:a13}),
and (\ref{eq:a16}) by replacing
${\bf k}_i \rightarrow -{\bf k}_i$ and/or
$\omega_{k_i} \rightarrow -\omega_{k_i}$ (of course, the
energy replacements are not to be carried out
in the pion-field normalization factors).  For example,
\begin{equation}
 \langle {\bf p}^\prime ,\chi^\prime;{\bf k}_1,a\mid 
 H_{\pi \pi NN}\mid {\bf p},\chi; {\bf k}_2,b  \rangle = 
-\frac{i}{F_\pi^2} \frac{\omega_{k_1}+\omega_{k_2}}
 {\sqrt{4 \, \omega_{k_1} \, \omega_{k_2}}} \, \epsilon_{abc} \tau_c \ .
 \end{equation}
\section{Configuration-space representation}
\label{app:fns}

We list here the configuration-space representation of the
two-body currents at NLO and N$^2$LO.  To this end, it is
convenient to define $z_\pi \equiv m_\pi \, r$,
$z_\Lambda \equiv \Lambda\, r$, $z_L \equiv r\, L(q;x)$,
\begin{equation}
 z_{\pm} \equiv m_\pi/\Lambda \pm z_\Lambda/2 \ , \qquad 
 z^*_{\pm} \equiv L(q;x)/\Lambda \pm z_\Lambda/2 \ ,
\end{equation}
and the complement error function
\begin{equation}
 \phi(z) \equiv \frac{2}{\sqrt{\pi}} \int_{z}^\infty {\rm d}t \, {\rm e}^{-t^2} \ .
\end{equation}
The dependence of $z_L$ and $z^*_{\pm}$ upon the variable $x$ is understood. 
Then the functions $f_\Lambda(r)$, Eq.~(\ref{eq:flb}), along with its first
derivative $f_\Lambda^\prime(r)$, and $\nabla g_\Lambda({\bf r},{\bf q})$,
Eq.~(\ref{eq:glb}), are given by
\begin{equation}
 f_\Lambda(r)=\frac{m_\pi}{8\, \pi}\, \frac{{\rm e}^{m^2_\pi/\Lambda^2}}{z_\pi}
 \left[\phi(z_-)\, {\rm e}^{-z_\pi} -\phi(z_+)\, {\rm e}^{z_\pi} \right] \ ,
 \end{equation}
 \begin{equation}
f_\Lambda^\prime(r)= 
\frac{ m_\pi^2}{8 \pi}\, \frac{{\rm e}^{m_\pi^2/\Lambda^2}}{\,z^2_\pi}
 \Big[\phi(z_+)\, {\rm e}^{ z_\pi}\left( 1-z_\pi \right)
  -\phi(z_-)\, {\rm e}^{-z_\pi}\left( 1+z_\pi \right) \Big]
  +\frac{ \Lambda^2 }{4\pi\sqrt{\pi}} \,\frac{ {\rm e}^{-z^2_\Lambda/4} }{z_\Lambda} \  , 
\end{equation}
\begin{equation} 
\nabla g_\Lambda({\bf r},{\bf q})\Big|_\perp =\hat{\bf r}\, \int_{-1}^{+1} {\rm d}x\, 
 {\rm e}^{-i\,x\, {\bf q}\cdot {\bf r}/2}\, E_q(x,r) \ , 
\end{equation}
where only the transverse part of $\nabla g_\Lambda({\bf r},{\bf q})$
(orthogonal to the photon momentum ${\bf q}$) is of interest, and
\begin{equation}
 E_q(x,r)\! =\!\frac{ {\rm e}^{L^2(q;x)/\Lambda^2} }{8 \pi\,z^2_\Lambda}
 \Big[\phi(z^*_+)\, {\rm e}^{ z_L}\left( 1-z_L-z_\Lambda^2/2 \right)
  -\phi(z^*_-)\, {\rm e}^{-z_L}\left( 1+z_L-z_\Lambda^2/2 \right) \Big]
  + \frac{ {\rm e}^{-z^2_\Lambda/4} }{4 \pi \sqrt{\pi}\, z_\Lambda} \ .
\end{equation}
In the limit $\Lambda \rightarrow \infty$ these functions reduce to:
\begin{equation}
f_\infty(r)=\frac{m_\pi}{4\, \pi}\, \frac{ {\rm e}^{-z_\pi} }{z_\pi} \ ,\qquad
f_\infty^\prime(r)=-\frac{ m_\pi^2}{4 \pi}\, \frac{ {\rm e}^{-z_\pi} }{z_\pi^2} 
\left( 1+z_\pi \right) \ , \qquad
E_{q,\infty} (r)= \frac{ {\rm e}^{-z_L} }{8\, \pi} \ .
\end{equation}
The complete NLO current---the sum of the two contributions
in Eqs.~(\ref{eq:nlor1}) and~(\ref{eq:nlor2})---is then written as
\begin{eqnarray}
 {\bf j}^{\rm NLO}({\bf q})\Big|_\perp\!\!&=&\!\!e\frac{g_A^2}{F_\pi^2} ({\bm \tau}_1\times{\bm \tau}_2)_z \Bigg\{
 {\rm e}^{i{\bf q}\cdot {\bf r}_1} f_\Lambda^\prime(r)\, 
 {\bm \sigma}_1 ({\bm \sigma}_2 \cdot \hat{\bf r})  +
 {\rm e}^{i{\bf q}\cdot {\bf r}_2} f_\Lambda^\prime(r)\,
 {\bm \sigma}_2 ({\bm \sigma}_1 \cdot \hat{\bf r}) \nonumber \\
 &+&\!\!{\rm e}^{i{\bf q}\cdot {\bf R}} \Bigg[ \frac{g_\Lambda^{(1)}({\bf r},{\bf q})}{r^2}
 \left[ {\bm \sigma}_1 ({\bm \sigma}_2 \cdot \hat{\bf r}) +
 {\bm \sigma}_2 ({\bm \sigma}_1 \cdot \hat{\bf r})+
 \hat{\bf r} ({\bm \sigma}_1 \cdot {\bm \sigma}_2) \right]
 +i\frac{g_\Lambda^{(2)}({\bf r},{\bf q})}{r}{\bm \sigma}_1 ({\bm \sigma}_2 \cdot {\bf q}) \nonumber \\
&-&\!\!i\frac{g_\Lambda^{(2)}(-{\bf r},{\bf q})}{r}{\bm \sigma}_2 ({\bm \sigma}_1 \cdot {\bf q})
-i\frac{g_\Lambda^{(3)}({\bf r},{\bf q})}{r}\hat{\bf r}\, ({\bm \sigma}_1 \cdot \hat{\bf r})
 ({\bm \sigma}_2 \cdot {\bf q})
 +i\frac{g_\Lambda^{(3)}(-{\bf r},{\bf q})}{r}\hat{\bf r}\, ({\bm \sigma}_1 \cdot {\bf q})
 ({\bm \sigma}_2 \cdot \hat {\bf r}) \nonumber \\
 &-&\!\!g_\Lambda^{(4)}({\bf r},{\bf q})\, \hat{\bf r}\, ({\bm \sigma}_1 \cdot {\bf q})
 ({\bm \sigma}_2 \cdot {\bf q}) 
 -\frac{g_\Lambda^{(5)}({\bf r},{\bf q})}{r^2}\, \hat{\bf r}\, ({\bm \sigma}_1 \cdot \hat{\bf r})
 ({\bm \sigma}_2 \cdot \hat {\bf r}) \Bigg] \Bigg\} \ ,
\end{eqnarray}
where ${\bf r}={\bf r}_1-{\bf r}_2$ and ${\bf R}=({\bf r}_1+{\bf r}_2)/2$, and
the functions $g_\Lambda^{(i)}$ with $i=1,\dots,5$ are defined as
\begin{eqnarray}
 g_\Lambda^{(1)}({\bf r},{\bf q})&=&\int_{-1}^{+1} {\rm d}x\, 
 {\rm e}^{-i\,x\, {\bf q}\cdot {\bf r}/2}\, \left( 1-r\frac{{\rm d}}{{\rm d}r}\right)E_q(x,r) \ ,\\
 g_\Lambda^{(2)}({\bf r},{\bf q})&=&\frac{1}{2}\int_{-1}^{+1} {\rm d}x\, 
 {\rm e}^{-i\,x\, {\bf q}\cdot {\bf r}/2}\,
 \left(1+x\right)\, E_q(x,r) \ ,\\
 g_\Lambda^{(3)}({\bf r},{\bf q})&=&\frac{1}{2}\int_{-1}^{+1} {\rm d}x\, 
 {\rm e}^{-i\,x\, {\bf q}\cdot {\bf r}/2}\,
 \left(1+x\right)\, \left( 1-r\frac{{\rm d}}{{\rm d}r}\right)\,E_q(x,r)  \ ,\\
 g_\Lambda^{(4)}({\bf r},{\bf q})&=&\frac{1}{4}\int_{-1}^{+1} {\rm d}x\, 
 {\rm e}^{-i\,x\, {\bf q}\cdot {\bf r}/2}\,
 \left(1-x^2\right)\,E_q(x,r)  \ ,\\
 g_\Lambda^{(5)}({\bf r},{\bf q})&=&\int_{-1}^{+1} {\rm d}x\, 
 {\rm e}^{-i\,x\, {\bf q}\cdot {\bf r}/2}\,
 \left( 3-3\,r\frac{{\rm d}}{{\rm d}r}+r^2\frac{{\rm d}^2}{{\rm d}r^2} \right)\,E_q(x,r)  \ .
\end{eqnarray}

The configuration-space representation of the N$^2$LO
current in Eq.~(\ref{eq:j1dlt}) reads
\begin{eqnarray}
  {\bf j}^{\rm N^2LO}_{\rm b-g}({\bf q})&=& i\,\frac{ e\, \mu^*}{9 \, m_N}\,
 \frac{g_A\, h_A}{\Delta\, F_\pi^2}\, {\rm e}^{i\,{\bf q}\cdot {\bf r}_1}\, {\bf q} \times
 \Bigg[ 4\, \tau_{2,z}\,
 \Big[h_S(r)\, {\bm \sigma}_2 +h_T(r)\, \hat{\bf r}\, ({\bm \sigma}_2 \cdot \hat{\bf r})\Big] 
 \nonumber \\
 &-&({\bm \tau}_1\times{\bm \tau}_2)_z 
 \Big[h_S(r)\, {\bm \sigma}_1\times{\bm \sigma}_2
  +h_T(r)\, ({\bm \sigma}_1 \times \hat{\bf r})
  ({\bm \sigma}_2 \cdot \hat{\bf r})\Big] \Bigg]
 +1 \rightleftharpoons 2 \ ,
 \end{eqnarray}
where $h_S(r) = f_\Lambda^\prime(r)/r$ and 
\begin{eqnarray}
h_T(r) &=& f_\Lambda^{\prime\prime}(r)-f_\Lambda^\prime(r)/r =
 \frac{ m_\pi^3}{8 \pi}\, \frac{{\rm e}^{m_\pi^2/\Lambda^2}}{\,z^3_\pi}
 \Big[\phi(z_-)\, {\rm e}^{ -z_\pi}\left( 3+3\,z_\pi+z_\pi^2 \right) \nonumber \\
  &&-\phi(z_+)\, {\rm e}^{z_\pi}\left( 3-3\,z_\pi+z_\pi^2 \right) \Big]
  -\frac{ \Lambda^3 }{8\pi\sqrt{\pi}} \,\frac{ {\rm e}^{-z^2_\Lambda/4} }{z^2_\Lambda}
  \left( 6+z_\Lambda^2\right) \ ,
\end{eqnarray}
and again in the limit $\Lambda \rightarrow \infty$,
\begin{equation}
h_{S,\infty}(r)=-\frac{ m_\pi^3}{4 \pi}\, \frac{ {\rm e}^{-z_\pi} }{z_\pi^3} \left( 1+z_\pi \right) \ , \qquad
h_{T,\infty}(r)= \frac{ m_\pi^3}{4 \pi}\, \frac{ {\rm e}^{-z_\pi} }{z_\pi^3} \left( 3+3\, z_\pi
+z_\pi^2 \right)  \ .
\end{equation}
\section{One-loop two-body currents with $\Delta$ isobars}
\label{app:l2pid}

We begin by including a single $\Delta$ isobar in the intermediate
states.  The relevant diagrams are shown in Fig.~\ref{fig:fig11}.
We find for type a) and b) diagrams:
\begin{eqnarray}
{\rm type\,\, a)}&=&-e\frac{h^2_A}{F^4_\pi} \, 
\left(2\, \tau_{2,z}-T^\dagger_{1,z} \, {\bf T}_1 \cdot {{\bm \tau}_2}\right)
 \int \frac{({\bf S}^\dagger_1 \cdot {\bf q}_2) \, 
 {\bf S}_1}{(\omega_1+\Delta)(\omega_2+\Delta) (\omega_1 +\omega_2) }  - {\rm h.c.} \ ,
 \label{eq:dia_ad} \\
{\rm type\,\, b)}\!\!&=&\!\! e\frac{h^2_A}{F^4_\pi} \,
\left(2\, \tau_{2,z}-T^\dagger_{1,z} \, {\bf T}_1 \cdot {{\bm \tau}_2}\right)
 \int ({\bf q}_1-{\bf q}_3)
 \, ({\bf S}^\dagger_1 \cdot {\bf q}_2) \, 
  ({\bf S}_1 \cdot {\bf q}_1)  \nonumber \\
 \!\!&\times&\!\!\frac{\omega_1+\omega_2+\omega_3+\Delta}
  {(\omega_1+\Delta)
  (\omega_2+\Delta) (\omega_3+\Delta)
   (\omega_1+\omega_2)(\omega_1+\omega_3)(\omega_2+\omega_3)} \,
  - {\rm h.c.} \ , 
\end{eqnarray}
where ${\bf S}$ and ${\bf T}$ are the spin- and isospin-transition operators
defined in Eq.~(\ref{eq:stran}), and $\Delta$ denotes $m_\Delta -m_N$.
The spin-isospin structures can be
further simplified and expressed in terms of the Pauli matrices ${\bm \sigma}$
and ${\bm \tau}$.

The contributions of type c)-e) diagrams can be written as
\begin{eqnarray}
{\rm type\,\, c)}&=& \int
\left[ -\frac{v^{\pi\, \dagger}_{\Delta N} ({\bf q}_2) \, {\bf j}^\pi_{\Delta N}
({\bf q}_1)}{\Delta} +{\bf j}^{(-)}_{\rm c}({\bf q}_1,{\bf q}_2) \right] - {\rm h.c.}  \ ,
 \label{eq:dia_cd} \\
{\rm type\,\, d)}&=& \int
\left[ -\frac{v^{\pi \, \dagger}_{N\Delta} ({\bf q}_2) \, {\bf j}^\pi_{N \Delta}
({\bf q}_1)}{\Delta} +{\bf j}^{(-)}_{\rm d}({\bf q}_1,{\bf q}_2) \right] - {\rm h.c.}  \ ,
 \label{eq:dia_dd} \\
 {\rm type\,\, e)}&=&\int
\left[ -\frac{v^{\pi\, \dagger}_{\Delta N} ({\bf q}_2) \, {\bf j}^{\pi \pi}_{\Delta N}
({\bf q}_1,{\bf q}_3)}{\Delta} +{\bf j}^{(-)}_{\rm e}({\bf q}_1,{\bf q}_2,{\bf q}_3) \right] 
- {\rm h.c.}  \ ,\label{eq:dia_ed}
 \end{eqnarray}
where we have defined the one-pion-exchange transition potential 
$NN \rightarrow \Delta N$ as
\begin{equation}
 v^\pi_{\Delta N}({\bf q}_2)=-\frac{g_A\, h_A}{F_\pi^2} 
 \frac{({\bf S}_1\cdot {\bf q}_2) \, ({\bm \sigma}_2 \cdot {\bf q}_2)}{q_2^2+m_\pi^2}\,
 {\bf T}_1\cdot {\bm \tau}_2 \ ,
 \label{eq:vnd}
\end{equation}
the transition currents $\gamma NN \rightarrow \Delta N$ associated
with the seagull $\gamma \pi N \Delta$-coupling and pion-in-flight terms
respectively as
\begin{eqnarray}
 {\bf j}^\pi_{\Delta N}({\bf q}_1)&=&-i\, e \frac{g_A\, h_A}{F_\pi^2} 
 \frac{{\bf S}_1 \, ({\bm \sigma}_2 \cdot {\bf q}_1)}{q_1^2+m_\pi^2} 
 \, ({\bf T}_1 \times {\bm \tau}_2)_z \ , \label{eq:jpid} \\
 {\bf j}^{\pi\pi}_{\Delta N}({\bf q}_1,{\bf q}_3)&=& i\, e\frac{g_A\, h_A}{F^2_\pi}\, 
\frac{{\bf q}_1-{\bf q}_3}{(q^2_1+m_\pi^2) (q_3^2+m_\pi^2)}
  ({\bf S}_1\cdot {\bf q}_1)\, ({\bm \sigma}_2\cdot {\bf q}_3)
  \, ({\bf T}_1 \times {\bm \tau}_2)_z \ , \label{eq:jpipid}
\end{eqnarray}
and $v^\pi_{N\Delta}$ and ${\bf j}^\pi_{N \Delta}$ are obtained from 
$v^\pi_{\Delta N}$ and ${\bf j}^\pi_{\Delta N}$ by the replacements
${\bf S} \rightleftharpoons {\bm \sigma}$ and ${\bf T} \rightleftharpoons {\bm \tau}$.
The contributions labeled ${\bf j}^{(-)}_{\rm c,d,e}$ are given by
\begin{eqnarray}
{\bf j}^{(-)}_{\rm c}({\bf q}_1,{\bf q}_2) &=&  
 - e\frac{(g_A \, h_A)^2}{F^4_\pi}\,f_1^\Delta(\omega_1,\omega_2)
  ({\bf S}^\dagger_1\cdot {\bf q}_2)\, {\bf S}_1 \, \bigg[ ({\bf T}^\dagger_1\times{\bf T}_1)_z\,
{\bm \sigma}_2\cdot({\bf q}_1\times{\bf q}_2) \nonumber \\ 
&+& \left(2\, \tau_{2,z}-T^\dagger_{1,z} \, {\bf T}_1 \cdot {{\bm \tau}_2}\right)\,
{\bf q}_1\cdot{\bf q}_2 \bigg] \ , \\
{\bf j}^{(-)}_{\rm d}({\bf q}_1,{\bf q}_2) &=& 
i\, e\frac{(g_A\, h_A)^2}{2\, F^4_\pi}\,f_1^\Delta(\omega_1,\omega_2)\,
({\bm \sigma}_1\cdot {\bf q}_2)\, {\bm \sigma}_1 \nonumber \\
&\times& \bigg[ \!({\bf S}^\dagger_2\cdot {\bf q}_1) \, ({\bf S}_2\cdot {\bf q}_2) 
\left[ ({\bf T}^\dagger_2\times{\bf T}_2)_z +i(2 \, \tau_{1,z}-{\bm \tau}_1\cdot{\bf T}^\dagger_2 \, T_{2,z}) \right]\!+\! {\rm h.c.}\bigg] \ , \\
{\bf j}^{(-)}_{\rm e}({\bf q}_1,{\bf q}_2,{\bf q}_3) &=&  e\frac{(g_A \, h_A)^2}{ F^4_\pi}\,f_2^\Delta(\omega_1,\omega_2,\omega_3)\,
 ({\bf q}_1-{\bf q}_3)\,({\bf S}^\dagger_1\cdot {\bf q}_2)\, ({\bf S}_1 \cdot {\bf q}_1) \nonumber \\
 &\times& \bigg[ ({\bf T}^\dagger_1\times{\bf T}_1)_z\,{\bm \sigma}_2\cdot({\bf q}_3\times{\bf q}_2) 
+\left(2\, \tau_{2,z}-T^\dagger_{1,z} \, {\bf T}_1 \cdot {{\bm \tau}_2}\right){\bf q}_2\cdot{\bf q}_3 \bigg] \ , 
\end{eqnarray}
where the functions $f_1^\Delta(\omega_1,\omega_2)$ 
and $f_2^\Delta(\omega_1,\omega_2,\omega_3)$
denote the following combinations of pion energies and $\Delta N$ mass differences:
 \begin{equation}
 f_1^\Delta(\omega_1,\omega_2) = \frac{(\omega_1+\omega_2+\Delta)
 (\omega_1+\omega_2)-\omega_1\, \omega_2}
{\omega_1^2\,\omega_2^2\,(\omega_1+\Delta)
(\omega_2+\Delta)(\omega_1+\omega_2)} \ ,
\end{equation}
 \begin{eqnarray}
 f_2^\Delta(\omega_1,\omega_2,\omega_3) &=&\left[\omega_1^2\,\omega_2^2\,\omega_3^2\,
 (\omega_1+\Delta) (\omega_2+\Delta)(\omega_3+\Delta) (\omega_1+\omega_2)(\omega_1+\omega_3)
 (\omega_2+\omega_3)\right]^{-1} \nonumber \\
 &\times& \Big\{ \omega_1\, \omega_2\, \omega_3 (\omega_1+\omega_2+\omega_3)^2
 +\Delta^2\,(\omega_1+\omega_2)(\omega_1+\omega_3)
 (\omega_2+\omega_3) \nonumber \\
 &+&\omega_1^2\, \omega_2^2\,(\omega_1+\omega_2) + 
 \omega_1^2\, \omega_3^2\,(\omega_1+\omega_3) +
 \omega_2^2\, \omega_3^2\,(\omega_2+\omega_3)  \nonumber \\
&+& \Delta \Big[ 3\, \omega_1\, \omega_2\, \omega_3\,(\omega_1+\omega_2+\omega_3)
+\omega_1\, \omega_2\,(\omega_1+\omega_2)^2 + 
 \omega_1\, \omega_3\,(\omega_1+\omega_3)^2 \nonumber \\
 &+& \omega_2\, \omega_3\,(\omega_2+\omega_3)^2\Big] \Big\} \ .
 \label{eq:dia_ed1}
\end{eqnarray}

\begin{figure}[bthp]
\centerline{
\includegraphics[width=4.5in]{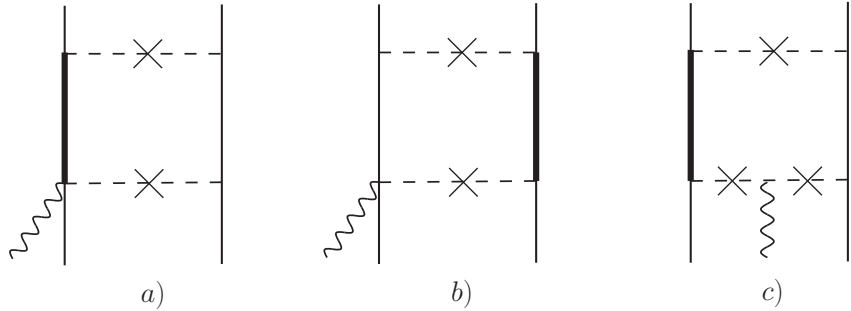}}
\caption{Direct box diagrams with a single $\Delta$ isobar in the intermediate
states.  See text for discussion.  Dashed lines with a cross represent
$v_{\Delta N}$ transition potentials, or ${\bf j}^\pi_{\Delta N}$ and 
${\bf j}^{\pi\pi}_{\Delta N}$ transition currents.}
\label{fig:fig12}
\end{figure}
At this stage, it is useful to comment on the structure of the
contributions in Eqs.~(\ref{eq:dia_cd})--(\ref{eq:dia_ed}).  The
first terms on the r.h.s.~of each of these equations are represented
by the diagrams in Fig.~\ref{fig:fig12}, {\it i.e.} (from top to
bottom) an instantaneous interaction mediated by a transition
potential, a $\Delta$-nucleon energy denominator taken in the
static limit, and a two-body current inducing excitation of a
single $\Delta$.  They have a simple interpretation~\cite{Schiavilla92}: for example,
the type c) in Fig.~\ref{fig:fig11} is the matrix element $\langle\Delta N\mid\!
{\bf j}_{\pi \Delta N}\!\mid NN\rangle$ evaluated by treating
the final $\mid \!\Delta N\rangle$ state in first order perturbation
theory,
\begin{equation}
 \mid \Delta N\rangle = \mid NN\rangle +
 \sum_{\Delta^\prime N^\prime} 
 \mid \Delta^\prime N^\prime\rangle 
 \frac{\langle \Delta^\prime N^\prime \mid v^{\Delta N}\mid
 NN\rangle}{E_{NN}-E_{\Delta^\prime N^\prime}} \simeq \left( 1 -\frac{v^\pi_{\Delta N}}
 {\Delta} \right) \mid NN\rangle  \ .
\end{equation}

One additional feature of these terms is that their configuration-space
representations are particularly simple, since they are given, for example for
type c), by the product $- v^\pi_{\Delta N}({\bf r}) \,\, {\bf j}^\pi_{\Delta N}({\bf q})/\Delta$,
and similarly for type d) and e).  Here $v^\pi_{\Delta N}({\bf r})$ and
${\bf j}^\pi_{\Delta N}({\bf q})$ are the configuration-space representations
of, respectively, the transition potential and the current in Eq.~(\ref{eq:jpid}).

The remaining terms ${\bf j}^{(-)}_{\rm c,\,d,\,e}$
represent corrections to this picture, arising from crossed-box
diagrams.  In particular, the functions $f_1^\Delta(\omega_1,\omega_2)$ and
$f_2^\Delta(\omega_1,\omega_2,\omega_3)$ give, up to pion-energy factors
from field normalizations, the sum of the energy denominators
for the six and thirty crossed-box diagrams from, respectively,
type c)-d) and e) contributions,
\begin{eqnarray}
 f_1^\Delta(\omega_1,\omega_2)&=&-\frac{1}{2\, \omega_1\, \omega_2}\, \left[ \,{\rm sum\,\, of\,\,
6\,\, type\,\, c)\,\, or \,\,d)\,\, crossed\!-\!box\,\, diagrams}\, \right] \ , 
\label{eq:f3}\\
f_2^\Delta(\omega_1,\omega_2,\omega_3)&=&\frac{1}{4\, \omega_1\, \omega_2\,\omega_3}\,
 \left[ \,{\rm sum\,\, of\,\,
30\,\, type\,\, e)\,\, crossed\!-\!box\,\, diagrams}\, \right] 
\label{eq:f4} \ .
\end{eqnarray} 

Lastly, the contributions of type f)-i) diagrams vanish, while
those of type j)-k) are written as
\begin{eqnarray}
{\rm type\,\, j)}&=& \int
\left[ -\frac{v^{{\rm c}\,\dagger}_{\Delta N} 
\, \,{\bf j}^{\pi\pi}_{\Delta N}({\bf q}_1,{\bf q}_2)}{\Delta}
+ {\bf j}^{(-)}_{\rm j}({\bf q}_1,{\bf q}_2) \right]  - {\rm h.c.}  \ , 
 \end{eqnarray}
  \begin{eqnarray}
{\rm type\,\, k)}&=&\!\!-i\,e\frac{g_A\, h_A\, D_T}{2\, F^2_\pi}\,
\int ({\bf q}_1-{\bf q}_2) \Bigg[f_1^\Delta(\omega_1,\omega_2)\, 
  ({\bm \sigma}_1\cdot {\bf q}_2) ({\bf S}_1^\dagger \cdot {\bm \sigma}_2)
  ({\bf S}_1\cdot {\bf q}_1)\nonumber \\
&\times& \bigg[{\bf T}^\dagger_1\cdot{\bm \tau}_2 \, ({\bf T}_1\times{\bm \tau}_1)_z 
-i\, \left[({\bf T}_1^\dagger\times{\bf T}_1) \times{\bm \tau}_2 \right]_z \bigg]
  + \frac{1}{\Delta\, \omega_1^2\, \omega_2^2}\,
{\bf S}^\dagger_1 \cdot {\bm \sigma}_2\, 
  ({\bf S}_1\cdot {\bf q}_1)\, ({\bm \sigma}_1\cdot {\bf q}_2) \nonumber \\
&\times& {\bf T}^\dagger_1 \cdot{\bm \tau}_2\,({\bf T}_1\times{\bm \tau}_1)_z \Bigg]\!
  -\!{\rm h.c.} \ , 
\end{eqnarray}
where the contact (momentum-independent)
transition potential $v_{\Delta N}^{\rm c}$ has been defined as
 \begin{equation}
  v^{\rm c}_{\Delta N}= D_T\, \,{\bf T}_1 \cdot{\bm \tau}_2\,\,
  {\bf S}_1 \cdot {\bm \sigma}_2 \ ,
 \end{equation}
 and the current $ {\bf j}^{(-)}_{\rm j}$ is given by
 \begin{eqnarray}
  {\bf j}^{(-)}_{\rm j}({\bf q}_1,{\bf q}_2)\!\!&=&\!\! i\,e\frac{g_A\, h_A}{2\, F^2_\pi}
   ({\bf q}_1-{\bf q}_2) \,f_1^\Delta(\omega_1,\omega_2)\Bigg[  v_{\Delta N}^{\rm c\, \dagger} \,
({\bf T}_1\times{\bm \tau}_2)_z ({\bm \sigma}_2\cdot{\bf q}_2)\! -\! D_{T}\Big[{\bf T}^\dagger_1\cdot{\bm \tau}_2
\,({\bf T}_1\times{\bm \tau}_2)_z \nonumber \\
&-&2\,i\,(2\, \tau_{2,z}-T^\dagger_{1,z}\, {\bf T}_1\cdot{\bm \tau}_2)\Big]
({\bm \sigma_2}\cdot{\bf q}_2)({\bf S}^\dagger_1\cdot{\bm \sigma}_2)\Bigg]({\bf S}_1\cdot{\bf q}_1)- {\rm h.c.} \ .
 \end{eqnarray}
The next set of contributions we consider includes two $\Delta$
isobars in the intermediate states, the relevant diagrams are displayed
in Fig.~\ref{fig:fig13}.  In analogy to Eqs.~({\ref{eq:dia_cd})--(\ref{eq:dia_ed}),
we write:
\begin{eqnarray}
{\rm type\,\, l)}&=& \int
\left[ -\frac{v^{\pi\, \dagger}_{\Delta \Delta} ({\bf q}_2) \, {\bf j}^\pi_{\Delta \Delta}
({\bf q}_1)}{2\, \Delta} +{\bf j}^{(-)}_{\rm l}({\bf q}_1,{\bf q}_2) \right] - {\rm h.c.}  \ ,
 \label{eq:dia_add} \\
 {\rm type\,\, m)}&=&\int
\left[ -\frac{v^{\pi\, \dagger}_{\Delta \Delta} ({\bf q}_2) \, {\bf j}^{\pi \pi}_{\Delta \Delta}
({\bf q}_1,{\bf q}_3)}{2\, \Delta} +{\bf j}^{(-)}_{\rm m}({\bf q}_1,{\bf q}_2,{\bf q}_3) \right] 
- {\rm h.c.}  \ ,\label{eq:dia_bdd}
 \end{eqnarray}
where the transition potential $v^\pi_{\Delta\Delta}$ and currents
${\bf j}^\pi_{\Delta \Delta}$ and $ {\bf j}^{\pi \pi}_{\Delta \Delta}$ are
obtained from Eqs.~(\ref{eq:vnd})--(\ref{eq:jpipid}) by the replacements
$g_A \rightarrow h_A$ and ${\bm \sigma}_2 
\,\, ({\bm \tau}_2) \rightarrow {\bf S}_2 \,\,({\bf T}_2)$, and
\begin{eqnarray}
 {\bf j}^{(-)}_{\rm l}({\bf q}_1,{\bf q}_2)&=&- v^{\pi\, \dagger}_{\Delta \Delta} ({\bf q}_2)\,  
 {\bf j}^{\pi }_{\Delta \Delta}({\bf q}_1) \left[ f_3^\Delta(\omega_1,\omega_2) -\frac{1}{2\, \Delta}\right]
-i\, e\frac{h_A^4}{2\, F_\pi^4}\,f_4^\Delta(\omega_1,\omega_2)
({\bf S}_1^\dagger \cdot {\bf q}_2)\, {\bf S}_1 \nonumber \\
 &\times& \Bigg[ ({\bf S}_2^\dagger\cdot{\bf q}_1) ({\bf S}_2 \cdot {\bf q}_2)\,
\left[({\bf T}_1^\dagger\times{\bf T}_2^\dagger)_z{\bf T}_1\cdot{\bf T}_2
-2\,({\bf T}_1^\dagger\times{\bf T}_1)_z+{\bf T}_2^\dagger\cdot({\bf T}_1^\dagger\times{\bf T}_1)\, T_{2,z}\right]\nonumber \\
&-& ({\bf S}_2^\dagger\cdot{\bf q}_2) ({\bf S}_2 \cdot {\bf q}_1)\, 
{\bf T}_1^\dagger\cdot{\bf T}_2^\dagger({\bf T}_1\times{\bf T}_2)_z \Bigg] \ ,
 \label{eq:jm-}\\
\!\!\!\!{\bf j}^{(-)}_{\rm m}({\bf q}_1,{\bf q}_2,{\bf q}_3)\!\!&=&\!\! -
  v^{\pi\, \dagger}_{\Delta \Delta}({\bf q}_2)\,  
 {\bf j}^{\pi \pi}_{\Delta \Delta}({\bf q}_1,{\bf q}_3)
 \left[ f_5^\Delta(\omega_1,\omega_2,\omega_3) -\frac{1}{2\, \Delta}\right] \nonumber \\
 &+&i\, e\frac{h_A^4}{2\, F_\pi^4}\,({\bf q}_1-{\bf q}_3)\,f_6^\Delta(\omega_1,\omega_2,\omega_3)\,
({\bf S}_1^\dagger \cdot {\bf q}_2)\, ({\bf S}_1 \cdot {\bf q}_1)\nonumber \\
&\times&\Bigg[ ({\bf S}_2^\dagger\cdot{\bf q}_3) ({\bf S}_2 \cdot {\bf q}_2)\,
\left[({\bf T}_1^\dagger\times{\bf T}_2^\dagger)_z{\bf T}_1\cdot{\bf T}_2
-2\,({\bf T}_1^\dagger\times{\bf T}_1)_z+{\bf T}_2^\dagger\cdot({\bf T}_1^\dagger\times{\bf T}_1)\, T_{2,z}\right]\nonumber \\
&-& ({\bf S}_2^\dagger\cdot{\bf q}_2) ({\bf S}_2 \cdot {\bf q}_3)\, 
{\bf T}_1^\dagger\cdot{\bf T}_2^\dagger({\bf T}_1\times{\bf T}_2)_z \Bigg]  \ .
\label{eq:jn-}
\end{eqnarray}
 The functions $f_3^\Delta(\omega_1,\omega_2)$ and $f_5^\Delta(\omega_1,\omega_2,\omega_3)$
 are defined as
\begin{eqnarray}
 f_3^\Delta(\omega_1,\omega_2)\!\!&=&\!\!-
\frac{\omega_1\, \omega_2}{4}\, \left[ \,{\rm sum\,\, of\,\,
12 \,\, type\,\, m)\,\, direct\,\, and\,\, crossed\!-\!box\,\, diagrams}\, \right] \ , \\
\!\!\!\!\!\!\! f_5^\Delta(\omega_1,\omega_2,\omega_3)\!\!&=&\!\!\
\frac{\omega_1\, \omega_2\,\omega_3}{8}\,
 \left[ \,{\rm sum\,\, of\,\,60 \,\, type\,\, n)\,\,  direct\,\, and\,\,
crossed\!-\!box\,\, diagrams}\, \right] \ ,
\end{eqnarray}
while $f_4^\Delta(\omega_1,\omega_2)$ and $f_6^\Delta(\omega_1,\omega_2,\omega_3)$ as in Eqs.~(\ref{eq:f3})
and~(\ref{eq:f4}), but for diagrams of type l) and m), respectively.  They are
explicitly given by
\begin{eqnarray}
 f_3^\Delta(\omega_1,\omega_2)&=&
  \frac{\omega_1\, \omega_2\, \left[2\, \Delta^3+(4\, \Delta^2+\omega_1\, \omega_2)
 (\omega_1+\omega_2)+2\, \Delta\, (\omega_1+\omega_2)^2\right]}
 {2\, \Delta\, (\omega_1+\omega_2)\, (\omega_1+\Delta)^2\, (\omega_2+\Delta)^2} \ , \\
 f_4^\Delta(\omega_1,\omega_2)&=&\frac{\Delta^2+\omega_1^2+\omega_1\,\omega_2+\omega_2^2
 +2\,\Delta\, (\omega_1+\omega_2)}
 {\omega_1\, \omega_2\, (\omega_1+\omega_2)\, (\omega_1+\Delta)^2\, (\omega_2+\Delta)^2}\ ,
\end{eqnarray}
and 
\begin{eqnarray}
\!\!\!\!\! f_5^\Delta(\omega_1,\omega_2,\omega_3)\!\!&=&\!\! \frac{\omega_1\, \omega_2\,\omega_3}{2}
 \Bigg[ \frac{(\omega_1+\omega_2+\omega_3)(\omega_1+\omega_2+\omega_3+\Delta)}
 {(\omega_1+\Delta)(\omega_2+\Delta)(\omega_3+\Delta)
 (\omega_1+\omega_2)(\omega_1+\omega_3)(\omega_2+\omega_3)} \nonumber \\
 &+&\frac{1}{\Delta\, (\omega_1+\Delta)(\omega_2+\Delta)(\omega_3+\Delta)}
 +\omega_1\, \omega_2\,\omega_3\, f_6^\Delta(\omega_1,\omega_2,\omega_3)\Bigg]\ , \\
\!\!\!\!\! f_6^\Delta(\omega_1,\omega_2,\omega_3)\!\!&=&\!\! \frac{\omega_1+\omega_2+\omega_3}
 {\omega_1\, \omega_2\, \omega_3\, (\omega_1+\omega_2)(\omega_1+\omega_3)(\omega_2+\omega_3)}
 \Bigg[ \frac{1}{(\omega_1+\Delta)^2}+\frac{1}{(\omega_2+\Delta)^2}\nonumber \\
 &+&\!\!\frac{1}{(\omega_3+\Delta)^2}\Bigg]   
 +\frac{\omega_1\, \omega_2\, \omega_3 -\Delta^2\,(\omega_1+\omega_2+\omega_3)}
 {\omega_1\, \omega_2\, \omega_3\, 
 (\omega_1+\omega_2)(\omega_1+\omega_3)(\omega_2+\omega_3)}
 \nonumber \\
&\times&\!\! \Bigg[ \frac{1}{(\omega_1+\Delta)^2(\omega_2+\Delta)^2} 
+ \frac{1}{(\omega_1+\Delta)^2(\omega_3+\Delta)^2} 
+\frac{1}{(\omega_2+\Delta)^2(\omega_3+\Delta)^2}\Bigg] \nonumber \\
\!\!&+&\!\!\frac{\Delta^4(\omega_1+\omega_2+\omega_3)+\omega_1\, \omega_2\,\omega_3
(\omega_1\, \omega_2+\omega_1\,\omega_3+\omega_2\,\omega_3-6\, \Delta^2)}
{\omega_1\, \omega_2\, \omega_3\,(\omega_1+\Delta)^2(\omega_2+\Delta)^2(\omega_3+\Delta)^2
 (\omega_1+\omega_2)(\omega_1+\omega_3)(\omega_2+\omega_3)}.
\end{eqnarray}
We observe that, in contrast to the case of a single $\Delta$,
the energy denominators for the direct and crossed box diagrams do not add up
to $1/(2\, \Delta)$---as one would have naively expected---indeed,
that is the reason for including this term
within the square brackets of Eqs.~(\ref{eq:jm-}) and~(\ref{eq:jn-}). 

Next, we consider diagrams n)-s) in Fig.~\ref{fig:fig13},
which involve contact terms with two $\Delta$'s.  The contributions of type
n)-p) vanish, while those of type q)-s) read:
\begin{eqnarray}
{\rm type \,\, q)}&=&i\,e\frac{h^2_A}{2\, F^2_\pi} 
\,({\bf T}_1^\dagger\times{\bf T}_1)_z \,
\int ({\bf q}_1-{\bf q}_2) f_4^\Delta(\omega_1,\omega_2) \Bigg[
C_{S}^\prime\, ({\bf S}^\dagger_1\cdot{\bf q}_2)\, ({\bf S}_1\cdot{\bf q}_1) \nonumber \\
&+& C_T^\prime\, ({\bf S}_1^\dagger\cdot{\bf q}_2)\,({\bm \Sigma}_1\cdot{\bm \sigma}_2)\,
({\bf S}_1\cdot{\bf q}_1)\Bigg] \ ,
\end{eqnarray}
\begin{eqnarray}
{\rm type \,\, r)}&=& -i\,e\frac{h^2_A}{2\, F^2_\pi}\frac{D_T^\prime}{\Delta}
({\bf T}_1^\dagger \cdot {\bf T}^\dagger_2) \, 
 ({\bf T}_1\times{\bf T}_2)_z\,({\bf S}_1^\dagger \cdot {\bf S}^\dagger_2)
\int ({\bf q}_1-{\bf q}_2)
 ({\bf S}_1\cdot {\bf q}_1) \, ({\bf S}_2\cdot {\bf q}_2) \nonumber \\
&\times& \frac{\omega_1+\omega_2+ \Delta}
 {\omega_1\, \omega_2 \, (\omega_1+\omega_2)(\omega_1+\Delta) (\omega_2+\Delta)} -{\rm h.c.} \ ,
 \end{eqnarray}
 \begin{eqnarray}
{\rm type \,\, s)}\!&=&\!-i\,e\frac{h^2_A\, D_T^{\prime\prime}}{2\, F^2_\pi}
 \Big[{\bf T}_1^\dagger\cdot{\bf T}_2^\dagger\, ({\bf T}_1\times{\bf T}_2)_z 
+2({\bf T}_1^\dagger\times{\bf T}_1)_z -T^\dagger_{2,z}{\bf T}_2\cdot({\bf T}_1^\dagger\times{\bf T}_1)\Big] \nonumber \\
&&\int ({\bf q}_1-{\bf q}_2) f_4^\Delta(\omega_1,\omega_2)\,
({\bf S}^\dagger_1 \cdot {\bf q}_1)\,
({\bf S}_1 \cdot {\bf S}^\dagger_2)\, ({\bf S}_2 \cdot {\bf q}_2) -{\rm h.c.} \ .
\end{eqnarray} 
\section{Currents from contact interactions}
\label{app:ct2d}

In this Appendix we list the four-nucleon contact interaction Hamiltonians involving
two gradients of the nucleon fields.  Minimal substitution,
\begin{equation}
\nabla N({\bf x}) \rightarrow \left[ \nabla
- i\, e \, e_N {\bf A}({\bf x}) \right] N({\bf x}) \ ,
\label{eq:minsub} 
\end{equation}
in the nucleon-derivative couplings then leads to the corresponding electromagnetic-interaction
Hamiltonians, which are listed as well.  In the last section of this Appendix we report the
expressions for the vertices induced by these Hamiltonians.

\subsection{Four-nucleon contact interaction Hamiltonians}
\label{sub:ct2d}

The four-nucleon contact interaction Hamiltonians with two gradients acting
on the nucleon fields have the following expressions
\cite{vanKolck94,Epelbaum98}:
\begin{eqnarray}
\label{eq:ct2d1}
H_{{\rm CT2D},1} &=& \, \, C_{1}^{\prime} \int {\rm d}{\bf x}\,
\left[ \left[N^\dagger({\bf x}) \nabla N({\bf x})\right]^{2} +
       \left[ \left[\nabla N({\bf x})\right]^\dagger N({\bf x})\right]^2  
       \right] \ , \\ 
H_{{\rm CT2D},2} &=& \, \, C_{2}^{\prime} \int {\rm d}{\bf x}\,
\left[N^\dagger({\bf x}) \nabla N({\bf x})\right] \cdot
\left[  \left[\nabla N({\bf x})\right]^\dagger N({\bf x})\right] \ ,\\
H_{{\rm CT2D},3} &=& \, \, C_{3}^{\prime} \int {\rm d}{\bf x}\, 
\left[ N^\dagger({\bf x})N({\bf x}) \right] \left[N^\dagger({\bf x}) \nabla^2 N({\bf x}) + 
\left[\nabla^2 N({\bf x})\right]^\dagger N({\bf x})\right] \ ,\\
H_{{\rm CT2D},4} &=& i \, C_{4}^{\prime} \int {\rm d}{\bf x}\, 
\bigg[\! \left[ N^\dagger({\bf x})\nabla N({\bf x}) \right] \cdot 
\left[ \left[\nabla N({\bf x})\right]^\dagger \times {\bm \sigma} N({\bf x}) \right] \nonumber \\
&&  \quad \quad \quad \, \,  + \left[ \left[\nabla N({\bf x})\right]^\dagger N({\bf x})\right] \cdot 
\left[ N^\dagger({\bf x}){\bm \sigma }\times \nabla N({\bf x})\right] \!\bigg] \ , \\
H_{{\rm CT2D},5} &=& i \, C_{5}^{\prime} \int {\rm d}{\bf x}\,
\left[ N^\dagger({\bf x}) N({\bf x}) \right] \left[ \left[\nabla N({\bf x})\right]^\dagger \cdot 
{\bm \sigma} \times \nabla N({\bf x}) \right] \ , \\
H_{{\rm CT2D},6} &=& i \, C_{6}^{\prime} \int {\rm d}{\bf x}\,
\left[ N^\dagger({\bf x}) {\bm \sigma} N({\bf x})\right] \cdot 
\left[ \left[\nabla N({\bf x})\right]^\dagger \times \nabla N({\bf x}) \right] \ , \\
H_{{\rm CT2D},7} &=& \, \, 
\left( C_{7}^{\prime} \delta_{ik}\delta_{jl}+C_{8}^{\prime}\delta_{il}\delta_{kj}+C_{9}^{\prime}\delta_{ij}\delta_{kl}\right) \nonumber\\
&& \quad \, \, \, \, \, \int {\rm d}{\bf x}\, \, \bigg[\!
\left[N^\dagger({\bf x}) \sigma_k \partial_i N({\bf x})\right] \left[N^\dagger({\bf x}) \sigma_l \partial_j N({\bf x}) \right]  \nonumber \\
&& \quad \quad \quad \, \, \, +
\left[ \left[\partial_i N({\bf x})\right]^\dagger \sigma_k N({\bf x})\right] 
\left[ \left[\partial_j N({\bf x})\right]^\dagger \sigma_l N({\bf x})\right]\!\bigg] \ , \\
H_{{\rm CT2D},8} &=& \, \,\left(
C_{10}^{\prime}\delta_{ik}\delta_{jl}+C_{11}^{\prime}\delta_{il}\delta_{kj}+C_{12}^{\prime}\delta_{ij}\delta_{kl}\right) \nonumber \\
&& \quad \, \, \, \, \, \int {\rm d}{\bf x}\,\, 
\left[N^\dagger({\bf x}) \sigma_k \partial_i N({\bf x})\right] \left[\left[\partial_j N({\bf x})\right]^\dagger \sigma_l N({\bf x}) \right] \ , \\
\label{eq:ct2d2}
H_{{\rm CT2D},9} &=& \left(\frac{1}{2}\,
C_{13}^{\prime}\left(\delta_{ik}\delta_{jl}+\delta_{il}\delta_{kj}\right)+
C_{14}^{\prime}\delta_{ij}\delta_{kl}\right)  \nonumber \\
&& \,\,\,  \int {\rm d}{\bf x}\,\, 
\left[ \left[\partial_i N({\bf x})\right]^\dagger \sigma_k \partial_j N({\bf x}) + 
       \left[\partial_j N({\bf x})\right]^\dagger \sigma_k \partial_i N({\bf x}) \right]
 \left[N^\dagger({\bf x})\sigma_lN({\bf x})\right]  \ .
\end{eqnarray}
\subsection{Contact electromagnetic-interaction Hamiltonians}
Minimal substitution leads to the following contact electromagnetic-interaction Hamiltonians:
\begin{eqnarray}
\label{eq:ct2d1g} 
H_{{\rm CT\gamma},1} &=& - i\, e\, C_{1}^{\prime} \int {\rm d}{\bf x}\, {\bf A}({\bf x})\cdot \bigg[\!
\left[N^\dagger({\bf x})\left(\overrightarrow{\nabla}-\overleftarrow{\nabla}\right)N({\bf x})\right]
\left[N^\dagger({\bf x}) e_N N({\bf x})\right]  \nonumber \\
&& \quad \quad \quad \quad \quad \quad \quad \quad \, 
+   \left[N^\dagger({\bf x}) e_N N({\bf x})\right]\left[N^\dagger({\bf x})\left(\overrightarrow{\nabla}-\overleftarrow{\nabla}\right)
N({\bf x})\right]\!\bigg] \ , \\ 
H_{{\rm CT\gamma},2} &=& - i\, e\, C_{2}^{\prime} \int {\rm d}{\bf x}\, {\bf A}({\bf x})\cdot \bigg[\!
\left[N^\dagger({\bf x}) e_N N({\bf x})\right]\left[ \left[\nabla N({\bf x})\right]^\dagger N({\bf x})\right]  \nonumber \\
&& \quad \quad \quad \quad \quad \quad \quad \quad \, 
-   \left[N^\dagger({\bf x}) \nabla N({\bf x})\right]\left[N^\dagger({\bf x}) e_N N({\bf x})\right]\!\bigg] \ , \\
H_{{\rm CT\gamma},3} &=& - i\, e\, C_{3}^{\prime} \int {\rm d}{\bf x}\, {\bf A}({\bf x})\cdot 
\left[2 \, N^\dagger({\bf x}) N({\bf x})\right]
\left[ N^\dagger({\bf x})\left(\overrightarrow{\nabla}-\overleftarrow{\nabla}\right) e_N N({\bf x})\right] \ , \\
H_{{\rm CT\gamma},4} &=& \quad \, \, e\, C_{4}^{\prime} \int {\rm d}{\bf x}\, {\bf A}({\bf x})\cdot \bigg[\!
\left[N^\dagger({\bf x})\left(\overrightarrow{\nabla}+\overleftarrow{\nabla}\right)N({\bf x})\right] \times
\left[N^\dagger({\bf x}) {\bm \sigma}\, e_N N({\bf x})\right]  \nonumber \\
&& \quad \quad \quad \quad \quad \quad \quad \quad \, 
+   \left[N^\dagger({\bf x})\, e_N N({\bf x})\right]\left[N^\dagger({\bf x})\left(\overrightarrow{\nabla}+\overleftarrow{\nabla}\right)
\times {\bm \sigma}N({\bf x})\right]\!\bigg] \ , \\
H_{{\rm CT\gamma},5} &=& \quad \, \, e\, C_{5}^{\prime} \int {\rm d}{\bf x}\, {\bf A}({\bf x})\cdot 
\left[N^\dagger({\bf x})N({\bf x})\right] 
\left[N^\dagger({\bf x})\left(\overrightarrow{\nabla}+\overleftarrow{\nabla}\right)\times {\bm \sigma}\, e_N N({\bf x})\right] \ \ , \\
H_{{\rm CT\gamma},6} &=& \quad \, \, e\, C_{6}^{\prime} \int {\rm d}{\bf x}\, {\bf A}({\bf x})\cdot 
\left[N^\dagger({\bf x}){\bm \sigma}N({\bf x})\right] \times
\left[N^\dagger({\bf x})\left(\overrightarrow{\nabla}+\overleftarrow{\nabla}\right) e_N N({\bf x})\right] \ \ , \\
H_{{\rm CT\gamma},7} &=& \, -i \, e \,
\left( C_{7}^{\prime} \delta_{ik}\delta_{jl}+C_{8}^{\prime}\delta_{il}\delta_{kj}+C_{9}^{\prime}\delta_{ij}\delta_{kl}\right) \nonumber\\
&& \quad \quad \, \, \, \, \, \, \, \,  \int {\rm d}{\bf x}\, \, \bigg[\!
A_j({\bf x})\left[N^\dagger({\bf x}) \left(\overrightarrow{\partial_i}-\overleftarrow{\partial_i}\right)\sigma_k N({\bf x})\right]
\left[N^\dagger({\bf x})\sigma_l\, e_N N({\bf x})\right]  \nonumber \\
&& \quad \quad \quad \quad \quad \, \, \, +
A_i({\bf x})\left[N^\dagger({\bf x})\sigma_k\, e_N N({\bf x})\right]
\left[N^\dagger({\bf x}) \left(\overrightarrow{\partial_j}-\overleftarrow{\partial_j}\right)\sigma_l N({\bf x})\right] \! \bigg] \ , \\
H_{{\rm CT\gamma},8} &=&  \, \, \, \, \, \, i \, e \,
\left(C_{10}^{\prime}\delta_{ik}\delta_{jl}+C_{11}^{\prime}\delta_{il}\delta_{kj}+C_{12}^{\prime}\delta_{ij}\delta_{kl}\right) \nonumber \\
&& \quad \quad \, \, \, \, \, \, \, \,  \int {\rm d}{\bf x}\, \, \bigg[\!
A_j({\bf x})\left[N^\dagger({\bf x}) \sigma_k \partial_i N({\bf x})\right]\left[N^\dagger({\bf x})\sigma_l\, e_N N({\bf x})\right]  \nonumber \\
&& \quad \quad \quad \quad \quad \, \, \, -
A_i({\bf x})\left[N^\dagger({\bf x})\sigma_k\, e_N N({\bf x})\right]
\left[\left[\partial_j N({\bf x})\right]^\dagger \sigma_l N({\bf x})\right] \! \bigg] \ ,\\
\label{eq:ct2d2g}
H_{{\rm CT\gamma},9} &=& \, \, \, \, \, \, i\, e \, \left(\frac{1}{2}\,
C_{13}^{\prime}\left(\delta_{ik}\delta_{jl}+\delta_{il}\delta_{kj}\right)+C_{14}^{\prime}\delta_{ij}\delta_{kl}\right)  \nonumber \\
&& \quad \, \, \, \quad \quad \int {\rm d}{\bf x}\,\, \bigg[\!
A_j({\bf x})\left[N^\dagger({\bf x}) \left(\overrightarrow{\partial_i}-\overleftarrow{\partial_i}\right)\sigma_k\, e_N N({\bf x})\right]
\left[N^\dagger({\bf x})\sigma_l N({\bf x})\right]  \nonumber \\
&& \quad \quad \quad \quad \quad \, \, \, \,  +
A_i({\bf x})\left[N^\dagger({\bf x}) \left(\overrightarrow{\partial_j}-\overleftarrow{\partial_j}\right) \sigma_k\, e_N N({\bf x})\right]
\left[N^\dagger({\bf x})\sigma_l N({\bf x})\right] \! \bigg] \ .
\end{eqnarray}

\subsection{Contact interaction vertices}

The vertices induced by the contact electromagnetic-interaction
Hamiltonians are listed below.  The notation is the same as in
Appendix~\ref{app:vert}, but for
\begin{equation}
\langle H_{{\rm CT}\gamma,i}\rangle
\equiv  \langle {\bf p}_1^\prime ,\chi_1^\prime;
         {\bf p}_2^\prime, \chi_2^\prime\mid H_{{\rm CT\gamma},i}\mid {\bf p}_1 ,\chi_1;
          {\bf p}_2 ,\chi_2 ; {\bf q},\lambda \rangle \ , \qquad i=1,\dots,9 \ ,
\end{equation}
and
\begin{eqnarray}
 \langle H_{{\rm CT\gamma},1}\rangle &=& \, \,
2 \, e\, C_1^{\prime}\bigg[e_1\left({\bf p}_2 +{\bf p}_2^{\prime}\right) 
+                        e_2\left({\bf p}_1 +{\bf p}_1^{\prime}\right)\bigg] \cdot 
\frac{\hat{\bf e}_{{\bf q} \lambda}} {\sqrt{2\, \omega}_q}\ , \\
\langle  H_{{\rm CT\gamma},2} \rangle &=&  
-\, e\, C_2^{\prime}\bigg[e_1\left({\bf p}_2 +{\bf p}_2^{\prime}\right) 
+                       e_2\left({\bf p}_1 +{\bf p}_1^{\prime}\right)\bigg] \cdot 
\frac{\hat{\bf e}_{{\bf q} \lambda}} {\sqrt{2\, \omega}_q}\ , \\
\langle H_{{\rm CT\gamma},3} \rangle &=& \, \,
2\, e\, C_3^{\prime}\bigg[e_1\left({\bf p}_1 +{\bf p}_1^{\prime}\right) 
+                       e_2\left({\bf p}_2 +{\bf p}_2^{\prime}\right)\bigg] \cdot 
\frac{\hat{\bf e}_{{\bf q} \lambda}} {\sqrt{2\, \omega}_q}\ , \\
\langle H_{{\rm CT\gamma},4} \rangle &=& 
-i\, e\, C_4^{\prime}\left({\bm \sigma}_1+{\bm \sigma}_2\right)\times 
\bigg[e_1\left({\bf p}_2 -{\bf p}_2^{\prime}\right) 
+   e_2\left({\bf p}_1 -{\bf p}_1^{\prime}\right)\bigg] \cdot 
\frac{\hat{\bf e}_{{\bf q} \lambda}} {\sqrt{2\, \omega}_q}\ , \\
\langle H_{{\rm CT\gamma},5} \rangle &=&  -i\, e\, C_5^{\prime}
\bigg[e_1 \,{\bm \sigma}_1\times \left({\bf p}_1 -{\bf p}_1^{\prime}\right) 
   +  e_2 \,  {\bm \sigma}_2 \times \left({\bf p}_2 -{\bf p}_2^{\prime}\right)\bigg] \cdot 
\frac{\hat{\bf e}_{{\bf q} \lambda}} {\sqrt{2\, \omega}_q}\ , \\
\langle H_{{\rm CT\gamma},6} \rangle &=& \quad  \!
  i\, e\, C_6^{\prime}
\bigg[e_1 \,{\bm \sigma}_2\times \left({\bf p}_1 -{\bf p}_1^{\prime}\right)   +  e_2 \, 
           {\bm \sigma}_1 \times \left({\bf p}_2 -{\bf p}_2^{\prime}\right)\bigg] \cdot 
\frac{\hat{\bf e}_{{\bf q} \lambda}} {\sqrt{2\, \omega}_q}\ , \\
\langle H_{{\rm CT\gamma},7} \rangle &=&
 2\, e\, \bigg[
C_7^{\prime} [ e_1\left({\bf p}_2 + {\bf p}_2^{\prime}\right)\cdot{\bm \sigma}_2 \, {\bm \sigma}_1 +
               e_2\left({\bf p}_1 + {\bf p}_1^{\prime}\right)\cdot{\bm \sigma}_1 \, 
	       {\bm \sigma}_2 ]\nonumber \\
&&\quad + C_8^{\prime} [ e_1\left({\bf p}_2 + {\bf p}_2^{\prime}\right)\cdot{\bm \sigma}_1 \, {\bm \sigma}_2 +
               e_2\left({\bf p}_1 + {\bf p}_1^{\prime}\right)\cdot{\bm \sigma}_2 \, 
	       {\bm \sigma}_1 ]\nonumber \\
&& \quad + C_9^{\prime}\,{\bm \sigma}_1\cdot{\bm \sigma}_2 
[e_1\left({\bf p}_2 + {\bf p}_2^{\prime}\right)+ 
 e_2\left({\bf p}_1 + {\bf p}_1^{\prime}\right) ]\bigg]
\cdot \frac{\hat{\bf e}_{{\bf q} \lambda}} {\sqrt{2\, \omega}_q}\ , \\
\langle H_{{\rm CT\gamma},8} \rangle &=& -  e\, \bigg[
C_{10}^{\prime} [ e_1\left({\bf p}_2 + {\bf p}_2^{\prime}\right)\cdot{\bm \sigma}_2 \, {\bm \sigma}_1 +
                  e_2\left({\bf p}_1 + {\bf p}_1^{\prime}\right)\cdot{\bm \sigma}_1 \, 
		  {\bm \sigma}_2 ]
  \nonumber \\
&& \quad + C_{11}^{\prime} [ e_1\left({\bf p}_2 + {\bf p}_2^{\prime}\right)\cdot{\bm \sigma}_1 \, {\bm \sigma}_2 +
                  e_2\left({\bf p}_1 + {\bf p}_1^{\prime}\right)\cdot{\bm \sigma}_2 \, 
		  {\bm \sigma}_1 ] 
\nonumber \\
&& \quad + C_{12}^{\prime}\,{\bm \sigma}_1\cdot{\bm \sigma}_2 
[e_1\left({\bf p}_2 + {\bf p}_2^{\prime}\right)+ 
 e_2\left({\bf p}_1 + {\bf p}_1^{\prime}\right) ]\bigg]
\cdot \frac{\hat{\bf e}_{{\bf q} \lambda}} {\sqrt{2\, \omega}_q}\ , \\
\langle H_{{\rm CT\gamma},9} \rangle &=& -  e\, \bigg[
C_{13}^{\prime}\big[ e_1\left({\bf p}_1 + {\bf p}_1^{\prime}\right)\cdot {\bm \sigma}_1 \,{\bm \sigma}_2 +
                      e_1\left({\bf p}_1 + {\bf p}_1^{\prime}\right)\cdot {\bm \sigma}_2 \,{\bm \sigma}_1 \nonumber \\
 &&\quad \quad \, \, \, \, \, \,  +            e_2\left({\bf p}_2 + {\bf p}_2^{\prime}\right)\cdot {\bm \sigma}_1 \,{\bm \sigma}_2 + 
                      e_2\left({\bf p}_2 + {\bf p}_2^{\prime}\right)\cdot {\bm \sigma}_2 \,{\bm \sigma}_1 \big] \nonumber \\
 && \quad \! + 2\, C_{14}^{\prime}\,{\bm \sigma}_1\cdot{\bm \sigma}_2 
[e_1\left({\bf p}_1 + {\bf p}_1^{\prime}\right)+ 
 e_2\left({\bf p}_2 + {\bf p}_2^{\prime}\right) ]\bigg]
\cdot \frac{\hat{\bf e}_{{\bf q} \lambda}} {\sqrt{2\, \omega}_q}\ .
\end{eqnarray}
\end{document}